\documentclass[fleqn,usenatbib]{mnras}

\usepackage{newtxtext,newtxmath}

\usepackage[T1]{fontenc}

\DeclareRobustCommand{\VAN}[3]{#2}
\let\VANthebibliography\thebibliography
\def\thebibliography{\DeclareRobustCommand{\VAN}[3]{##3}\VANthebibliography}

\usepackage{graphicx}	
\usepackage{amsmath}	
\usepackage{etoolbox}
\usepackage{xspace}
\usepackage{booktabs}
\usepackage{array}
\usepackage[capitalise]{cleveref}

\newcommand*{\newacronym}[2]{#2 (#1)}
\newcommand*{\acronym}[1]{#1}

\newcommand*{\swn}[2][]{\texttt{#2}\ifblank{#1}{}{\footnote{\url{#1}}}}
\newcommand*{\Nbody}{\hbox{$N$-body}\xspace}

\newcommand*{\LOSVD}[2][]{\ensuremath{\mathcal{L}\ifblank{#1}{}{^{(\text{#1})}}_{#2}}}
\newcommand*{\AICp}{\ensuremath{\text{AIC}_p}\xspace}

\newcommand*{\pder}[2]{\ensuremath{\frac{\partial#1}{\partial #2}}}
\newcommand*{\hess}[3]{\ensuremath{\frac{\partial^{2}#1}{\partial #2 \partial #3}}}
\newcommand*{\vphiII}{\ensuremath{\bigl\langle v_\phi^2\bigr\rangle^{1/2}}\xspace}

\title[Pattern speeds of face-on barred galaxies]{Recovering pattern speeds of simulated face-on barred galaxies via Schwarzschild modelling}

\author[I. Tikhonenko et al.]{
I.~S.~Tikhonenko,$^{1}$\thanks{E-mail: tikhonenko@mpe.mpg.de (MPE)}
J.~Thomas,$^{1,2}$
R.~P.~Saglia$^{1,2}$
\\
$^{1}$Max-Planck-Institut für Extraterrestriche Physik, Giessenbach-Str. 1, D-85748 Garching, Germany\\
$^{2}$Universitäts-Sternwarte München, Scheinerstrasse 1, D-81679 München, Germany\\
}

\date{Accepted XXX. Received YYY; in original form ZZZ}

\pubyear{2026}

\begin{document}
\label{firstpage}
\pagerange{\pageref{firstpage}--\pageref{lastpage}}
\maketitle

\begin{abstract}
Stellar bars are a major driving force in the secular evolution of their host galaxies.
{To better understand the connections between the 3D bar density structure, its orbital composition, stellar populations, and underlying dark matter distribution, it is desirable to construct detailed dynamical models of barred galaxies.
However, only a few external barred galaxies have been studied in this way so far.}
We present a new Schwarzschild orbit superposition code for triaxial potentials with figure rotation and test it extensively using mock data from an \Nbody simulation of a strongly barred galaxy.
We investigate the recovery of model parameters in the nearly face-on case {which was not previously considered}. In particular, we demonstrate a $10\%$ accuracy of both the pattern speed and mass-to-light ratio and a $20\%$ accuracy for the dark matter halo mass scaling at the inclination $20^\circ$. 
Surprisingly, we obtain a similarly {accurate} result for the pattern speed in an exact face-on limit, where 
conventional methods such as Tremaine-Weinberg are no longer applicable. 
This result suggests that varying the pattern speed at fixed bar length, corresponding to the transition between slow and fast bars, alters the distribution function in a way that produces a systematic change in the vertical velocity distribution, which can not be compensated by the in-plane velocity components.
\end{abstract}



\section{Introduction}

Stellar bars are ubiquitous in the Local Universe. They are found in between $1/3$ and $2/3$ of all nearby disc galaxies \citep[for example,][]{Sheth_etal2008,Buta_etal2015,Erwin2018}. 
At close to edge-on orientations, they manifest themselves as so-called \newacronym{B/PS}{boxy/peanut shaped} bulges,
which are the inner parts of  bars that have grown in the vertical direction due to {the} buckling instability (see \citealp{Athanassoula2016} and references therein).
Bars play a key role in the secular evolution of galactic discs which they inhabit.
In particular, \Nbody simulations revealed that bars influence their host galaxy via the outward
redistribution of {the} angular momentum by particles trapped at Lindblad resonances
\citep{Athanassoula2003,Weinberg_Katz2007a,Weinberg_Katz2007b,Ceverino_Klypin2007,Sellwood2014}:
as the bar passes its angular momentum to the halo, its pattern speed decreases, while its
amplitude grows (however this is not always the case when the dissipative component is present in the simulation, for example see \citealp{Bland-Hawthorn_etal2024}).
How exactly this dynamical friction mechanism works is still a very active area of research (e.g.~\citealp{Chiba_Schonrich2022}, see also \citealp{Hamilton_etal2023} for an interesting analogy with plasma physics). 

One of the most important resonances in barred galaxies is the corotation, where the pattern speed of the bar matches the circular velocity of stars. {A corotation radius $R_c$ is defined as the radius where $\Omega_p(R_c) = v_{\text{circ}}/R_c$}. Using the dimensionless ratio $\mathcal{R}$ of the corotation radius to the size of the bar, bars are usually divided into `slow' ($\mathcal{R} > 1.4$) and `fast' ($\mathcal{R} < 1.4$). 
While observations of Local Universe galaxies suggest bars to be usually fast, most cosmological simulations (EAGLE, \citealp{Schaye_etal2015}; TNG100 
\citealp{Marinacci_etal2018,Naiman_etal2018,Nelson_etal2018,Pillepich_etal2018,Springel_etal2018}; TNG50, \citealp{Nelson_etal2019,Pillepich_etal2019}) rather puzzlingly have much larger numbers of slow bars \citep{Aguerri_etal2015,Algorry_etal2017,Zhao_etal2020,Frankel_etal2022,Cuomo_etal2019,Roshan_etal2021}. A notable exception from this trend is the Auriga simulations suite \citep{Grand_etal2017}, where bars remain fast throughout their evolution, because the galaxies are more baryon-dominated compared to other simulations \citep{Fragkoudi_etal2021}.

In any case, to classify {a} bar as `slow' or `fast', one needs to measure its pattern speed.  In \Nbody simulations, the distribution function is essentially known at each moment of time, therefore there are a lot of good ways to determine it (see \citealp{Pfenniger_etal2023} for a review of methods and \citealp{Dehnen_etal2022} for a practical and robust implementation).
For {external galaxies}, however, usually only surface brightness and the \newacronym{LOS}{line-of-sight} velocity component distribution {are} available at a subset of a sky plane.  Therefore, the \newacronym{TW}{Tremaine-Weinberg} 
method
{is widely used because it is relatively straightforward to implement and largely independent of model assumptions}
\citep{Tremaine_Weinberg1984,Merrifield_Kuijken1995,Meidt_etal2008}.
However, its performance relies on how well the tracer population obeys {the} continuity equation, and how well the slits are positioned over it. {This means the tracer’s total mass must be conserved, and it cannot change its form or brightness drastically as it rotates.} For example, \cite{Williams_etal2021} found that using clumpy ISM tracers results in {a} 20\%-40\% overestimate of the pattern speed depending on the tracer. The method is very sensitive with respect to the correct determination of the disc {position} angle, and produces best results for intermediate inclinations ($i$) and intermediate {position} angles of the bar \citep{Guo_etal2019,Zou_etal2019,Garma-Oehmichen_etal2020,Garma-Oehmichen_etal2022,Geron_etal2023}. For example, \citet{Zou_etal2019} show that in their simulations the disc PA misalignments {as little as $5^\circ$} can lead to 35\% systematic errors in the $\Omega_p$, and uncertainties can be generally kept below $10\%$ only when $15^{\circ} \lesssim i \lesssim 70^{\circ}$ and $10^\circ \lesssim \text{PA}_{\text{bar}} \lesssim 75^{\circ}$. \citet{Borodina_etal2023} argue that an even narrower range of disc inclinations $[5^{\circ}; 50^{\circ}]$ is necessary for an accurate recovery.

The \acronym{TW} method is limited to the first moments of the \acronym{LOS} velocity distribution measured along the slits.
At the same time, 
the advance in instrumentation in the past decades resulted in development of \newacronym{IFU}{integral-field units}, which can measure the spectra in patches covering the entire image plane and there is a wealth of data coming from the surveys such as $\text{ATLAS}^\text{3D}$ \citep{Cappellari_etal2011}, SAMI \citep{Croom_etal2021}, CALIFA \citep{Sanchez_etal2012}, MaNGA \citep{Bundy_etal2015}, MASSIVE \citep{Ma_etal2014}, and TIMER \citep{Gadotti_etal2019}.
Therefore, an attractive alternative to the \acronym{TW} method which is also free from {position} angle misalignment biases {is} to construct a dynamical model of the object in question using the full available information about stellar kinematics and then recover the pattern speed as one of the model parameters.

\citet{Schwarzschild1979} proposed a very generic method to construct distribution functions of stationary systems. Since the integrals of motion are conserved along the orbit, following Jeans' theorem one can produce the desired distribution function as a weighted sum of orbits by solving an optimisation problem \citep{Pfenniger1984a,Richstone_Tremaine1985,Zhao1996b}. The models obtained in this way can then be compared in terms of how well {their} luminosity density and kinematics reproduce the observational data \citep{Rix_etal1997}. The initial spherical-symmetry assumption was later generalized to be able to model axisymmetric systems \citep{Marel_etal1998,Cretton_etal1999,Gebhardt_etal2000a,Thomas_etal2004,Valluri_etal2004}.
Using full \newacronym{LOSVD}{line-of-sight velocity distributions} or their high-order moments in the modelling breaks the mass-anisotropy degeneracy, which together with high-resolution photometry data from e.g.~{Hubble Space Telescope} and long-slit or, more recently, \acronym{IFU} spectroscopy, allowed these codes to be used for dynamical measurements of central \newacronym{SMBH}{supermassive black hole} masses of the galaxies beyond the Local Group
{\citep{%
Gebhardt_etal2003,Gebhardt_etal2007,Valluri_etal2005,Nowak_etal2007,Nowak_etal2008a,Cappellari_etal2009,Krajnovic_etal2009,Gultekin_etal2009a,Rusli_etal2013}}.
{However, assuming axial symmetry can introduce a significant bias in dynamical mass measurements, if the system is actually triaxial \citep{Thomas_etal2007,denBrok_etal2021,Thater_etal2026}.} 
{Therefore a newer ``generation'' of Schwarzschild codes is currently applied, which have in common that they all lift the assumption of axial symmetry to avoid the respective biases, but differ significantly in the employed methods, such as deprojection technique, orbit sampling, and model evaluation  
\citep{vandenBosch_etal2008,vandenBosch_vandeVen2009,Neureiter_etal2021,deNicola_etal2022,Thater_etal2023,Pilawa_etal2024}.}

A barred galaxy can be considered a triaxial system in the rotating frame of reference aligned with the bar. Hence, 
a triaxial Schwarzschild modelling code {can be extended} to include figure rotation of the potential by taking into account the fictitious forces, acting in the non-inertial frame of the rotating bar. {Contrary to Milky Way studies, where triaxial dynamical models of the bar have been constructed for at least past 30 years \citep{Zhao1996b,Hafner_etal2000,Wang_etal2012,Wang_etal2013,Khoperskov_etal2025}, the Schwarzschild modelling of external barred galaxies is still a relatively new field.
  \citet{Vasiliev_Valluri2020} and, subsequently, \citet{Tahmasebzadeh_etal2022} presented codes 
which are capable to recover the parameters of \Nbody simulation mock data of galaxies at edge-on \citep{Dattathri_etal2024} and intermediate inclinations \citep{Tahmasebzadeh_etal2022}. Using the latter of these codes,} \cite{Jin_etal2025a} were able to recover the pattern speed of edge-on barred galaxies from Auriga simulation suite with $\approx 10\%$ accuracy. \citet{Tahmasebzadeh_etal2024} presented an estimate of the bar pattern speed in NGC 4371 (inclination $\sim 60^\circ$, bar {position} angle $\sim 14^{\circ}$) with a comparable precision.

However, up to now, there was little discussion of the possibility of the pattern speed recovery {of galaxies in} nearly face-on orientations, as well as how much the pattern speed recovery depends on the actual kinematics and how much it is constrained by the density alone. This discussion is, nevertheless, necessary in the light of large photometric surveys, such as Euclid or LSST, which are expected to deliver large amounts of high-resolution data for many nearly face-on barred galaxies, detected via automatic morphology classifications tools \citep[see e.g.][]{ECQ1bars_etal2025}.

In this work, we address this question and present a third dynamical modelling code suitable for potentials with figure rotation, based on \swn{SMART} \citep{Neureiter_etal2021}.
We construct dynamical models using mock \Nbody data of a strongly barred galaxy at a nearly face-on orientation and demonstrate the sufficiently accurate recovery of model parameters by the new code.


\section{Triaxial code adaption for rotating potentials}
\label{sec:code}
\subsection{Orbit integration}
The total {gravitational} potential of the dynamical model in the inertial frame is a sum of stellar potential,
dark matter potential, and Keplerian potential of the central black hole
\begin{equation}
  \Phi = \Phi_* + \Phi_\text{DM} + \Phi_\text{BH}.
\end{equation}
The stellar potential is computed by numerically solving Poisson equations using an expansion into spherical harmonics of the stellar density provided on a 3D spherical grid. $\Phi_\text{DM}$ is either one of parametric dark halo potentials, or is computed in a similar way to the stellar potential if dark matter density is provided (for example, when analysing N-body simulations, see \citealp{Neureiter_etal2021}). The total acceleration is given by a sum of gradients of the component potentials
\begin{equation}
	\mathbf{a} = - \nabla\Phi_* - \nabla\Phi_{\text{DM}} - \nabla\Phi_{\text{BH}},
\end{equation}
which are either analytic expressions, or derived by the Poisson solver in case of non-parametric densities.

A more practical frame for modelling barred galaxies is a rotating frame where the $z$-axis is perpendicular to the disc plane, while the $x$-axis is aligned with the bar major axis.
Assuming a constant pattern speed $\Omega_p$, a barred galaxy can be considered a stationary triaxial system, therefore, a triaxial Schwarzschild code would be able to produce a model of it given that inertial forces are correctly taken into account.
The total potential would get an additional centrifugal component
\begin{align}
  \Phi_{\text{eff}} &= \Phi_* + \Phi_\text{DM} + \Phi_\text{BH} + \Phi_{\text{cf}}, \\
  \Phi_\text{cf} &= - \tfrac12 \, \Omega_p^2\, R^2,
\end{align}
where $R^2 = x^2 + y^2$, while the total acceleration would change to
\begin{align}
  \mathbf{a} &= - \nabla\Phi_* - \nabla\Phi_{\text{DM}} - \nabla\Phi_{\text{BH}} + \mathbf{a}_{\text{cf}}  + \mathbf{a}_{\text{cor}}, \span\span \\
  \mathbf{a}_{\text{cf}} &= -\nabla \Phi_{\text{cf}} = \Omega_p^2 \mathbf{R}, &&\text{(centrifugal component)}\\
  \mathbf{a}_{\text{cor}} &= - 2\mathbf{\Omega}_p \times \mathbf{v}. &&\text{(Coriolis component)}
\end{align}
Instead of the total energy, the Jacobi integral, a combination of inertial frame total energy $E$ and inertial frame angular momentum component $L_z$, is conserved
\begin{equation}
  E_J = E - \Omega_p L_z = \tfrac12|\mathbf{v}|^2 + \Phi_{\text{eff}}.
\end{equation}
The pattern speed $\Omega_p$ thus becomes a free model parameter.

We sample the orbital initial conditions in the inertial frame, following \citet[appendix C]{Neureiter_etal2021}, which itself is a generalization of the routine described in \citet{Thomas_etal2004}.
This algorithm sets up $N_E$ energy shells and subdivides each shell into $i_E^2$ sequences (where $i_E = 1\mathop{..}N_E$), sampling $L_z$ between its minimal and maximal allowed values. Thus, each sequence is indexed by a pair $(i_E, i_{L_z})$ and has $(E^{(\text{seq})},L_z^{(\text{seq})})$ attached to it. 
{The orbital sequences themselves are integrated in parallel, however all orbits in each sequence are integrated sequentially}.
Compared to \citet{Neureiter_etal2021} we entirely omit the initial pseudorandom sampling stage and directly start filling the \newacronym{SoS}{surfaces of section}. Inside each sequence, we bin the azimuthal angle ($\phi$) into $N_\phi$ sectors, {where $N_\phi = 10$, matching the number of azimuthal bins of the library grid (see \cref{ssec:mock-data})}.
Then, we generate a new starting point with $z=0$ such that it would be placed inside a randomly selected azimuthal bin on a maximal possible distance from the already existing $z=0$ \acronym{SoS} crossings of all previously integrated orbits in $(x, y, v_r)$ space, tracked in the same $\phi$ bin.
The Cartesian coordinates of a new point $(x^{(0)}, y^{(0)}, 0)$ can be trivially converted to its spherical coordinates $(r^{(0)}, 0, \phi^{(0)})$, while the remaining velocity components are computed as following
\begin{align}
	v_\phi^{(0)} &= \frac{L_z^{(\text{seq})}}{r^{(0)}}, \\
	v_\theta^{(0)} &= \sqrt{2\, \Bigl(E - \Phi(r^{(0)}, 0, \phi^{(0)} - \bigl(v_\phi^{(0)}\bigr)^2 -\bigl(v_r^{(0)}\bigr)^2 \Bigr)}.
\end{align}
In case $\bigl(v_\theta^{(0)}\bigr)^2$ is negative, the new point is discarded.

{In total, we use $93000$ orbits to construct a dynamical model}.
{We do not use orbital dithering and each of these $93000$ orbits represents an independent building block of the final dynamical model.}
The orbital integration is done in the rotating frame via a \cite{Cash_Karp1990} adaptive Runge-Kutta recipe. We correct the initial azimuthal velocities of orbit initial conditions for the frame rotation, and integrate the orbits until they cross the $z=0$ \acronym{SoS} 100 times.
After building the orbital library, we transform all velocities back to the inertial frame, store the model luminosity density and kinematics and proceed to the next step.

\subsection{Solving for orbital weights}
The core part of any Schwarzschild code is to find a set of orbital weights $w_l$, such that the observational data is closely matched with the comparable characteristics of the final dynamical model. The data used for Schwarzschild modelling is typically a luminosity density on the grid (computed by a deprojection routine, or known exactly in case of mock data obtained from simulations), and a set of {values}, representing the \acronym{LOSVD}s on a tesselation of the field of view. The common \acronym{LOSVD} {representations} are \newacronym{GH}{Gauss-Hermite} moments (\citealp{vandenBosch_etal2008} code and its derivatives), histogram bin values on a fixed velocity grid (\citealp{Gebhardt_etal2000a,Thomas_etal2004}), or B-splines \citep{Vasiliev_Valluri2020}. Another possible non-parametric approach was suggested by \citet{Falcon-Barroso_Martig2021}. We use the representation of LOSVDs by histograms, because we find it to be the most straightforward.
Hence
\begin{align}
	p^{\text{(data)}}_j = p^{(\text{mod})}_j &= \sum_{l=1}^{N_\text{orbits}} p_{jl}^{(\text{mod})}w_l \label{eq:photocond}, \\
	\LOSVD[data]{jk} \approx \LOSVD[mod]{jk} &= \sum_{l=1}^{N_\text{orbits}} \LOSVD[mod]{jkl} w_l,
\end{align}
where $p_j$ stands for the luminosity density in a given bin, while $\LOSVD{jk}$ is a value in the $k$th velocity channel of the $j$th Voronoi bin. The index $l$ hereinafter is reserved for summation over orbits.
For kinematics, we assume a sum of the squared model residuals as a measure of a goodness of the fit
\begin{equation}
	\chi^2 = \chi^2_{\text{kin}} = \sum_{j=1}^{N_\text{tess}}\sum_{k=1}^{N_\text{vel}} \left(\frac{\LOSVD[data]{jk} - \LOSVD[mod]{jk}}{\Delta \LOSVD[data]{jk}}\right)^2,\
\end{equation}
where $N_{\text{tess}}$ is the number of Voronoi bins in the kinematic
data tessellation, while $N_{\text{vel}}$ is the number of histogram
bins in each \acronym{LOSVD}. The total number of datapoints is then
$N_{\text{tess}}\times N_{\text{vel}}$. 
To ensure we accurately reconstruct the rather complex bar 3D geometry,
we explicitly require the model photometry (luminosity density) to be exactly matching the data in
\cref{eq:photocond}. 
This condition implies that the total luminosity is conserved, thus
\begin{equation}
	\sum_{l=1}^{N_{\text{orbits}}} w_l = 1.
\end{equation}

In most applications the number of data points is smaller than the number of orbital
weights (i.e. the number of free parameters in the orbit model). The weight determination is therefore formally under-determined and in particular when the data is noisy, the distribution function associated to the orbital weights can develop stronger variations than expected for a typical galaxy. This problem is much reduced by the requirement that orbital weights have to be positive to be physically meaningful but it's hard to quantify this rigorously. In practice, regularisation has turned out very useful to obtain reasonable distribution functions \citep{Thomas_etal2005}.
We therefore employ the maximum-entropy method of \cite{Richstone_Tremaine1988} with
\begin{equation}
	S = \sum_{l=1}^{N_{\text{orbits}}} w_l \, \ln {w_l}. \label{eq:entropy}
\end{equation}
and maximize the following quantity instead of {minimizing} $\chi^2$
\begin{equation}
  \hat S = S - \alpha\chi^2, \label{eq:optfun}
\end{equation}
where $\alpha$ factor controls the amount of regularization.  The models with higher $\alpha$ better reproduce the data, but can be prone to overfitting the noise, while lower $\alpha$ results in too smooth and isotropic \acronym{DF}s. Therefore, there is no good a priori value for $\alpha$, and it has to be determined on a case-by-case basis depending on the available data and the objects under study \citep{Thomas_etal2005}.

The optimization itself is carried out via the Newton method, solving
$\nabla \hat S = 0$ with luminosity density constraints enforced via
Lagrange multipliers.
The entropy enforces the weights to be strictly positive
\begin{equation}
	\forall l=1\mathop{..}N_{\text{orbits}} \:\: w_l > 0.
\end{equation}
which guarantees that they are physically meaningful. During the numerical weight optimisation, however, negative
weights sometimes occur. While this can be avoided with a very small step-size used to update the weights
a more efficient way is to check after each Newton step if any of the weights are negative, and, set these weights to a very small positive number. 
This effectively excludes the corresponding orbits from the model, which is relatively easy to show by computing the Hessian of the target function with respect to orbital weights
\begin{equation}
	H_{l_1\,l_2} = \hess{\hat S}{w_{l_1}}{w_{l_2}} = -\frac{\delta_{l_{1}\,l_2}}{w_{l_1}} - 2\alpha\, \frac{\LOSVD[mod]{jkl_1}\LOSVD[mod]{jkl_2}}{\bigl(\Delta\LOSVD[data]{jk}\bigr)^2}\label{eq:Shathessian}
\end{equation}
The second term in the equation is bounded, therefore for a very small $w_{l_1}$ the entire Hessian would be dominated by the first term. The update step in Newton method for $w_l$ would have a form of
\begin{equation}
	H_{l'l}\Delta w_{l} = -h \pder{\hat S}{w_l'},
\end{equation}
which for a small value of $w_l'$ is equivalent to
\begin{equation}
	\Delta w_{l} \approx h w_l\pder{\hat S}{w_l} \approx 0,
\end{equation}
meaning that the weight would stay at the same small positive value it was assigned. Changing some of the weights outside of the update step does however mean that the luminosity density constraint is no longer exactly satisfied, therefore we renormalise all weights, and reduce the next Newton step size.

We start with an entropy-only model with $\alpha = 0$ and completely uniform weights distribution with each $w_l$ set to $(N_\text{orbits})^{-1}$. After a convergence is reached, we increase the regularization factor (thus increasing the contribution of the data) and redo the fit using the current solution for the weights as initial conditions for the new fit.
The justification for this approach can be found in \citet[section~5.1 and appendix~D]{Neureiter_etal2021}.
The weights obtained in the fit, the intrinsic model kinematics, as well as the final $\chi^2$ are stored for every regularization strength factor value (29 in total, the first one is pure entropy ($\alpha = 0$), therefore has no meaningful $\chi^2$ value), allowing us to compare the models with different amount of regularization in \cref{sec:results}.



\section{Mock data preparation}

\begin{figure}
  \centering
  \includegraphics[width=0.7\linewidth]{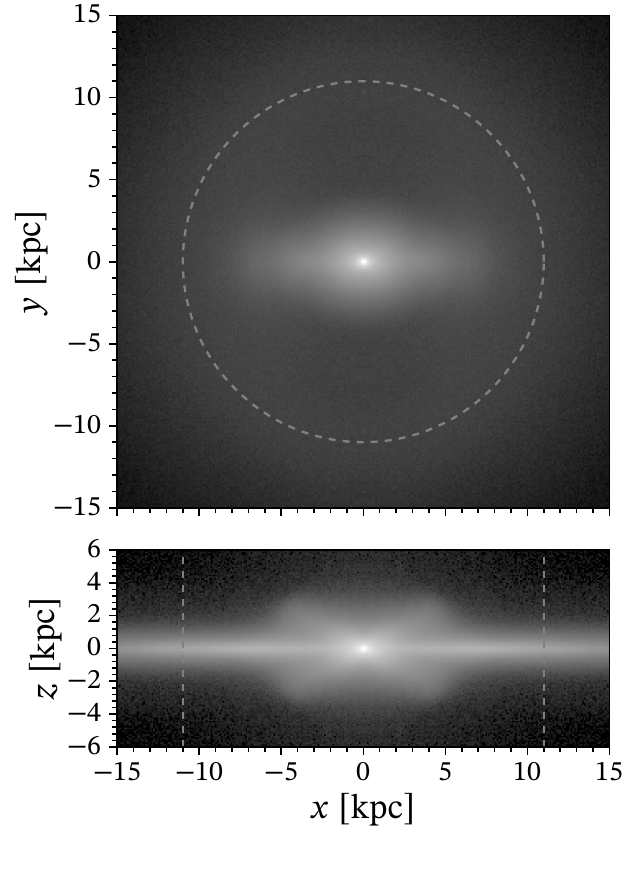}
  \caption{Face-on (\emph{top}) and side-on (\emph{bottom}) projections of the N-body system that we modelled (BLx model, snapshot at $t=450$). The gray dashed circle on the top plot and the vertical lines on the bottom plot indicate the corotation radius.}
  \label{fig:BLx-projections}
\end{figure}

\begin{figure}
  \centering
  \includegraphics[width=\linewidth]{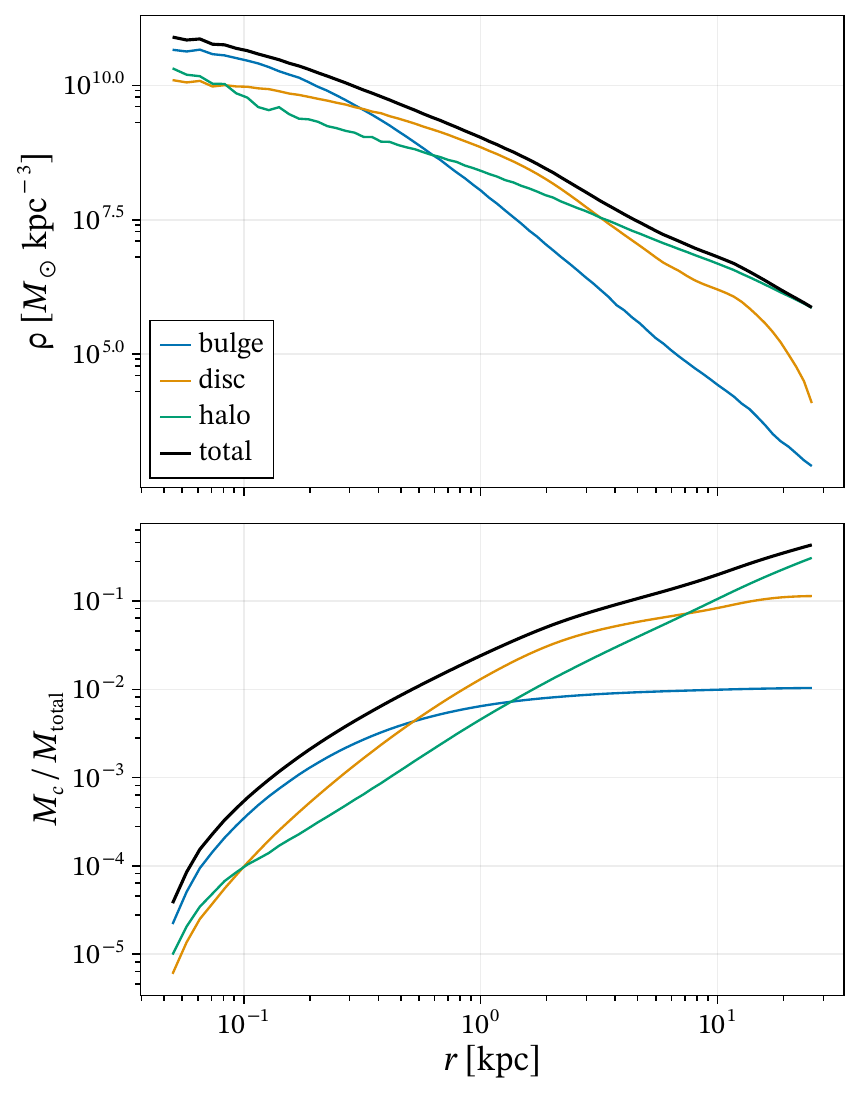}
  \caption{\emph{Top}: radial density profiles of the various \Nbody components, separated by colour; \emph{bottom}: respective cumulative mass profiles, normalized to the total mass of the system.}
  \label{fig:BLx-dens-mass-profiles}
\end{figure}

\subsection{\texorpdfstring{\Nbody}{N-body} simulation}
We pick a snapshot from the model BLx \citep{Smirnov_etal2021} for testing the code. This model has an advantage that its orbital substructures as well as their kinematic signatures were studied in great detail \citep{Smirnov_etal2021,Zakharova_etal2023,Zakharova_etal2024}. The full details of the setup and its rationale can be found in \citet{Smirnov_Sotnikova2018} and \citet{Smirnov_etal2021},
while here we provide only a brief description.

This model is initialized self-consistently using the \swn{mkgalaxy} script of \citet{McMillan_Dehnen2007} with a dynamically cold axisymmetric disc with a bulge, embedded into a generalized \newacronym{NFW}{\citealp*{NFW1997}} halo with the inner slope $\gamma_0 = 7/9$, scale radius $r_s=6 R_d$, halo transition exponent $\eta = 4/9$, and outer slope $\gamma_\infty = 31/9$. The disc has an exponential radial profile with the scale length $R_d =\ensuremath {3.5\ \text {kpc}}$. In the vertical direction, the disc is an isothermal sheet with the constant vertical scale height $z_d = 0.05\, R_d = \ensuremath {0.175\ \text {kpc}}$. The radial velocity dispersion profile also obeys an exponential law with the scaling length $R_\sigma = 2R_d$. The central dispersion is set in such a way that the Toomre parameter $Q$ value is $1.2$ at $R = 2R_d$ (see \citealp{McMillan_Dehnen2007} for details).
The total mass of the disc ($M_d$) is assumed to be \ensuremath {5\times 10^{10}\ \text {M}_\odot }. 
The bulge is described
by a \citet{Hernquist1990} profile with the scale length $r_b = 0.1 R_d = \ensuremath {0.35\ \text {kpc}}$ and mass $M_b = 0.1 M_d$.
The ``light'' component of the system (disc + bulge) is represented by 4.4 million particles in total with a softening length  $\varepsilon_d$ set to $\approx \ensuremath {13\ \text {pc}}$. The dark halo has 4.5 million particles with $\varepsilon_h\approx \ensuremath {45\ \text {pc}}$
and is allowed to self-consistently evolve with the system to facilitate the angular-momentum exchange (see e.g.~\citealt{Athanassoula2003}).
The halo profile scaling was adjusted in such a way that the total halo mass within $4R_d$ is $1.5 M_d$ (see \citealt{Smirnov_etal2021} for details).

After setting up the initial snapshot, the model was integrated for~\hbox{$\approx \ensuremath {8\ \text {Gyr}}$} with \swn{gyrfalcON} \hbox{$N$-body} integrator \citep{Dehnen2002} from the \swn{NEMO} stellar dynamics toolbox \citep{Teuben1995}. By $t \sim \ensuremath {6\ \text {Gyr}}$ the model develops a strong bar as well as a prominent \acronym{B/PS} bulge in the side-on projection (see \cref{fig:BLx-projections}).
We estimate the bar pattern speed in the simulation by the means of \swn{patternSpeed.py} script \citep{Dehnen_etal2022} and find it to be equal to \ensuremath {17.7\ \text {km}\,\text {s}^{-1}\,\text {kpc}^{-1}} (the uncertainty of this value, measured by computing the pattern speed for a few adjacent snapshots, is below \ensuremath {0.1\ \text {km}\,\text {s}^{-1}\,\text {kpc}^{-1}}). Assuming this value of pattern speed, the corotation radius $R_c$~is~$\approx \ensuremath {11\ \text {kpc}}$. {The outer radius of the bar region, measured by the \swn{patternSpeed.py}, is $\approx 10\ \text{kpc}$ (see \citealp{Semczuk_etal2024} for the definition, because it differs from the original description of \citealp{Dehnen_etal2022}).}
The density and the cumulative mass profiles of the components of the evolved simulation are displayed in \cref{fig:BLx-dens-mass-profiles}: the bulge completely dominates inside $r = \ensuremath {0.3\ \text {kpc}}$, while the enclosed mass of the halo starts to be comparable with the enclosed mass of the disc and bulge for $r > \ensuremath {7\ \text {kpc}}$.

\subsection{Mock data}
\label{ssec:mock-data}
\subsubsection{Photometry}
\label{sssec:mock-photometry}
We use the smoothed 3D luminosity density of the simulation as input
for the dynamical models. We will discuss non-parametric deprojections of the barred stellar system in a companion paper (Ding et al., in prep). The density is binned on a spherical grid with the longitude ($\theta$) and the azimuth ($\phi$) being sampled uniformly, while the radius follows a quasi-logarithmic binning:
$$
\begin{aligned}
	\text{boundaries}&: &r_e &= \tfrac{b}{a}\left(e^{a\, (i + 1/2)} - c\right),& i &= 0\mathop{..}N \\
	\text{midpoints}&: &r_\text{mid} &= \tfrac{b}{a}\left(e^{a\, i} - c\right),& i &= 1\mathop{..}N
\end{aligned}
$$
where $c$ controls the deviation of the binning from the logarithmic one, while $a$ and $b$ are chosen in such a way that the midpoints of the innermost bin ($i=1$) and the outermost bin ($i=N$) match $r_{\min}$ and $r_{\max}$ respectively. In practice, for the data described in the beginning of this section, we adopt $c=0.5$, $r_{\min} = \ensuremath {0.2\ \text {kpc}}$, $r_{\max} = \ensuremath {25\ \text {kpc}}$, $N_r = 100$, $N_\phi = 36$, $N_\theta=20$.

To produce a density from the \Nbody simulation that is as smooth as possible, we employ the following approach. Since a bar can be well approximated as a triaxial system in the corotating frame, we fold all particles into a single octant and mirror them back with respect to coordinate planes, thus effectively increasing the number of particles 8 times. For each particle we then find the radius of the ball, containing its 60 nearest neighbours. We distribute the mass of the particle into a few hundred points (``droplets'') contained within this ball. The mass of each droplet follows
\begin{equation}
	m(r) = \begin{cases}
		A - C r^2,& r < \frac{b}2\\
		C (r-b)^2,& \frac{b}{2} < r < b
	\end{cases}
\end{equation}
where $r$ is the distance from the ball centre (original particle location), $b$ is the size of the ball, and the other parameters are designed for the sum of the droplets to match the original particle mass within the ball. The mass in each cell of a spherical grid thus corresponds to the sum of the masses of all droplets which fall into this cell. 

We carry out this procedure for each model component (bulge, disc, halo) independently.
For the disc, we enforce triaxial symmetry at all radii and axial symmetry beyond the bar length to reduce the noise in the outer regions of the density {(i.e. the disc component is triaxial within the rotation radius and axisymmetric outside)}. For the bulge and the halo, we enforce spherical symmetry. 
The resulting densities are internally interpolated by the code on a ``data'' spherical grid, used inside the Poisson solver to compute the potential, and a much coarser ``library'' spherical grid, used in the Schwarzschild modelling itself. In this work, we adopt $N_r = 25$, $N_\theta = 7$, and $N_\phi = 10$ for the library grid.

\subsubsection{Kinematics}

\begin{figure}
	\centering
	\includegraphics[width=0.9\linewidth]{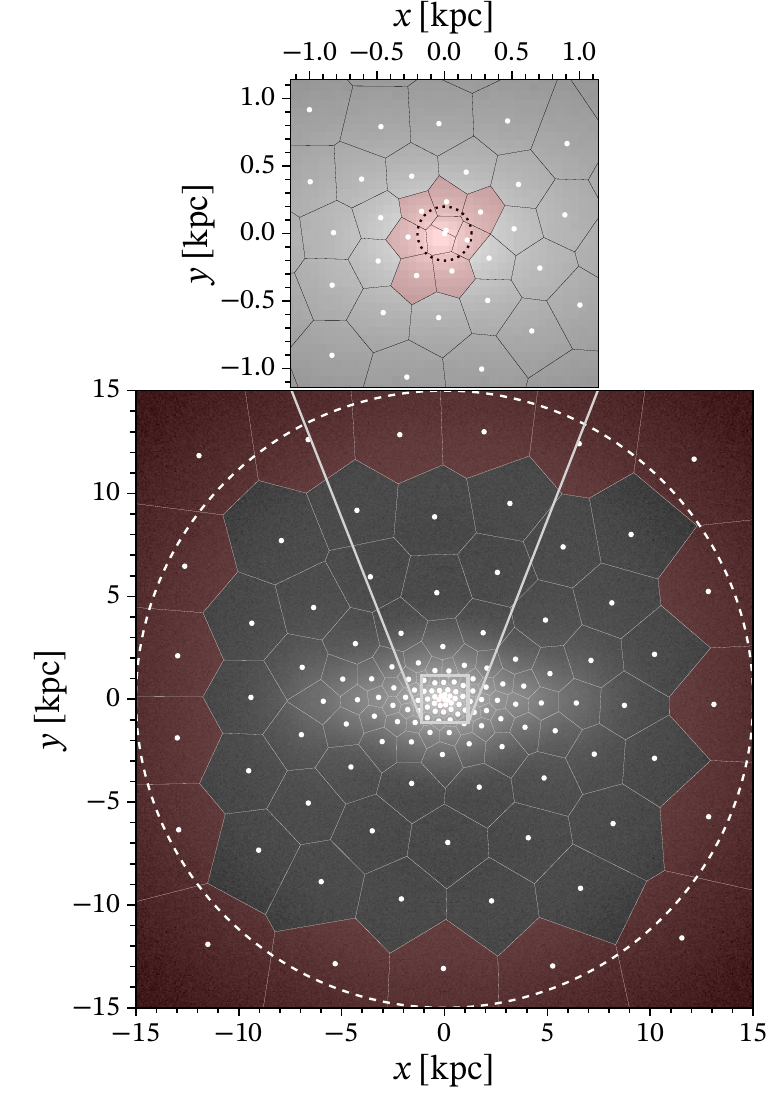}
\caption{%
A tessellation of the \acronym{FoV}, used in the dynamical modelling of the exactly face-on run ($i=0$). For each Voronoi bin, we show the cell boundaries (thin grey lines) and the generators (white dots).The bins which are excluded from the modelling are indicated with the red colour.
The white dashed circle corresponds to $r_{\max} = \ensuremath {15\ \text {kpc}}$, and every Voronoi bin which does not lie entirely within it is excluded from the modelling. An \emph{inset} plot above shows the very central part of the \acronym{FoV}. The black dashed circle indicates the value of $r_{\min} = \ensuremath {0.2\ \text {kpc}}$, and every Voronoi bin which does not lie completely outside of this circle is likewise excluded from the modelling.}
	\label{fig:finaltess}
\end{figure}
\label{sssec:mock-kinematics}
The kinematic data used by the code is a set of \acronym{LOSVD}s in the bins of a sky plane tessellation, represented as histograms.
We prepare the data for different disc inclinations (the bar viewing\footnote{Since position angles are typically measured from the north, we call the angle between the sky plane $x$-axis and the bar major axis in counter-clockwise direction bar \emph{viewing} angle (bar VA).} angle is always kept zero, i.e. the projected major axes of the bar and the disc coincide) and consider a perfectly face-on case ($i=0^{\circ}$) and two almost face-on inclinations ($i=10^{\circ}$ and $i=20^\circ$). This choice is justified by the fact that the bars are much easier to detect in close to face-on orientations ($< 20^{\circ}$), which are unsuitable for the \acronym{TW}, while Schwarzschild is potentially more flexible.
For comparison, we also test a case closer to edge-on inclination ($i=75^{\circ}$) to probe how the inclination impacts the recovery of the \Nbody model parameters.

\begin{table}
	\centering
	\begin{tabular}{rccccc}
		\toprule
		   label &     $i$    &   bar VA  & $N_{\text{tess}}$ & $N_{\text{vel}}$ & $N_{\text{datapoints}}$\\
		\midrule
		 \tt  I0 &  $0^\circ$ & $0^\circ$ &              $57$ &       $31$       &      $1767$            \\
		 \tt I10 & $10^\circ$ & $0^\circ$ &              $58$ &       $31$       &      $1798$            \\
		 \tt I20 & $20^\circ$ & $0^\circ$ &              $57$ &       $31$       &      $1767$            \\
		\bottomrule
	\end{tabular}
	\caption{The summary of mock data sets. For each inclination, we independently produce a surface brightness map and re-run \swn{vorbin} to obtain the desired tessellation. Therefore, despite the fact that we select inner and outer radii in the same way, the number of bins can slightly differ between the cases. }
	\label{tab:test-runs}
\end{table}
We set the size of the \newacronym{FoV}{field of view} to be $[-15; 15]  \times [-15; 15]\ \ensuremath {\text {kpc}}^2$ to fully contain the bar and the region near the corotation (see \cref{ssec:mock-data}).
At 60 Mpc distance this would correspond to a $100 \times 100$ $\text{arcsec}^2$ FoV with a spatial sampling of $0.3''/\ensuremath {\text {pixel}}$ similar to an \acronym{IFU} like MUSE. 
The Voronoi tessellation is produced by applying the \swn{vorbin} code of \citet{Cappellari_Copin2003} to the pixelated surface brightness independently for each considered inclination.
The \newacronym{SN}{signal-to-noise} ratio for \swn{vorbin} is set to 450 across the entire \acronym{FoV}.
We then use the Voronoi tessellation generators to assign each "light" \Nbody particle to its Voronoi bin. The \acronym{LOS} velocities are binned from \ensuremath {-400\ \text {km}\,\text {s}^{-1}} to \ensuremath {400\ \text {km}\,\text {s}^{-1}} into 31 bins with the resulting resolution of $\approx \ensuremath {26\ \text {km}\,\text {s}^{-1}}$. 
Following \citet{Neureiter_etal2021}, we add synthetic errors to the data, set to $3\%$ of the maximal value of each \acronym{LOSVD}. 
Since the N-body model does not resolve the central kinematics well enough, 
we entirely exclude the bins falling inside the circle with $r_{\min} = \ensuremath {0.2\ \text {kpc}}$. We also exclude the bins which fall outside $r_{\max} = \ensuremath {15\ \text {kpc}}$ circle or have less than $2000$ particles inside. An example of the resulting tessellation (for $i=0^{\circ}$) is shown in \cref{fig:finaltess}; for parameter grid search described later we commonly use only the right half of the galaxy ($x>0$), recomputing the left half for a much smaller subset of all tested models.

A summary of the resulting mock data sets is given in \cref{tab:test-runs}. We label these sets according to the inclination (e.g.~\texttt{I20}).


\section{Results}
\label{sec:results}

\subsection{Finding the best-fitting model}
\label{ssec:gridsearch}

\begin{figure*}
	\centering
	\includegraphics[width=0.9\linewidth]{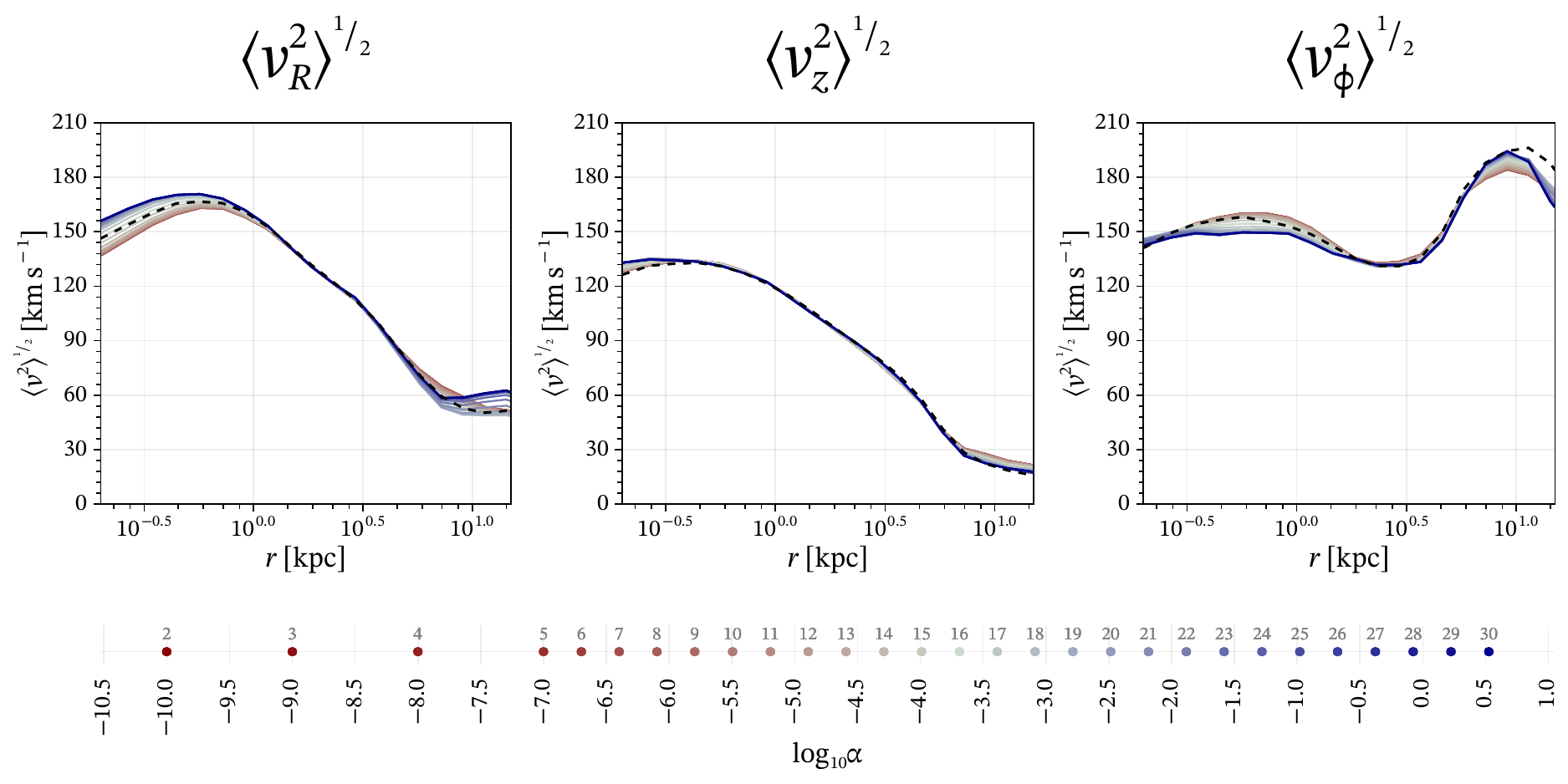}
	\caption{{Radial profiles of the intrinsic second-order velocity moments in the cylindrical frame. Dashed: N-body. Coloured lines: Schwarzschild model with true parameters but different regularization strengths ($\alpha$ indicated below the main panels of the plot)}. We show the index of each $\alpha$ value above the corresponding point, and use these indices ($i_\alpha$) in the following plots instead of $\log_{10} \alpha$ for convenience.}
	\label{fig:intrinsic-kin-moment-profiles-regularization}
\end{figure*}

We perform a grid search in the three dimensional space of models, spanned by the following parameters:
the bar pattern speed ($\Omega_p$), mass-to-light ratio ($\Upsilon$), and a constant scaling, applied to the N-body halo ($s_\text{DM}$, similar to \citealp{Neureiter_etal2021}). Since the we do not attempt the deprojection, the bar orientation angles are known exactly, which allows us to reuse the same 3D luminosity density in every run. 
We assume a true mass-to-light ratio of one and the halo scaling factor in the simulation obviously equals one as well:
\begin{equation}\label{eq:trueparams}
	\begin{aligned}
		\Omega_{p, \text{true}} &= \ensuremath {17.7\ \text {km}\,\text {s}^{-1}\,\text {kpc}^{-1}} \\
		\Upsilon_{\text{true}} &= 1 \\
		s_{\text{DM}, \text{true}} &= 1
	\end{aligned}	
\end{equation}

The grid search itself proceeds in two iterations. First, we evaluate $\chi^2$ on a large coarse grid with a stepsize corresponding to about $10\%$ of true parameter value in each dimension{: $1\ \text{km}\, \text{s}^{-1}\, \text{kpc}^{-1}$ for $\Omega_p$, $0.1$ for both $\Upsilon$ and $s_{\text{DM}}$}.
This effectively gives us the estimate of the $\chi^2$ response to the change in each model parameter. In the next iteration, we generate a new relatively large uniform grid in the vicinity of the $\chi^2$ minimum  with a stepsize depending on the magnitude of $\chi^2$ change in the corresponding parameter space dimension.
This process potentially can be iteratively repeated in a manner described in \citet[section~3.4]{Tahmasebzadeh_etal2022} many times, however this procedure is not guaranteed to converge to the true minimum in our case, in particular, as would be shown later, for face-on inclination, where the noise introduced by discreteness effects is comparable to the change in $\chi^2$.
In addition to that, sampling grid points independently can be done in a very time-efficient way on our {computing} cluster, because it does not involve any communication between the jobs.

The regularization strength factor $\alpha$ itself should ideally be considered a free model dimension since different values of $\alpha$ produce different physical distribution functions. A purely data-driven way to find the optimal distribution function based on information theory is described in \cite{Lipka_Thomas2021} and \cite{Thomas_Lipka2022} and involves estimating the effective degrees of freedom $m_{\text{eff}}$ and comparing the models by the value of the generalized Akaike information criterion, \AICp, instead of $\chi^2$. 
However, evaluating the $m_\text{eff}$ is rather computationally expensive {and since we do not vary the viewing angles here (which often lead to the strongest variations in $m_\text{eff}$) we approximate the model selection by focussing on the $\chi^2$ term only.}

In the special case of our N-body system, the true mass distributions as well as all velocity moments are  known, therefore we 
evaluate the difference between the intrinsic second-order velocity moments in the bins of the library grid and the N-body system to estimate a global
optimised value of $\alpha$ rather than using the more time-consuming data-driven approach. 
{We compute the luminosity-averaged radial profiles of second order moments
of the velocity components ($v_R$, $v_z$, $v_\phi$) 
for each probed regularization value, as shown in \cref{fig:intrinsic-kin-moment-profiles-regularization}.
The first-order moments in $R$ and $z$} are very close to zero due to symmetries of the system, therefore, their second-order momenta would just be equal to the velocity dispersion of the corresponding component. The azimuthal velocity and dispersion, however, are hard to recover independently of one another in our near face-on setup (see \cref{ssec:azimuthal_velocity}), therefore we combine them into a single value \vphiII which can be recovered accurately.

\begin{figure}
	\centering
	\includegraphics[width=\linewidth]{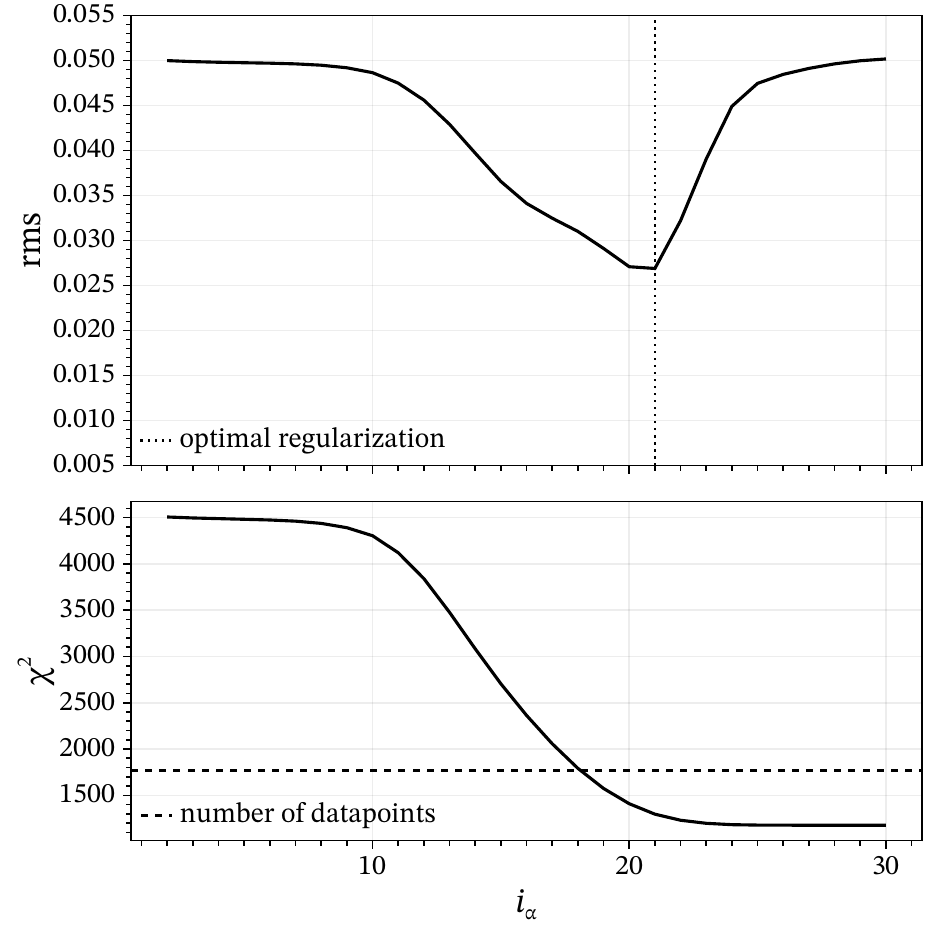}
	\caption{\emph{Top:} total relative difference in the instrinsic second-order velocity moments ($\text{rms}_\sigma$) between the model with true parameters ($i=20^{\circ}$) and the N-body depending on the regularization strength. The vertical dotted line indicates the location of the $\text{rms}_\sigma$ minimum, which for this case corresponds to $i_\alpha = 21$. \emph{Bottom:} the $\chi^2$ of the same model, with horizontal dashed line showing the number of kinematic data points.}
	\label{fig:intrinsic-dispersion-relerrors-optimal-regularization-chi2}
\end{figure}

The final metric that is used to {quantify the match of model and N-body system in terms of intrinsic kinematic moments} {and to select the optimal regularization} is defined as follows:
\begin{equation}
	\begin{aligned}
	\text{rms}_\sigma &= \tfrac13 \, (\text{rms}_{\sigma_R} + \text{rms}_{\sigma_z} + \text{rms}_{\sigma_\phi}), \\
		\text{rms}_{\sigma_k}^2  &= \frac{1}{N_r}\sum_{j=1} \Biggl( 1 -
			\frac{\bigl\langle v_{\text{model},\, k}^2(r_j)\bigr\rangle^{1/2}}%
			     {\bigl\langle v_{\text{data},\, k}^2(r_j)\bigr\rangle^{1/2}}
		\Biggr)^2,
	\end{aligned}
\end{equation}
where  $k = (R, \theta, \phi)$, $j$ is the index of a radial bin of the library grid and the mean squared errors are computed over the radial profiles of the corresponding component. In \Cref{fig:intrinsic-dispersion-relerrors-optimal-regularization-chi2} we show the resulting curve for the \texttt{I20} case as well at the $\chi^2$ curve for the best-fit model. The $\chi^2$ is monotonically decreasing with increasing $\alpha$-value: {for the strongest regularization, the relative change in $\chi^2$ is rather small, as the data have little influence on the model, then for intermediate $\alpha$ the fit rapidly improves, reaching a plateau again for large $\alpha$ (weak regularisation), when the model reaches the lowest achievable $\chi^2$ (and typically overfits the data).} The moments deviation reach a minimum at $i_\alpha = 21$, corresponding to $\alpha \approx \ensuremath {7\times 10^{-3}}$. This value is surprisingly quite close to \ensuremath {10^{-2}}, which \citet{Thomas_etal2005,Thomas_etal2007,Neureiter_etal2021} find for \acronym{ETG}s, despite the phase-space and orbital structure of barred galaxies being quite different from ellipticals. 

A model is typically described as a good fit to the data once the reduced $\chi^2$ {is close to 1}. {The number of degrees of freedom is smaller than the number of datapoints, thus, it only makes sense to consider the models for which $\chi^2_{\min} < N_\text{datapoints}$. In the case of the \texttt{I20} mock} data set this would correspond to $i_\alpha \geq 18$. These models are shown in \Cref{tab:3dsearch-i20}, where we also present the corresponding parameters values $(\Omega_p, \Upsilon, s_{\text{DM}})$ for the best-fit models as well as $\chi^2$ for the model with true parameter values listed in \cref{eq:trueparams}. 
For the optimal regularization, we get the following best-fit parameter values estimates:
\begin{equation}
	\begin{aligned}
		\Omega_{p, \text{best}} &= \ensuremath {19.5\pm 2\ \text {km}\,\text {s}^{-1}\,\text {kpc}^{-1}} \\
		\Upsilon_{\text{best}} &= 1.1 \pm 0.12 \\
		s_{\text{DM}, \text{best}} &= 0.8 \pm 0.35.
	\end{aligned}
\end{equation}
This corresponds to $\approx 10\%$ accuracy for the pattern speed and the mass-to-light ratio, while the halo scaling factor is recovered within $\sim 20\%$.

\begin{table}
	\centering
	\begin{tabular}{rlcccccc}
\toprule
$i_\alpha$ & $\alpha$ & $\chi^2_{\min, \text{red}}$ & $\chi^2_{\min}$ & $\chi^2_{\text{true}}$  & $\Omega_{p}$ & $\Upsilon$ & $s_{\text{DM}}$\\
\midrule
$18$ & $\ensuremath {8.19\times 10^{-4}}$ & $0.9496$ & $1678$ & $1791$ & $19.00$ & $1.04$ & $1.05$\\
$19$ & $\ensuremath {1.64\times 10^{-3}}$ & $0.8449$ & $1493$ & $1574$ & $19.50$ & $1.10$ & $0.80$\\
$20$ & $\ensuremath {3.28\times 10^{-3}}$ & $0.7595$ & $1342$ & $1409$ & $19.50$ & $1.10$ & $0.80$\\
$21$ & $\ensuremath {6.55\times 10^{-3}}$ & $0.7023$ & $1241$ & $1296$ & $\mathbf{19.50}$ & $\mathbf{1.10}$ & $\mathbf{0.80}$\\
$22$ & $\ensuremath {1.31\times 10^{-2}}$ & $0.6695$ & $1183$ & $1230$ & $19.50$ & $1.10$ & $0.80$\\
$23$ & $\ensuremath {2.62\times 10^{-2}}$ & $0.6525$ & $1153$ & $1197$ & $19.50$ & $1.10$ & $0.80$\\
$24$ & $\ensuremath {5.24\times 10^{-2}}$ & $0.6435$ & $1137$ & $1182$ & $19.50$ & $1.10$ & $0.80$\\
$25$ & $\ensuremath {1.05\times 10^{-1}}$ & $0.6401$ & $1131$ & $1178$ & $19.50$ & $1.10$ & $0.80$\\
$26$ & $\ensuremath {2.10\times 10^{-1}}$ & $0.6395$ & $1130$ & $1177$ & $19.50$ & $1.10$ & $0.80$\\
$27$ & $\ensuremath {4.19\times 10^{-1}}$ & $0.6389$ & $1129$ & $1177$ & $19.50$ & $1.10$ & $0.80$\\
$28$ & $\ensuremath {8.39\times 10^{-1}}$ & $0.6389$ & $1129$ & $1177$ & $19.50$ & $1.10$ & $0.80$\\
$29$ & $\ensuremath {1.68}$ & $0.6389$ & $1129$ & $1177$ & $19.50$ & $1.10$ & $0.80$\\
$30$ & $\ensuremath {3.36}$ & $0.6389$ & $1129$ & $1177$ & $19.50$ & $1.10$ & $0.80$\\
\bottomrule
\end{tabular}
\caption{The reduced minimum $\chi^2$ value, the absolute minimum $\chi^2$ value, the $\chi^2$ of the model with ``true'' parameters, and the values of the best-fit model parameters depending on the regularization for $i = 20^{\circ}$ case (\texttt{I20}). We only show the models which produce a fit to the data, i.e. those with $\chi^2_{\min, \text{red}} = \chi_{\min}^2 / N_{\text{datapoints}} < 1$. The parameter values corresponding to the optimal regularization are highlighted with the bold font.}
\label{tab:3dsearch-i20}
\end{table}

\begin{figure}
	\centering
	\includegraphics[width=0.95\linewidth]{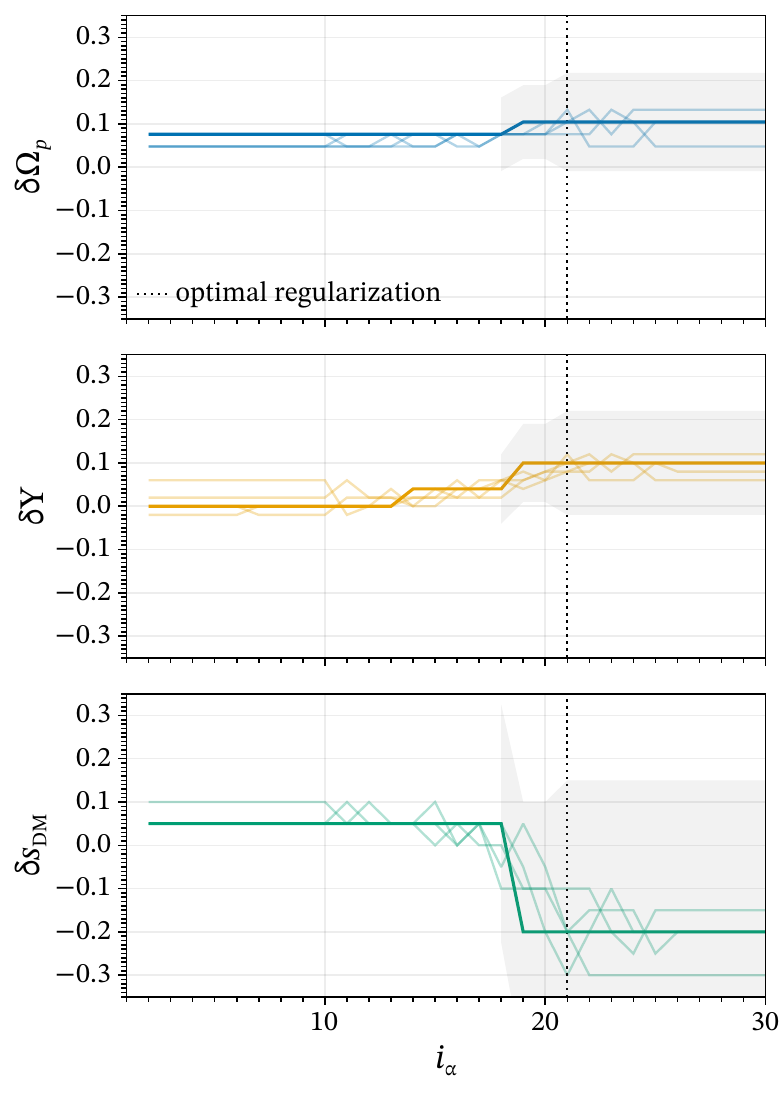}
  \caption{The relative errors of individual free parameters for each regularization strength factor, and their {uncertainties}, shown with grey bands (\texttt{I20}). The ``ghost'' lines (in the same colour but with {a} reduced opacity) are used to show the similar quantity but for a few neighbouring points to the $\chi^2$ minimum in the parameter space.  The vertical dotted black line indicates the optimal regularization.}
	\label{fig:3dsearch-relerr-i20}
\end{figure}

{The uncertainties for Schwarzschild model parameters can be defined either via confidence intervals at a fixed $\Delta \chi^2$ confidence level \citep{Zhu_etal2018,Tahmasebzadeh_etal2022,Dattathri_etal2024}, via a Bayesian approach by emulating the likelyhood \citep{Pilawa_etal2024}, or directly by the scatter of models across independent data subsets \citep[e.g.][]{Mehrgan_etal2019}}.
{The value of a $\Delta \chi^2$ threshold is rather arbitrary
\citep[see e.g.~the discussion in][]{Vasiliev_Valluri2020}, with a square root of the variance of the $\chi^2$ distribution with $K$ degrees of freedom being one of the most common suggestions \citep{vandenBosch_etal2008}. Since we work with simulated data, for which true parameter values are known, we adopt the following $\Delta \chi^2$ as a threshold to assess parameter recovery:
}
\begin{equation}
	\Delta \chi^2_0 = \chi^2_{\text{true}}(\alpha_{\text{opt}}) - \chi^2_{\min}(\alpha_{\text{opt}}) \approx 55\ 
    \text{(for \texttt{I20})},
\end{equation}
{i.e. we consider all of the models which are not further from the best-fit model then the one with true parameters.}
{We note that this value is quite close to, but slightly smaller than the $\sqrt{2N_{\text{datapoints}}}$ threshold, used e.g. by \citet{Tahmasebzadeh_etal2022}, which for the \texttt{I20} mock data set case would be equal to~$\approx 60$}. Since we do not estimate the $m_\mathrm{eff}$, the interpretation {of the difference in $\chi^2$} 
is ambiguous. But if the change in $m_\mathrm{eff}$ between the true and best-fit values is negligible then the large value of $\Delta{\chi^2}$ would suggest that the uncertainty in the recovered model parameters is dominated by discreteness effects in the model itself and that the data would allow an even more precise recovery of, e.g., the pattern speed, if the noise in the model could be further reduced. 
In any case, {we define a set of ``consistent'' models $M_c(\alpha)$, for which} a) $\chi^2$ value itself smaller than the number of datapoints b) the $\chi^2$ difference from the best-fit model smaller than the $\Delta\chi^2_0$, 
Since we have an {effectively} pre-defined uniform parameter grid instead of sampling from the posterior parameter distribution, we use the half-range of parameter values of models belonging to $M_c(\alpha)$ as {an} estimate of the scatter instead to the standard deviation:
\begin{equation}
	\Delta \theta = \frac12 \left(\max_{j\in M_c(\alpha)} \theta_j - \min_{j\in M_c(\alpha)} \theta_j\right).
\end{equation}
{where} $\theta$ is one of $(\Omega_p, \Upsilon, s_{\text{DM}})$. 
The relative parameter errors {(effectively telling how biased are the estimates)} are defined as $\delta\theta = \theta_{\text{best}}/\theta_{\text{true}} - 1$. We show this errors along with parameter uncertainties divided by the true value of the corresponding parameter in \cref{fig:3dsearch-relerr-i20}.

\subsection{Marginal distributions}
\label{ssec:marginal-dists}

\begin{figure*}
	\centering
	\includegraphics[width=\linewidth]{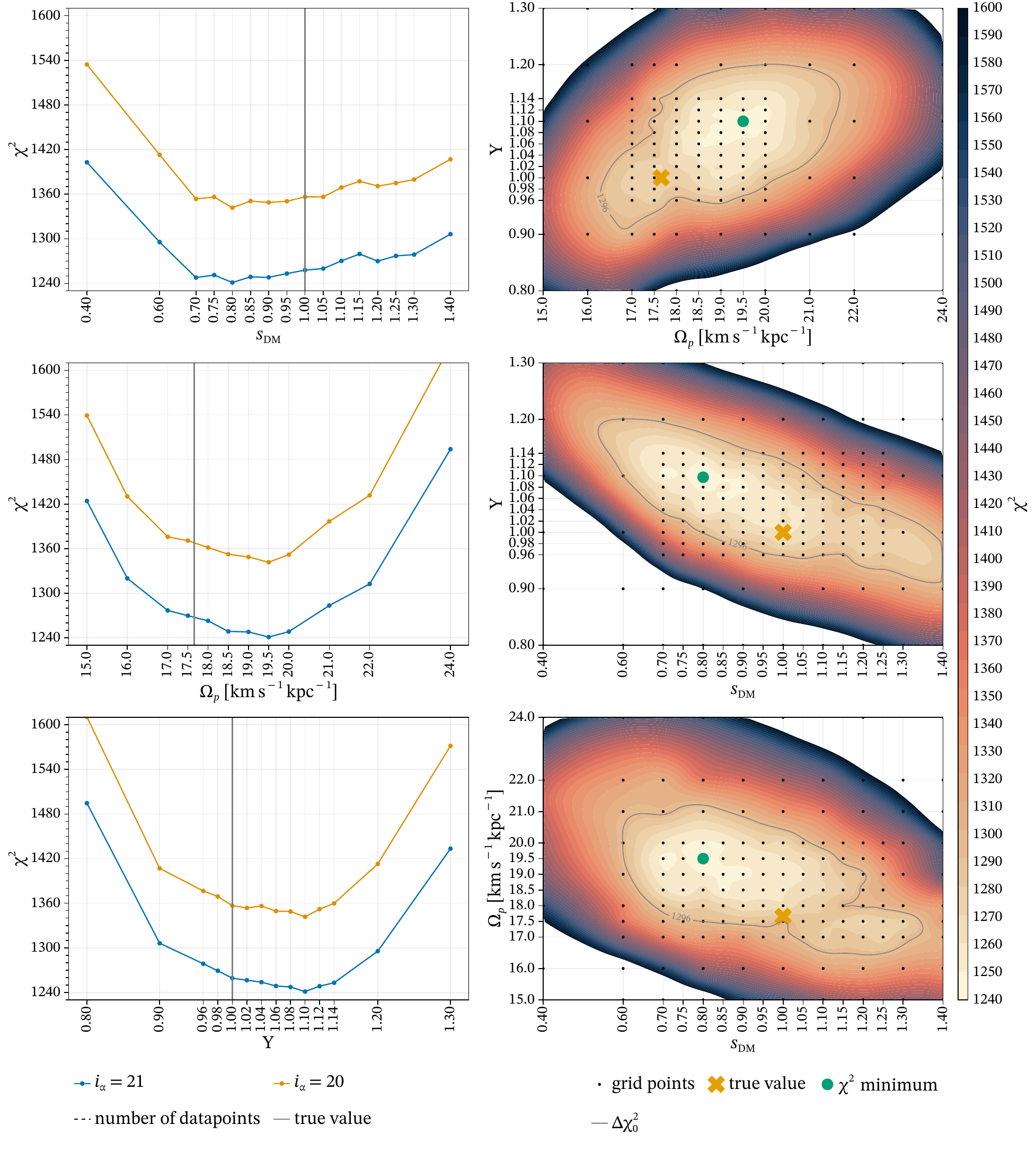}
	\caption{Marginal parameter distributions for the \texttt{I20} case. {The \emph{left} column shows 1D $\chi^2$ curves for each fit parameter $(s_{\text{DM}}, \Omega_p, \Upsilon)$ (solid grey line: true parameters; dashed line: number of data points). The \emph{right} columns illustrate the 2D distributions for the two complementary parameters (orange cross: true parameters; green dots: best-fit values;  grey contours: $\chi^2 < \chi_{\min} + \Delta \chi^2_0$).} 
    }
	\label{fig:marginaldists-i20}
\end{figure*}

Here we mostly focus on the \texttt{I20} case, leaving similar analysis for the rest of inclinations to Appendix~\ref{sec:i0i10-par-recovery}.

To access the quality of the parameter recovery, we plot the 2D and 1D parameter distributions for the optimal regularization in \cref{fig:marginaldists-i20}. 
Since we do not sample the parameter values from the real posterior distribution, we use a local $\chi^2$ minimum along the corresponding grid dimensions to marginalize, since this method
does not shift the position of the true 3D minimum.

We find the mass-to-light ratio to be the best-constrained and most accurately recovered parameter: it is slightly overestimated by $10\%$, but the best-fit value lies within the $12\%$ bounds, determined by our uncertainty estimation method, and would certainly lie within the $20\%$ bounds that we would have obtained if we were using the $\chi^2$ variance criterion of \citet{vandenBosch_etal2008}.
The precision of the pattern-speed recovery is only slightly worse ($11\%$). Its uncertainty bounds are comparable with the uncertainty of the mass-to-light ratio.

While the halo scaling factor has the largest uncertainty, it is actually surprising how well it is recovered. The \texttt{I20} mock data set has only $18$ out of 57 Voronoi bins beyond \ensuremath {7\ \text {kpc}}, where the halo starts to dominate (see bottom panel of \cref{fig:BLx-dens-mass-profiles}). Only $10$ bins lie entirely outside of the $R=\ensuremath {\text {7 kpc}}$ circle. In addition, the $s_{\text{DM}}$ is correlated with $\Upsilon$ as expected (middle left panel of \cref{fig:marginaldists-i20}). Nevertheless, the density profiles and the potentials of the light-emitting matter and the dark halo are sufficiently different to be determined independently which is particularly challenging when the galaxy inclination approaches pure face-on orientation, like in our case. It could be that this is related to the fact that both mass components have very different 3D shapes and that we implicitly assume these shapes to be known.

\subsection{Kinematics of the best-fit model}
\label{ssec:best-fit-kinematics}

\begin{figure*}
  \centering
  \includegraphics[width=\linewidth]{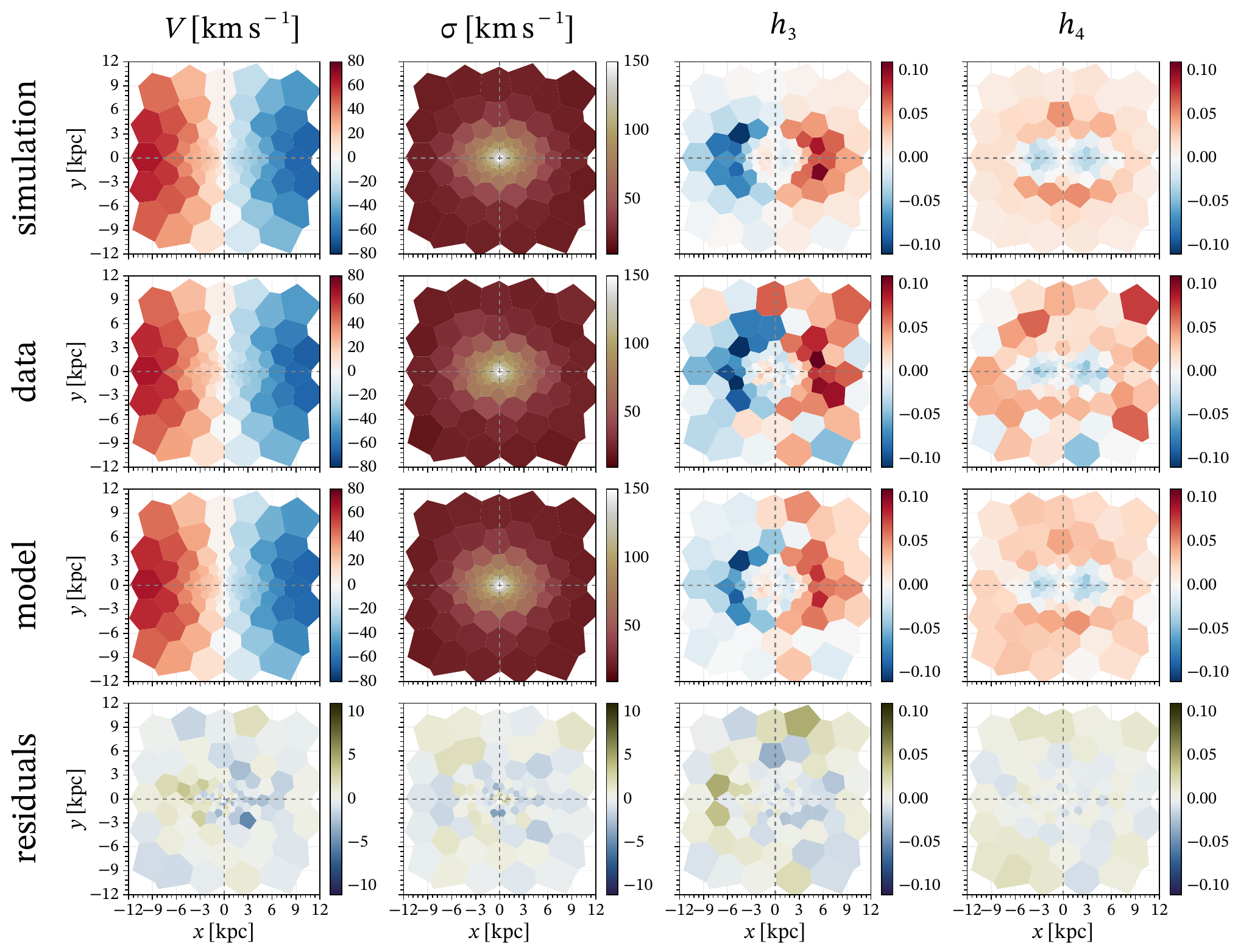}
  \caption{Maps of the Gauss-Hermite parameters ($V$, $\sigma$, $h_3$, $h_4$) for inclination $20^{\circ}$ (\texttt{I20}). \emph{Top} row: N-body simulation data without noise, \emph{second} row: noisy mock data fitted by the code, \emph{third} row: dynamical model at the optimal regularization strength factor ($i_\alpha = 21$), \emph{last} row: the residuals of the corresponding parameter of the dynamical model with respect to simulation data.}
  \label{fig:kinematic-maps-bestfit-i20}
\end{figure*}

\begin{figure*}
  \centering
  \includegraphics[width=\linewidth]{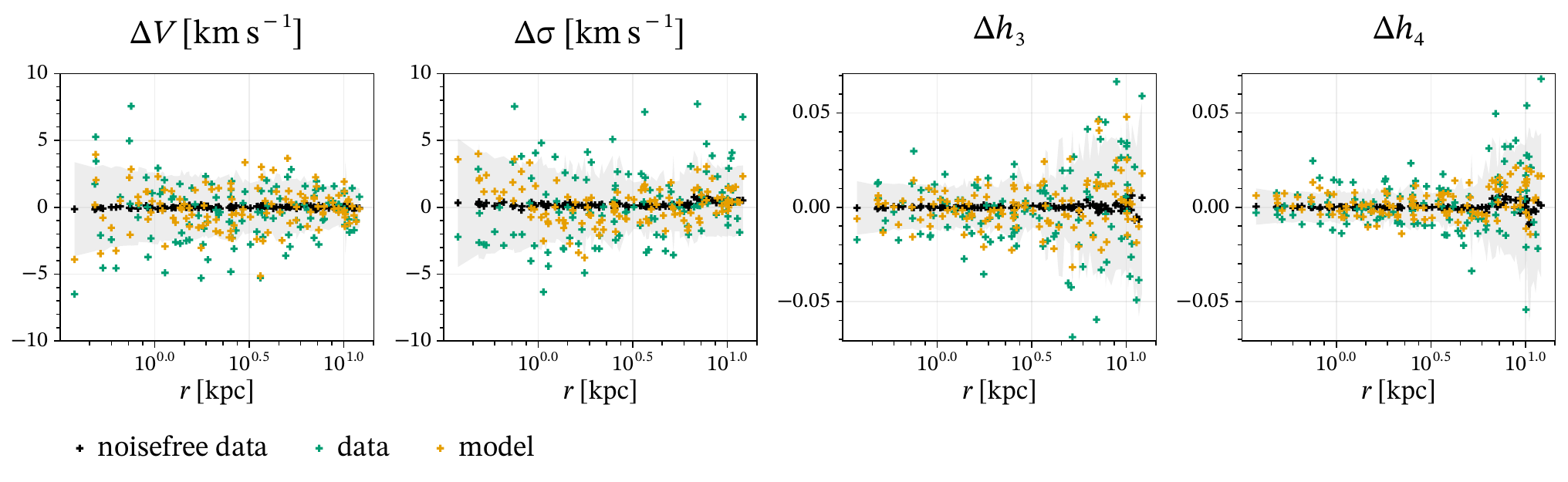}
	\caption{{Profiles of the Gauss-Hermite residuals ($\Delta V$, $\Delta \sigma$, $\Delta h_3$, $\Delta h_4$) with respect to the true N-body values for inclination $20^{\circ}$ (\texttt{I20}). The shaded error bands indicate the 1$\sigma$ uncertainties of the \acronym{GH} parameters corresponding to the adopted \acronym{LOSVD} errors for the mock data. The mock data (green crosses) and the best-fit model with optimal regularisation (orange cross) are within the shaded area as expected. At large radii the \acronym{LOSVD}s are not well sampled and the offsets slightly increase (see the main text for details).}}
  \label{fig:kinematic-profiles-errors-bestfit-i20}
\end{figure*}

\citet{Debattista_etal2005} show that a ``flat-topped'' vertical density distribution of a \acronym{B/PS} bulge
is well traced by an $h_4$ Gauss-Hermite moment of the \acronym{LOSVD}s along the bar major axis for nearly face-on inclinations. The \acronym{LOS} kinematic of the BLx model snapshot, which we use in this work, was studied in detail by \citet{Zakharova_etal2023}, where this result was confirmed, given that the disc inclination is $\leq 20^{\circ}$ for bar {viewing} angle equal to $0$.  It is therefore interesting to compare the performance of orbital superposition modelling in reproducing the \acronym{LOS} maps of the original \Nbody model.
In \cref{fig:kinematic-maps-bestfit-i20} we show the mean velocity, its standard deviation, and $h_3$ and $h_4$ \acronym{GH} moments on the full \acronym{FoV} tessellation adopted for \texttt{I20} mock data set. We compute the \acronym{LOSVD} moments for the original N-body data (first row of \cref{fig:kinematic-maps-bestfit-i20}), the actual data with the synthetic noise used in the modelling (second row) and the dynamical model with best-fit parameters at optimal regularization. Since the grid search was carried out  only for the right half of the data ($x > 0$), we compute the left half LOS kinematics ($x<0$) from the model with the same best-fit parameter values.

To compute the \acronym{GH} moments and their errors, we employ a following approach. For each \acronym{LOSVD} bin, we generate 1000 realizations of the normally distributed noise with \newacronym{std}{standard deviation} matching the assumed synthetic noise \acronym{std} in this bin (see \cref{sssec:mock-kinematics} for details). We add this new noise on top of the original \acronym{LOSVD} and then fit it with the truncated Gauss-Hermit series assuming $N_{\text{trunc}} = 4$:
\begin{equation}
	\mathcal{L}(v) = \frac{I}{\sqrt{2\pi}}\, e^{-w^2/2}\, \sum_{k=0}^{N_{\text{trunc}}} h_k\, H_k(v), \quad w = \frac{v - V}{\sigma}.
\end{equation}
We obtain 1000 realisations of kinematic moments ($V, \sigma, h_3, h_4$) and compute their mean values and standard deviations. This routine is performed for every studied \acronym{LOSVD}, allowing us to eliminate a potential source of bias from the suboptimal Gauss-Hermite fits to the inherently noisy data.

The velocity maps show the rotation pattern, expected for a disc galaxy at intermediate inclinations, with a rather characteristic pinch in the central part, corresponding to a peanut-shape bulge observed nearly face-on \citep{Saha_etal2018}. 
The LOS velocity dispersion maps roughly trace the bar shape (cf. fig.~30 in \citealp{Iannuzzi_Athanassoula2015}) and have a peak in the central part of the model.
The absolute errors for both quantities are $\approx 3\text{--}5\, \ensuremath {\text {km}\,\text {s}^{-1}}$ showing a slight downwards trend with increasing distance to the centre of the \acronym{FoV} and the first two moment deviations of the dynamical model generally lie within this error band (\cref{fig:kinematic-profiles-errors-bestfit-i20} {and bottom row of \cref{fig:kinematic-maps-bestfit-i20}}).

The noise-free $h_4$ maps show the region of negative values, matching the shape of the B/PS bulge of the model, surrounded by a broken ring-like structure of positive $h_4$ with a maximum at the bar minor axis, corresponding to the bar-disc boundary \citep[see][]{Zakharova_etal2023}. The addition of synthetic noise preserves the structure, but results in a many outer bins gaining large absolute values of $h_4$, which are essentially the artefacts of adding noise to the very narrow \acronym{LOSVD}s of the outer disc. This also explains the growing $\Delta h_4$ with increasing distance to the centre of the \acronym{FoV}. However, due to the selection of the optimal regularization value in \cref{ssec:gridsearch}, the dynamical model is close to the original noise-free structure within the error bounds. A slight systematic bias occurs in the outermost Voronoi bins of the \acronym{FoV}. However, these \acronym{LOSVD}s are not well sampled (since they have low velocity dispersions) and the noise in the wings is exaggerated by the fact that we set the \acronym{LOSVD} errors proportional to the maximum of the \acronym{LOSVD}, which implies larger errors for narrower \acronym{LOSVD}s. 

The $h_3$ maps show a rather complicated pattern of the regions with alternating signs (cf.~fig.~9 and fig.~B1 in \citealp{Zakharova_etal2024}\footnote{The BLx model used in this work was not analysed in detail there, but its kinematics closely match the BL model with $\text{bl}_\text{u}$ orbital group removed.}). Following the bar major axis ($x$-axis), from left to right, one first finds $V > 0$ and $h_3 < 0$, both increasing in the absolute value towards the centre of the \acronym{FoV}. This can be explained from the fact that \acronym{LOSVD}s in this region are dominated by particles near tangent points, producing a high-velocity peak, contaminated by a wing of slower-rotating particles, resulting in an $h_3$ {anti-correlated} with $V$ \citep[as explained by][]{Li_etal2018}.
Inside \ensuremath {4\ \text {kpc}}, the situation is reversed with both $V > 0$ and $h_3 > 0$: the peak of the \acronym{LOSVD} is formed by bar orbits, precessing with the same angular velocity, while the surrounding disc rotates faster, generating a high-velocity tail in the distribution and producing a positive value of $h_3$ (this matches the description for region B of fig.~8 in \citealp{Zakharova_etal2024}, with region C not existing in this model due to a smaller population of box orbits in the bar).
In the other half of the model, the situation with $h_3$ and $V$ correlation or anti-correlation regions is symmetric, except that both values have opposite signs with respect to the left half.
The regularized model is again close to the original data, as expected. The outermost Voronoi bins suffer from the largest offsets but those \acronym{LOSVD}s are not well sampled and the noise in the wings is exaggerated as stated above.

\section{Discussion}
\label{sec:discussion}

\subsection{Face-on pattern speed recovery}
\label{ssec:face-on-patter-speed-recovery}
\begin{figure}
	\centering
	\includegraphics[width=0.95\linewidth]{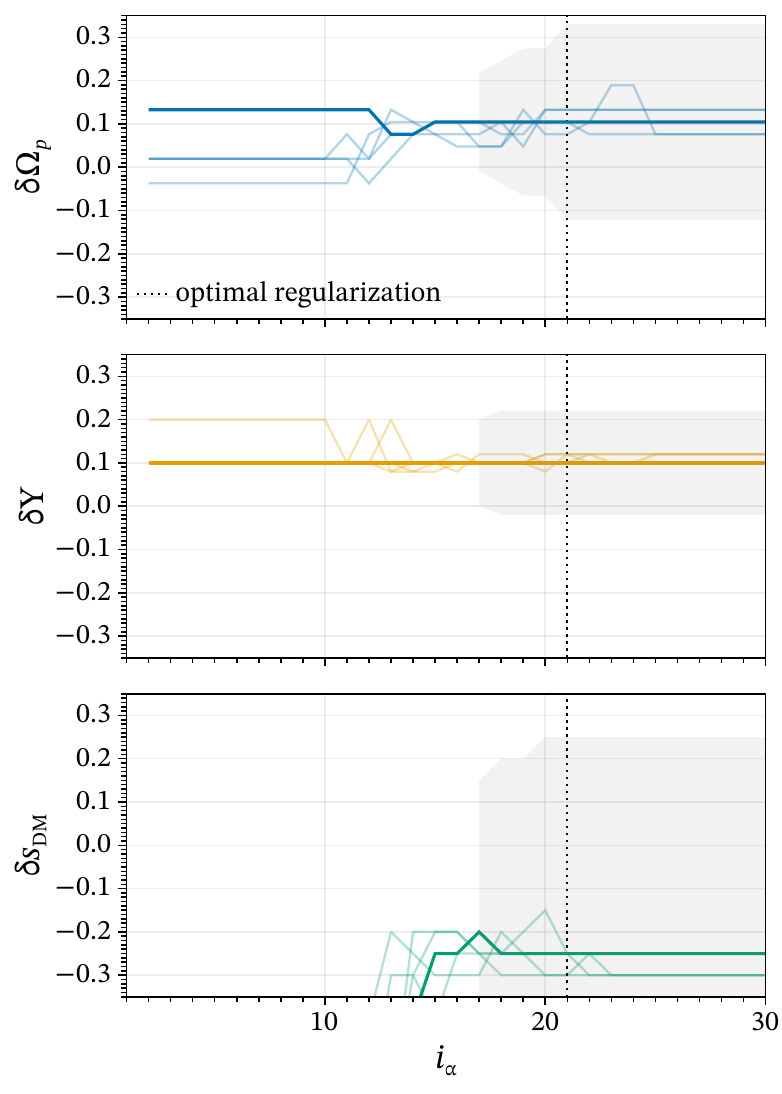}
  \caption{The relative errors of parameters for each regularization strength factor and their uncertainties, shown in a similar way to \cref{fig:3dsearch-relerr-i20} but for $i=0^\circ$.}
	\label{fig:3dsearch-relerr-i0}
\end{figure}
\begin{figure}
	\centering
	\includegraphics[width=0.7\linewidth]{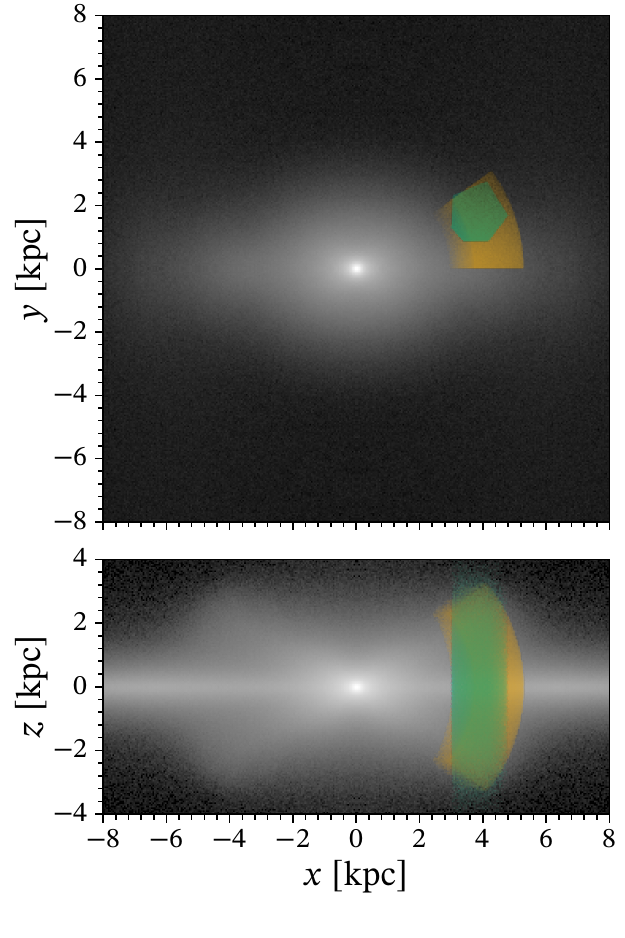}
	\caption{{Face-on and side-on projection of the 3d intrinsic density with Voronoi bin \#111 of the \texttt{I0} mock data set highlighted in green. The library bins, associated with the considered Voronoi bin, are highlighted in orange}.}
	\label{fig:select-libgrid-ibin111}
\end{figure}

\begin{figure*}
	\centering
	\includegraphics[width=0.4\linewidth]{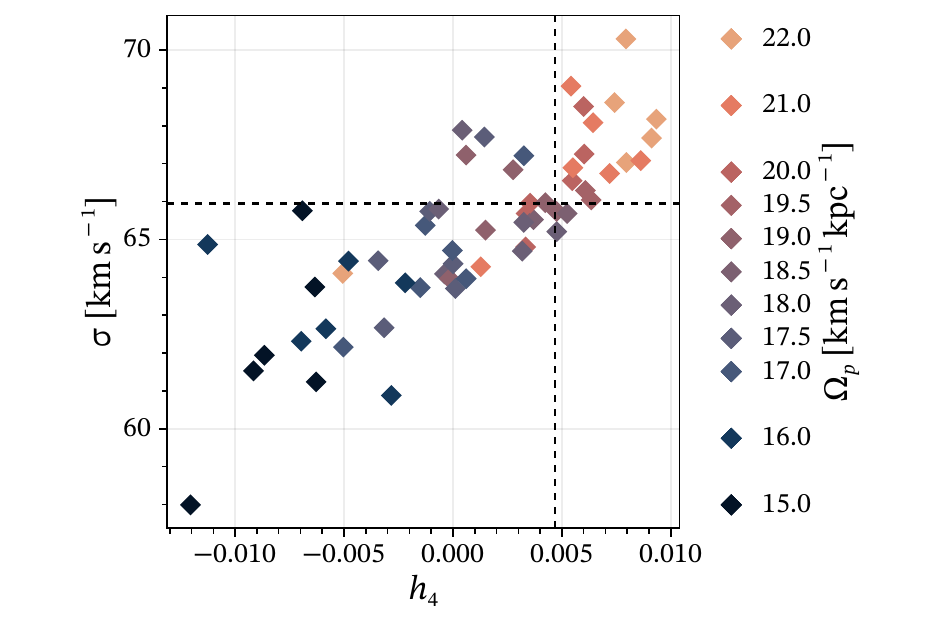}
	\includegraphics[width=0.56\linewidth]{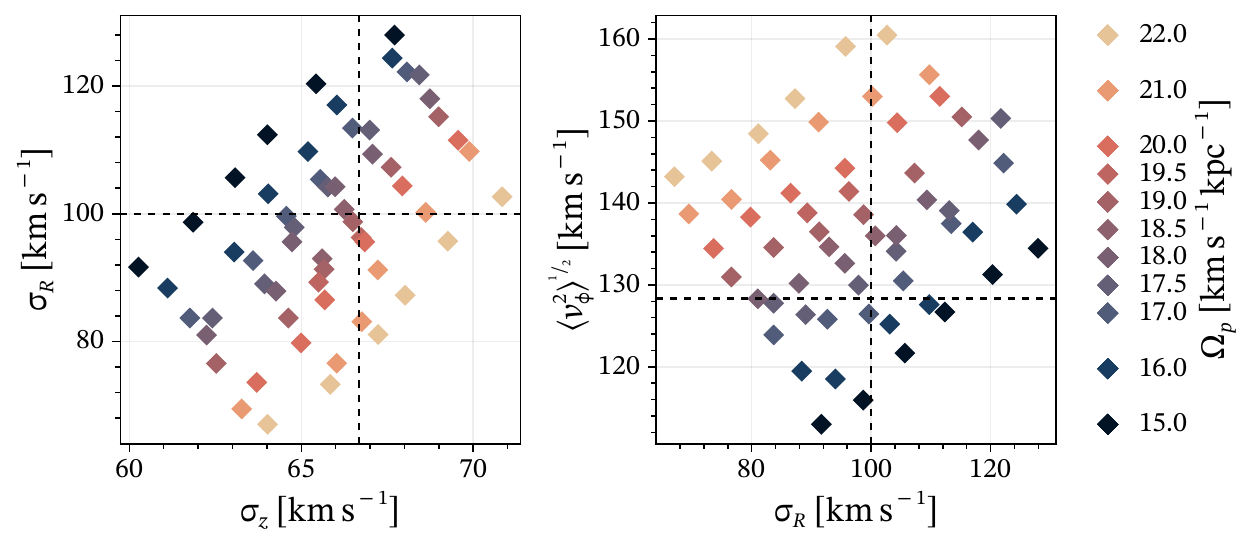}
	\includegraphics[width=0.4\linewidth]{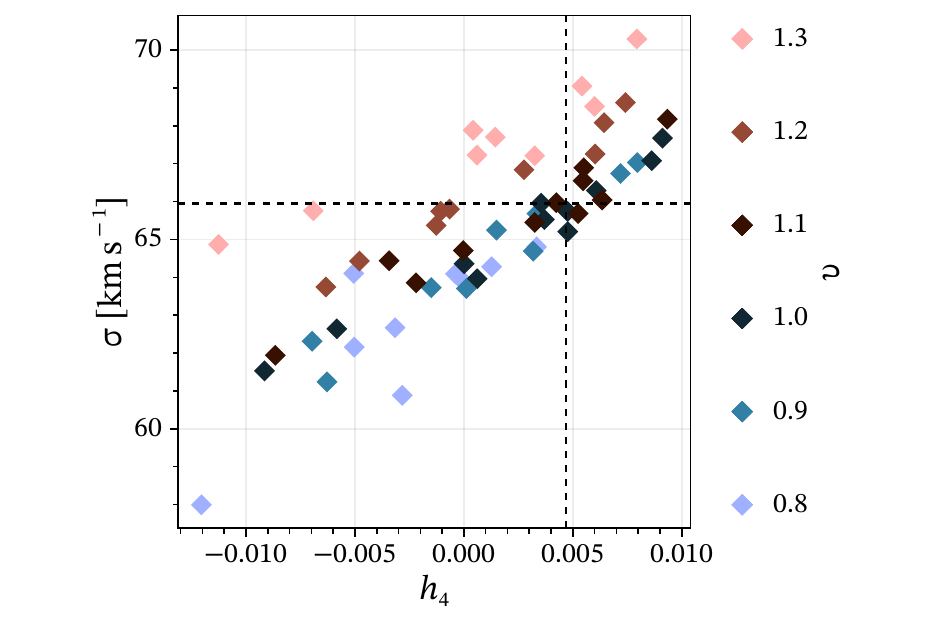}
	\includegraphics[width=0.56\linewidth]{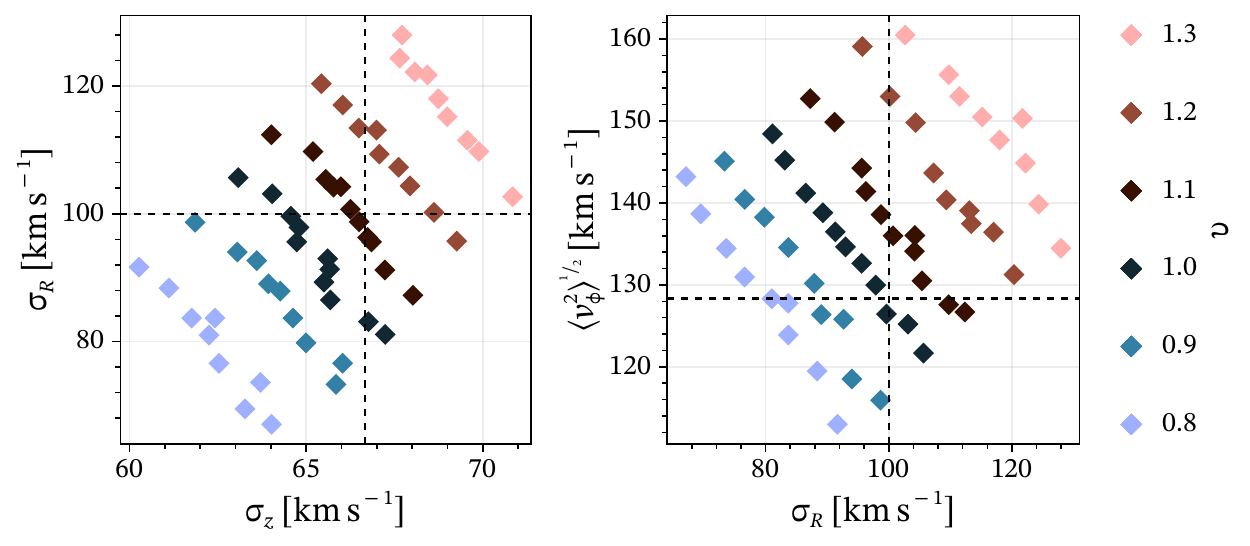}
	\caption{\emph{Left column}: $h_4$ and $\sigma$ of the bin \#111 \acronym{LOSVD} for the reduced 2-dimensional model grid (see main text). The colour on the \emph{top} plot shows the values of the pattern speed corresponding to the points, while on the \emph{bottom} plot it is used to indicate the mass scaling values. The dashed lines correspond to original \Nbody kinematics. \emph{Middle and right columns}: intrinsic second moments of the grid subset indicated with orange colour in \cref{fig:select-libgrid-ibin111}. Similarly to the \acronym{LOS} kinematics, the colour gradient in the \emph{top} plot is used to show the values of the pattern speed, while in the \emph{bottom} plot it indicates the mass scaling. The dashed lines correspond to the same values computed from the original \Nbody simulation.}
	\label{fig:intr-kinematics-face-on-recovery-in-one-bin}
\end{figure*}

\begin{figure}
	\centering
	\includegraphics[width=0.8\linewidth]{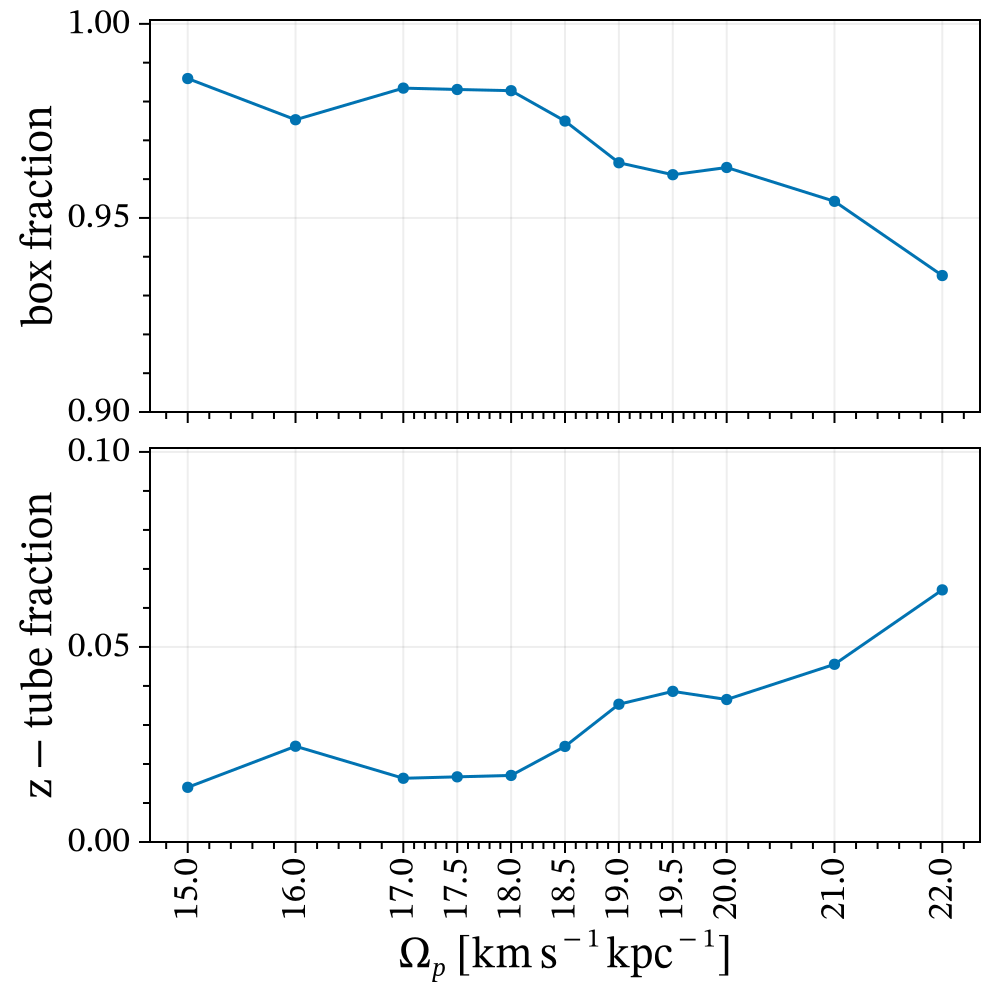}
	\caption{Orbital fractions of box orbits (\emph{top}) and $z$-tube orbits (\emph{bottom}) inside the Voronoi bin \#111 of \texttt{I0} with changing pattern speed.}
	\label{fig:orbit_fractions_inside_bin111}
\end{figure}

\begin{figure*}
	\centering
	\includegraphics[width=0.95\linewidth]{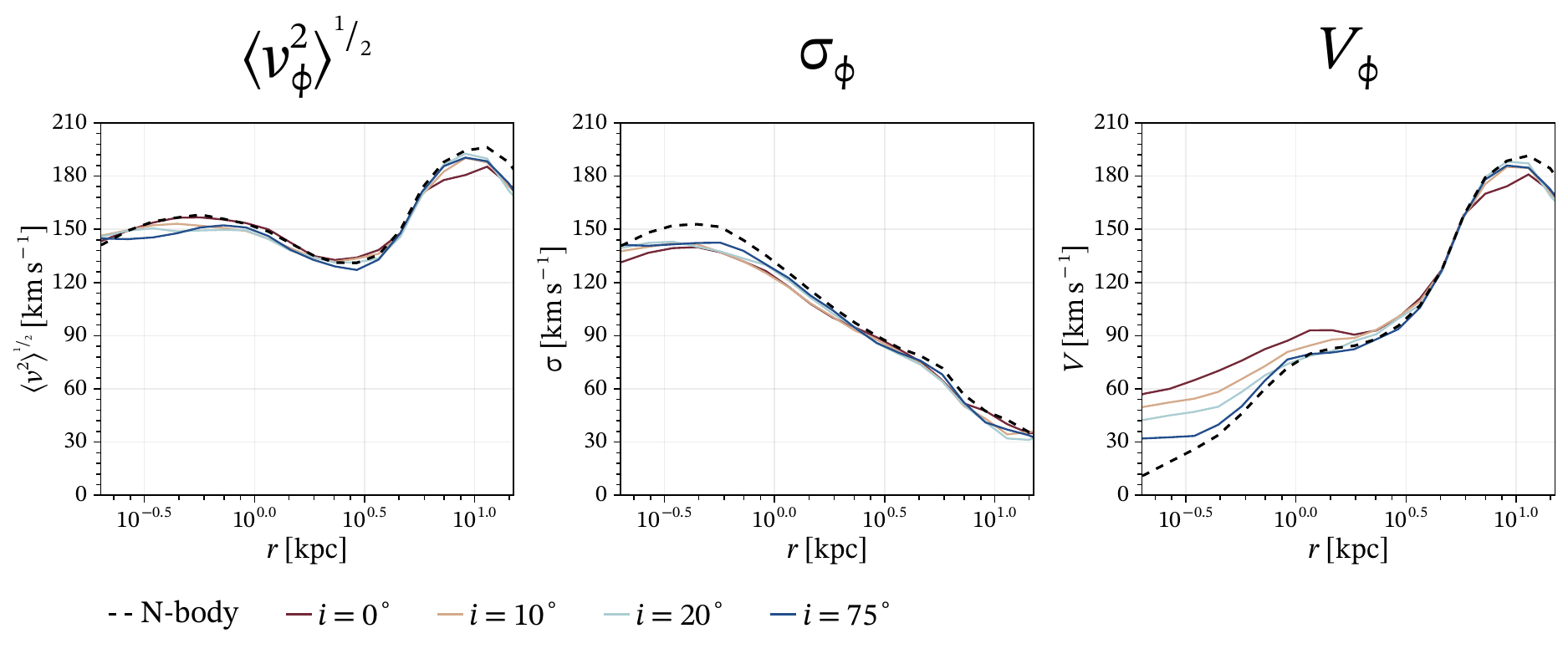}
	\caption{The total second moment radial profile of the intrinsic azimuthal velocity \vphiII and its components ($V_\phi$ and $\sigma_\phi$). We show the true moments computed in \Nbody simulation by a dashed black line and indicate with colour the dynamical models produced from mock data sets at different inclinations.}
	\label{fig:intr-kin-vphi2-decomposition}
\end{figure*}

In \cref{ssec:gridsearch} we demonstrated that for $i=20^\circ$ the bar pattern speed is recovered with $\approx 10\%$ accuracy. The same result, albeit with larger uncertainties also holds for {an exact} face-on case (see \cref{fig:3dsearch-relerr-i0}), where the first \acronym{LOSVD} moments are zero (see \cref{fig:kinematic-maps-bestfit-i0}). 
{We should note that the exact precision of the measurements will further depend on the observational errors, on uncertainties in the deprojection and halo shapes that we did not consider here. Since parametric bar models struggle to reproduce the in-plane bar density \citep{Dattathri_etal2024}, we are currently developing a non-parametric deprojection code for barred galaxies (Ding et al., in prep) and plan to discuss these effects in related future papers.}
{Nevertheless, the possibility of the face-on pattern speed recovery even in our idealized case with the true densities known both for the light-emitting matter and the dark halo still looks quite surprising at first. The Tremaine-Weinberg method, for example, would not be applicable to zero inclination at all.}

Contrary to the \acronym{TW} method, {instead of just the first moment}, we use the residuals of full \acronym{LOSVD}s to compare the models. 
To understand which kinematic information is responsible for the change in $\chi^2$ between the models with different pattern speeds, we consider the Voronoi bin near the very edge of the B/PS bulge, indicated with a green area in \cref{fig:select-libgrid-ibin111}. To simplify the reasoning, we replace the mass-to-light ratio and the halo scaling with a single mass scaling parameter $\upsilon$ by {selecting} the subset of the models for which the $s_{\text{DM}}$ was equal to $\Upsilon$. 

In the left column of \cref{fig:intr-kinematics-face-on-recovery-in-one-bin} we show the even moments\footnote{$V$ and $h_3$ for the face-on case are determined by noise only (see \cref{fig:kinematic-maps-bestfit-i0}).}: $\sigma$ and $h_4$ of the \acronym{LOSVD} in the selected Voronoi bin. The change of the mass scaling creates a clear vertical gradient, corresponding to a change in the velocity dispersion. The pattern speed change at any fixed mass scaling value produces a less pronounced but still significant increase in $h_4$. Since we consider an exactly face-on case, these moments are essentially the moments of the vertical velocity distribution and the code has no information about the radial or the azimuthal component of the velocity during the fit.

In \cref{ssec:gridsearch} we already showed that the dynamical model recovers the intrinsic kinematic moments accurately (see also \cref{ssec:azimuthal_velocity} for a discussion about the first moment of $v_\phi$), but we only considered the model with the true parameter values so far. Thus, we select the cells of the spherical grid which contain \Nbody particles falling into the Voronoi bin we are interested in (highlighted by orange colour in \cref{fig:select-libgrid-ibin111}), and compute the luminosity-averaged second moments of this subset for all models. As one can see in \cref{fig:intr-kinematics-face-on-recovery-in-one-bin}, the change in parameters results in the almost orthogonal colour gradients: the mass scaling increases every velocity component, corresponding to a change of the total kinetic energy, while the pattern speed change redistributes the total kinetic energy between velocity components.

Since the density is required to be exactly the same for every constructed dynamical model, the bar size stays constant, while the corotation radius changes according to the assumed value of the pattern speed. Thus, we effectively simulate the transition from the `slow' bar regime to the `fast' one. This transition would result in a change of the orbital structure of the bar. In \cref{fig:orbit_fractions_inside_bin111} we show the relative contributions of the two main types of orbits to a Voronoi bin \#111: box orbits and $z$-tubes.
A box orbit in the $x-y$ plane is characterized by a zero average value of angular momentum along the $z$ axis, while tubes have non-zero $L_z$.
With the increase of the pattern speed, we observe that the fraction of $z$-tube type orbits increases, which also correspond to an increase of the azimuthal components of intrinsic kinematics. The radial component, contrary, is largely influenced by a fraction of boxes, which decreases with increasing pattern speed.
A rather curious consequence is a redistribution of the small fraction of the radial component into the vertical component, which is seen as an increase of the LOS dispersion with increasing pattern speed at a fixed mass scaling value (left column of \cref{fig:intr-kinematics-face-on-recovery-in-one-bin}). This results in the change in $\chi^2$, which in turns makes the recovery possible.

The exact reason why the tube orbits have larger velocity dispersions and larger $h_4$ is still not well understood, and, to the best of our knowledge, there was no previous study with a comparable setup which we might refer to.

\subsection{Disentangling azimuthal velocity components}
\label{ssec:azimuthal_velocity}

In \cref{ssec:gridsearch} we  mentioned that we use \vphiII for selecting the optimal regularization. 
\Cref{fig:intr-kin-vphi2-decomposition} shows both, the radial profiles of \vphiII and of $\sigma_\phi$ and $V_\phi$ individually for the dynamical models {with the true parameter values for all studied mock data sets (\cref{tab:test-runs}), as well as the reference value from the \Nbody simulation (dashed black lines).}

{As expected, with increasing inclination, the recovery of the azimuthal velocities improves since the \acronym{LOSVD}s {contain more and more information about the detailed kinematics in the $\phi$ direction. To verify this trend even more, we also model a highly inclined case ($i=75^\circ$) for comparison. At this inclination, the recovery of the intrinsic kinematics in the $\phi$ direction becomes quite accurate.}}

{Surprisingly,}
outside 
$\approx \ensuremath {3\ \text {kpc}}$ (left panel of \cref{fig:intr-kin-vphi2-decomposition}) the azimuthal velocities match even in the face-on case. In contrast to the phase-space of systems without figure rotation, the orbit libraries in the barred case are not symmetric in $v_\phi$. This means that a model that is only constrained to reproduce the density might necessarily have to be rotating around the $z$ axis. Investigating by how much the distribution function of a barred disc is already determined by the density constraints alone is beyond the scope of this paper. But it seems to be much stronger constrained than in spheroidal galaxies without figure rotation. These density constraints on the distribution function might be connected to the surprising result that the pattern speed can be recovered even in the face-on case. We plan to investigate the tightness of the connection between density and distribution functions in barred discs in a future paper.

\section{Summary and conclusions}
\label{sec:conclusion}
A new Schwarzschild modelling code for potentials with figure rotation was developed and extensively tested on mock data sets constructed from an \Nbody simulation. This code is able to reproduce all features in the line-of-sight kinematic maps as well as recover the intrinsic kinematics of the \Nbody data. We then use the new code to investigate how well the model parameters: bar pattern speed $\Omega_p$, mass-to-light ration $\Upsilon$, and dark matter halo scaling factor $s_{\text{DM}}$ can be recovered for nearly face-on inclinations ($i<20^{\circ}$).

\begin{enumerate}
	\item For mock data at inclination $i=20^{\circ}$, we are able to recover the pattern speed and the mass-to-light ratio within $10\%$ and with~$\approx 12\%$ uncertainty. The halo scaling has a larger uncertainty (up to $35\%$), but is still recovered on a $20\%$ accuracy level.

	\item Even in exactly face-on case, the pattern speed is still recovered with a $10\%$ accuracy, albeit with a larger uncertainty ($23\%$). The dark matter halo scaling factor recovery becomes increasingly difficult with the inclination approaching $0^{\circ}$ due to a degeneracy with the mass-to-light ratio.

\end{enumerate}
To understand the mechanism which allows to recover pattern speeds in almost face-on orientation, we study the \acronym{LOS} and intrinsic kinematics dependence on the model parameters.
We find that for a point near the end of the B/PS bulge, the increasing pattern speed results in a redistribution of the components of intrinsic kinematics from radial component to azimuthal component. This trend is further corroborated by analysing the observed change in orbital fractions: the contribution of $z$-tube orbits which have a large $L_z$ values increases with increasing pattern speed. The increase of the pattern speed translates to the corotation moving inwards, corresponding to a change from a slow bar to a fast bar.
The change in the distribution function is however not entirely compensated by the angular component and also results in a small change of the vertical velocity distribution, introducing a difference in the $\chi^2$ of the \acronym{LOS} kinematics of the purely face-on case.

The stability of the pattern {speed} recovery with respect to the inclination and the accuracy of other parameters provokes an interesting question: could it even be possible to constrain $\Omega_p$ with a dynamical modelling code without any kinematic data at all? This is particularly relevant in the light of projects like Euclid or LSST, which are expected to deliver huge amounts of very high-resolution photometry data for many barred galaxies that are not covered by the existing spectroscopic surveys. We intend to investigate this matter in a future paper.

\section*{Acknowledgements}

The computations for this work were performed on the HPC system Viper at the Max Planck Computing and Data Facility.
We acknowledge a substantial use of the following software in analysing the results of our work: \swn{julia} \citep{Bezanson_etal2017}, \swn{Makie.jl}, \citep{DanischKrumbiegel2021}, and \swn{NaturalNeighbours.jl} \citep{VandenHeuvel2024}.

The authors thank the anonymous referee for their review and appreciate the comments which contributed to improving the quality of the paper.
IT would also like to express gratitude to Aakash Pandey for many late-night coffee conversations about the Milky Way bar and to Stavros Pastras for briefing on pattern speed measurements in real galaxies.

\section*{Data Availability}
The data underlying this article will be shared on reasonable request to the corresponding author.



\bibliographystyle{mnras}
\bibliography{refs,refs-euclid}

@article{ECQ1bars_etal2025,
  title = {Euclid {{Quick Data Release}} ({{Q1}}), {{A}} First Look at the Fraction of Bars in Massive Galaxies at $z<1$},
  author = {{Euclid Collaboration: Huertas-Company}, M. and Walmsley, M. and Siudek, M. and {Iglesias-Navarro}, P. and Knapen, J. H. and Serjeant, S. and Dickinson, H. J. and Fortson, L. and Garland, I. and Géron, T. and Keel, W. and Kruk, S. and Lintott, C. J. and Mantha, K. and Masters, K. and O'Ryan, D. and Popp, J. J. and Roberts, H. and Scarlata, C. and Makechemu, J. S. and Simmons, B. and Smethurst, R. J. and Spindler, A. and Baes, M. and Corsini, E. M. and Sánchez, H. Domínguez and {Duran-Camacho}, E. and Fu, H. and Junais, J. and {Mendez-Abreu}, J. and Nersesian, A. and Shankar, F. and Le, M. N. and {Vega-Ferrero}, J. and Wang, L. and Aghanim, N. and Altieri, B. and Amara, A. and Andreon, S. and Auricchio, N. and Baccigalupi, C. and Baldi, M. and Balestra, A. and Bardelli, S. and Basset, A. and Battaglia, P. and Bernardeau, F. and Biviano, A. and Bonchi, A. and Branchini, E. and Brescia, M. and Brinchmann, J. and Camera, S. and Capobianco, V. and Carbone, C. and Carretero, J. and Casas, S. and Castellano, M. and Castignani, G. and Cavuoti, S. and Chambers, K. C. and Cimatti, A. and {Colodro-Conde}, C. and Congedo, G. and Conselice, C. J. and Conversi, L. and Copin, Y. and Courbin, F. and Courtois, H. M. and Cropper, M. and Silva, A. Da and Degaudenzi, H. and Lucia, G. De and Giorgio, A. M. Di and Dolding, C. and Dole, H. and Dubath, F. and Duncan, C. A. J. and Dupac, X. and Dusini, S. and Ealet, A. and Escoffier, S. and Fabricius, M. and Farina, M. and Farinelli, R. and Faustini, F. and Ferriol, S. and Finelli, F. and Fotopoulou, S. and Frailis, M. and Galeotta, S. and George, K. and Gillard, W. and Gillis, B. and Giocoli, C. and {Gracia-Carpio}, J. and Grazian, A. and Grupp, F. and Gwyn, S. and Haugan, S. V. H. and Hoekstra, H. and Holmes, W. and Hook, I. M. and Hormuth, F. and Hornstrup, A. and Hudelot, P. and Jahnke, K. and Jhabvala, M. and Keihänen, E. and Kermiche, S. and Kubik, B. and Kuijken, K. and Kümmel, M. and Kunz, M. and {Kurki-Suonio}, H. and Boulc'h, Q. Le and Brun, A. M. C. Le and Mignant, D. Le and Ligori, S. and Lilje, P. B. and Lindholm, V. and Lloro, I. and Maino, D. and Maiorano, E. and Mansutti, O. and Marcin, S. and Marggraf, O. and Martinelli, M. and Martinet, N. and Marulli, F. and Massey, R. and McCracken, H. J. and Medinaceli, E. and Melchior, M. and Mellier, Y. and Meneghetti, M. and Merlin, E. and Meylan, G. and Mora, A. and Moresco, M. and Moscardini, L. and Neissner, C. and Nichol, R. C. and Niemi, S.-M. and Nightingale, J. W. and Padilla, C. and Paltani, S. and Pasian, F. and Pedersen, K. and Percival, W. J. and Pettorino, V. and Pires, S. and Polenta, G. and Poncet, M. and Popa, L. A. and Pozzetti, L. and Raison, F. and Renzi, A. and Rhodes, J. and Riccio, G. and Romelli, E. and Roncarelli, M. and Saglia, R. and Sakr, Z. and Sapone, D. and Sartoris, B. and Schirmer, M. and Schneider, P. and Scodeggio, M. and Secroun, A. and Seidel, G. and Seiffert, M. and Serrano, S. and Simon, P. and Sirignano, C. and Sirri, G. and Stanco, L. and Steinwagner, J. and {Tallada-Crespí}, P. and Taylor, A. N. and Tereno, I. and Toft, S. and {Toledo-Moreo}, R. and Torradeflot, F. and Tutusaus, I. and Valenziano, L. and Valiviita, J. and Vassallo, T. and Kleijn, G. Verdoes and Wang, Y. and Weller, J. and Zacchei, A. and Zamorani, G. and Zerbi, F. M. and Zinchenko, I. A. and Zucca, E. and Allevato, V. and Ballardini, M. and Bolzonella, M. and Bozzo, E. and Burigana, C. and Cappi, A. and Ferdinando, D. Di and Vigo, J. A. Escartin and Gabarra, L. and {Martín-Fleitas}, J. and Matthew, S. and Mauri, N. and Metcalf, R. B. and Pezzotta, A. and Pöntinen, M. and Porciani, C. and Risso, I. and Scottez, V. and Sereno, M. and Tenti, M. and Viel, M. and Wiesmann, M. and Akrami, Y. and Andika, I. T. and Anselmi, S. and Archidiacono, M. and {Atrio-Barandela}, F. and Benoist, C. and Benson, K. and Bertacca, D. and Bethermin, M. and Bisigello, L. and Blanchard, A. and Blot, L. and Böhringer, H. and Borgani, S. and Brown, M. L. and Bruton, S. and Calabro, A. and Quevedo, B. Camacho and Caro, F. and Carvalho, C. S. and Castro, T. and Charles, Y. and Cogato, F. and Contini, T. and Cooray, A. R. and Cucciati, O. and Davini, S. and Paolis, F. De and Desprez, G. and {Díaz-Sánchez}, A. and Diaz, J. J. and Domizio, S. Di and Diego, J. M. and Duc, P.-A. and Enia, A. and Fang, Y. and Ferrari, A. G. and Finoguenov, A. and Fontana, A. and Fontanot, F. and Franco, A. and Ganga, K. and {García-Bellido}, J. and Gasparetto, T. and Gautard, V. and Gaztanaga, E. and Giacomini, F. and Gozaliasl, G. and Guidi, M. and Gutierrez, C. M. and Hall, A. and Hartley, W. G. and Hemmati, S. and {Hernández-Monteagudo}, C. and Hildebrandt, H. and Hjorth, J. and Kajava, J. J. E. and Kang, Y. and Kansal, V. and Karagiannis, D. and Kiiveri, K. and Kirkpatrick, C. C. and Graet, J. Le and Legrand, L. and Lembo, M. and Lepori, F. and Leroy, G. and Lesci, G. F. and Lesgourgues, J. and Leuzzi, L. and Liaudat, T. I. and Loureiro, A. and {Macias-Perez}, J. and Maggio, G. and Magliocchetti, M. and Magnier, E. A. and Mannucci, F. and Maoli, R. and Martins, C. J. A. P. and Maurin, L. and Miluzio, M. and Monaco, P. and Moretti, C. and Morgante, G. and Murray, C. and Naidoo, K. and {Navarro-Alsina}, A. and Nesseris, S. and Passalacqua, F. and Paterson, K. and Patrizii, L. and Pisani, A. and Potter, D. and Quai, S. and Radovich, M. and Rodighiero, G. and Sacquegna, S. and Sahlén, M. and Sanders, D. B. and Sarpa, E. and Schaye, J. and Schneider, A. and Schultheis, M. and Sciotti, D. and Sellentin, E. and Smith, L. C. and Tanidis, K. and Testera, G. and Teyssier, R. and Tosi, S. and Troja, A. and Tucci, M. and Valieri, C. and Venhola, A. and Vergani, D. and Verza, G. and Vielzeuf, P. and Walton, N. A. and Soubrie, E. and Scott, D.},
  year = 2025,
  month = mar,
  journal = {A\&A, submitted (Euclid Q1 SI)},
  eprint = {2503.15311},
  primaryclass = {astro-ph},
  doi = {10.48550/arXiv.2503.15311},
  archiveprefix = {arXiv},
  langid = {english}
}

@article{Aguerri_etal2015,
  title = {Bar Pattern Speeds in {{CALIFA}} Galaxies. {{I}}. {{Fast}} Bars across the {{Hubble}} Sequence},
  author = {Aguerri, J. A. L. and {Méndez-Abreu}, J. and {Falcón-Barroso}, J. and Amorin, A. and {Barrera-Ballesteros}, J. and Cid Fernandes, R. and {García-Benito}, R. and {García-Lorenzo}, B. and González Delgado, R. M. and Husemann, B. and Kalinova, V. and Lyubenova, M. and Marino, R. A. and Márquez, I. and Mast, D. and Pérez, E. and Sánchez, S. F. and {van de Ven}, G. and Walcher, C. J. and Backsmann, N. and {Cortijo-Ferrero}, C. and {Bland-Hawthorn}, J. and {del Olmo}, A. and {Iglesias-Páramo}, J. and Pérez, I. and {Sánchez-Blázquez}, P. and Wisotzki, L. and Ziegler, B.},
  year = 2015,
  month = apr,
  journal = {A\&A},
  volume = {576},
  pages = {A102},
  publisher = {EDP},
  issn = {0004-6361},
  doi = {10.1051/0004-6361/201423383}
}

@article{Algorry_etal2017,
  title = {Barred Galaxies in the {{EAGLE}} Cosmological Hydrodynamical Simulation},
  author = {Algorry, David G. and Navarro, Julio F. and Abadi, Mario G. and Sales, Laura V. and Bower, Richard G. and Crain, Robert A. and Dalla Vecchia, Claudio and Frenk, Carlos S. and Schaller, Matthieu and Schaye, Joop and Theuns, Tom},
  year = 2017,
  month = jul,
  journal = {MNRAS},
  volume = {469},
  pages = {1054--1064},
  publisher = {OUP},
  issn = {0035-8711},
  doi = {10.1093/mnras/stx1008}
}

@article{Athanassoula2003,
  title = {What Determines the Strength and the Slowdown Rate of Bars?},
  author = {Athanassoula, E.},
  year = 2003,
  month = jun,
  journal = {MNRAS},
  volume = {341},
  pages = {1179--1198},
  publisher = {OUP},
  issn = {0035-8711},
  doi = {10.1046/j.1365-8711.2003.06473.x}
}

@inproceedings{Athanassoula2016,
  title = {Boxy/{{Peanut}}/{{X Bulges}}, {{Barlenses}} and the {{Thick Part}} of {{Galactic Bars}}: {{What Are They}} and {{How Did They Form}}?},
  shorttitle = {Boxy/{{Peanut}}/{{X Bulges}}, {{Barlenses}} and the {{Thick Part}} of {{Galactic Bars}}},
  booktitle = {Galactic {{Bulges}}},
  author = {Athanassoula, E.},
  year = 2016,
  month = jan,
  volume = {418},
  pages = {391},
  address = {eprint: arXiv:1503.04804},
  doi = {10.1007/978-3-319-19378-6_14}
}

@article{Bezanson_etal2017,
  title = {Julia: {{A Fresh Approach}} to {{Numerical Computing}}},
  shorttitle = {Julia},
  author = {Bezanson, Jeff and Edelman, Alan and Karpinski, Stefan and Shah, Viral B.},
  year = 2017,
  month = jan,
  journal = {SIAM Rev.},
  volume = {59},
  number = {1},
  pages = {65--98},
  publisher = {{Society for Industrial and Applied Mathematics}},
  issn = {0036-1445},
  doi = {10.1137/141000671}
}

@misc{Bland-Hawthorn_etal2024,
  title = {Turbulent Gas-Rich Disks at High Redshift: Bars \& Bulges in a Radial Shear Flow},
  shorttitle = {Turbulent Gas-Rich Disks at High Redshift},
  author = {{Bland-Hawthorn}, Joss and {Tepper-Garcia}, Thor and Agertz, Oscar and Federrath, Christoph},
  year = 2024,
  month = feb,
  number = {arXiv:2402.06060},
  eprint = {2402.06060},
  primaryclass = {astro-ph},
  publisher = {arXiv},
  doi = {10.48550/arXiv.2402.06060},
  archiveprefix = {arXiv}
}

@article{Borodina_etal2023,
  title = {On the {{Tremaine-Weinberg}} Method: How Much Can We Trust Gas Tracers to Measure Pattern Speeds?},
  shorttitle = {On the {{Tremaine-Weinberg}} Method},
  author = {Borodina, Olga and Williams, Thomas G. and Sormani, Mattia C. and Meidt, Sharon and Schinnerer, Eva},
  year = 2023,
  month = sep,
  journal = {MNRAS},
  volume = {524},
  pages = {3437--3445},
  publisher = {OUP},
  issn = {0035-8711},
  doi = {10.1093/mnras/stad2068}
}

@article{Bundy_etal2015,
  title = {Overview of the {{SDSS-IV MaNGA Survey}}: {{Mapping}} Nearby {{Galaxies}} at {{Apache Point Observatory}}},
  shorttitle = {Overview of the {{SDSS-IV MaNGA Survey}}},
  author = {Bundy, Kevin and Bershady, Matthew A. and Law, David R. and Yan, Renbin and Drory, Niv and MacDonald, Nicholas and Wake, David A. and Cherinka, Brian and {Sánchez-Gallego}, José R. and Weijmans, Anne-Marie and Thomas, Daniel and Tremonti, Christy and Masters, Karen and Coccato, Lodovico and {Diamond-Stanic}, Aleksandar M. and {Aragón-Salamanca}, Alfonso and {Avila-Reese}, Vladimir and Badenes, Carles and {Falcón-Barroso}, Jésus and Belfiore, Francesco and Bizyaev, Dmitry and Blanc, Guillermo A. and {Bland-Hawthorn}, Joss and Blanton, Michael R. and Brownstein, Joel R. and Byler, Nell and Cappellari, Michele and Conroy, Charlie and Dutton, Aaron A. and Emsellem, Eric and Etherington, James and Frinchaboy, Peter M. and Fu, Hai and Gunn, James E. and Harding, Paul and Johnston, Evelyn J. and Kauffmann, Guinevere and Kinemuchi, Karen and Klaene, Mark A. and Knapen, Johan H. and Leauthaud, Alexie and Li, Cheng and Lin, Lihwai and Maiolino, Roberto and Malanushenko, Viktor and Malanushenko, Elena and Mao, Shude and Maraston, Claudia and McDermid, Richard M. and Merrifield, Michael R. and Nichol, Robert C. and Oravetz, Daniel and Pan, Kaike and Parejko, John K. and Sanchez, Sebastian F. and Schlegel, David and Simmons, Audrey and Steele, Oliver and Steinmetz, Matthias and Thanjavur, Karun and Thompson, Benjamin A. and Tinker, Jeremy L. and {van den Bosch}, Remco C. E. and Westfall, Kyle B. and Wilkinson, David and Wright, Shelley and Xiao, Ting and Zhang, Kai},
  year = 2015,
  month = jan,
  journal = {ApJ},
  volume = {798},
  pages = {7},
  issn = {0004-637X},
  doi = {10.1088/0004-637X/798/1/7}
}

@article{Buta_etal2015,
  title = {A {{Classical Morphological Analysis}} of {{Galaxies}} in the {{Spitzer Survey}} of {{Stellar Structure}} in {{Galaxies}} ({{S4G}})},
  author = {Buta, Ronald J. and Sheth, Kartik and Athanassoula, E. and Bosma, A. and Knapen, Johan H. and Laurikainen, Eija and Salo, Heikki and Elmegreen, Debra and Ho, Luis C. and Zaritsky, Dennis and Courtois, Helene and Hinz, Joannah L. and {Muñoz-Mateos}, Juan-Carlos and Kim, Taehyun and Regan, Michael W. and Gadotti, Dimitri A. and {Gil de Paz}, Armando and Laine, Jarkko and {Menéndez-Delmestre}, Karín and Comerón, Sébastien and Erroz Ferrer, Santiago and Seibert, Mark and Mizusawa, Trisha and Holwerda, Benne and Madore, Barry F.},
  year = 2015,
  month = apr,
  journal = {ApJS},
  volume = {217},
  pages = {32},
  publisher = {IOP},
  issn = {0067-0049},
  doi = {10.1088/0067-0049/217/2/32}
}

@article{Cappellari_Copin2003,
  title = {Adaptive Spatial Binning of Integral-Field Spectroscopic Data Using {{Voronoi}} Tessellations},
  author = {Cappellari, Michele and Copin, Yannick},
  year = 2003,
  month = jun,
  journal = {MNRAS},
  volume = {342},
  pages = {345--354},
  publisher = {OUP},
  issn = {0035-8711},
  doi = {10.1046/j.1365-8711.2003.06541.x}
}

@article{Cappellari_etal2009,
  title = {The Mass of the Black Hole in {{Centaurus A}} from {{SINFONI AO-assisted}} Integral-Field Observations of Stellar Kinematics},
  author = {Cappellari, Michele and Neumayer, N. and Reunanen, J. and {van der Werf}, P. P. and {de Zeeuw}, P. T. and Rix, H.-W.},
  year = 2009,
  month = apr,
  journal = {MNRAS},
  volume = {394},
  number = {2},
  pages = {660--674},
  issn = {0035-8711},
  doi = {10.1111/j.1365-2966.2008.14377.x}
}

@article{Cappellari_etal2011,
  title = {The {{ATLAS3D}} Project - {{VII}}. {{A}} New Look at the Morphology of Nearby Galaxies: The Kinematic Morphology-Density Relation},
  shorttitle = {The {{ATLAS3D}} Project - {{VII}}. {{A}} New Look at the Morphology of Nearby Galaxies},
  author = {Cappellari, Michele and Emsellem, Eric and Krajnović, Davor and McDermid, Richard M. and Serra, Paolo and Alatalo, Katherine and Blitz, Leo and Bois, Maxime and Bournaud, Frédéric and Bureau, M. and Davies, Roger L. and Davis, Timothy A. and {de Zeeuw}, P. T. and Khochfar, Sadegh and Kuntschner, Harald and Lablanche, Pierre-Yves and Morganti, Raffaella and Naab, Thorsten and Oosterloo, Tom and Sarzi, Marc and Scott, Nicholas and Weijmans, Anne-Marie and Young, Lisa M.},
  year = 2011,
  month = sep,
  journal = {MNRAS},
  volume = {416},
  pages = {1680--1696},
  issn = {0035-8711},
  doi = {10.1111/j.1365-2966.2011.18600.x}
}

@article{Cash_Karp1990,
  title = {A Variable Order {{Runge-Kutta}} Method for Initial Value Problems with Rapidly Varying Right-Hand Sides},
  author = {Cash, J. R. and Karp, Alan H.},
  year = 1990,
  month = sep,
  journal = {ACM Trans. Math. Softw.},
  volume = {16},
  number = {3},
  pages = {201--222},
  issn = {0098-3500},
  doi = {10.1145/79505.79507}
}

@article{Ceverino_Klypin2007,
  title = {Resonances in Barred Galaxies},
  author = {Ceverino, D. and Klypin, A.},
  year = 2007,
  month = aug,
  journal = {MNRAS},
  volume = {379},
  pages = {1155--1168},
  issn = {0035-8711},
  doi = {10.1111/j.1365-2966.2007.12001.x}
}

@article{Chiba_Schonrich2022,
  title = {Oscillating Dynamical Friction on Galactic Bars by Trapped Dark Matter},
  author = {Chiba, Rimpei and Schönrich, Ralph},
  year = 2022,
  month = apr,
  journal = {MNRAS},
  volume = {513},
  number = {1},
  pages = {768--787},
  issn = {0035-8711, 1365-2966},
  doi = {10.1093/mnras/stac697},
  copyright = {https://creativecommons.org/licenses/by/4.0/},
  langid = {english}
}

@article{Cretton_etal1999,
  title = {Axisymmetric {{Three-Integral Models}} for {{Galaxies}}},
  author = {Cretton, N. and de Zeeuw, P. Tim and van der Marel, Roeland P. and Rix, Hans-Walter},
  year = 1999,
  month = oct,
  journal = {ApJS},
  volume = {124},
  number = {2},
  pages = {383},
  publisher = {IOP Publishing},
  issn = {0067-0049},
  doi = {10.1086/313264},
  langid = {english}
}

@article{Croom_etal2021,
  title = {The {{SAMI Galaxy Survey}}: The Third and Final Data Release},
  shorttitle = {The {{SAMI Galaxy Survey}}},
  author = {Croom, Scott M and Owers, Matt S and Scott, Nicholas and Poetrodjojo, Henry and Groves, Brent and {van~de~Sande}, Jesse and Barone, Tania M and Cortese, Luca and D’Eugenio, Francesco and {Bland-Hawthorn}, Joss and Bryant, Julia and Oh, Sree and Brough, Sarah and Agostino, James and Casura, Sarah and Catinella, Barbara and Colless, Matthew and Cecil, Gerald and Davies, Roger L and Drinkwater, Michael J and Driver, Simon P and Ferreras, Ignacio and Foster, Caroline and {Fraser-McKelvie}, Amelia and Lawrence, Jon and Leslie, Sarah K and Liske, Jochen and {López-Sánchez}, Ángel R and Lorente, Nuria P F and McElroy, Rebecca and Medling, Anne M and Obreschkow, Danail and Richards, Samuel N and Sharp, Rob and Sweet, Sarah M and Taranu, Dan S and Taylor, Edward N and Tescari, Edoardo and Thomas, Adam D and Tocknell, James and Vaughan, Sam P},
  year = 2021,
  month = jul,
  journal = {MNRAS},
  volume = {505},
  number = {1},
  pages = {991--1016},
  issn = {0035-8711},
  doi = {10.1093/mnras/stab229}
}

@article{Cuomo_etal2019,
  title = {Bar Pattern Speeds in {{CALIFA}} Galaxies. {{II}}. {{The}} Case of Weakly Barred Galaxies},
  author = {Cuomo, Virginia and Lopez Aguerri, J. Alfonso and Corsini, Enrico Maria and Debattista, Victor P. and {Méndez-Abreu}, Jairo and Pizzella, Alessandro},
  year = 2019,
  month = dec,
  journal = {A\&A},
  volume = {632},
  pages = {A51},
  publisher = {EDP},
  issn = {0004-6361},
  doi = {10.1051/0004-6361/201936415}
}

@article{DanischKrumbiegel2021,
  title = {Makie.Jl: {{Flexible}} High-Performance Data Visualization for {{Julia}}},
  author = {Danisch, Simon and Krumbiegel, Julius},
  year = 2021,
  journal = {J. Open Source Softw.},
  volume = {6},
  number = {65},
  pages = {3349},
  publisher = {The Open Journal},
  doi = {10.21105/joss.03349}
}

@article{Dattathri_etal2024,
  title = {Deprojection and Stellar Dynamical Modelling of Boxy/Peanut Bars in Edge-on Discs},
  author = {Dattathri, Shashank and Valluri, Monica and Vasiliev, Eugene and Wheeler, Vance and Erwin, Peter},
  year = 2024,
  month = may,
  journal = {MNRAS},
  volume = {530},
  pages = {1195--1217},
  publisher = {OUP},
  issn = {0035-8711},
  doi = {10.1093/mnras/stae802}
}

@article{Debattista_etal2005,
  title = {The {{Kinematic Signature}} of {{Face-On Peanut-shaped Bulges}}},
  author = {Debattista, Victor P. and Carollo, C. Marcella and Mayer, Lucio and Moore, Ben},
  year = 2005,
  month = aug,
  journal = {ApJ},
  volume = {628},
  pages = {678--694},
  issn = {0004-637X},
  doi = {10.1086/431292}
}

@article{Debattista2003,
  title = {On Position Angle Errors in the {{Tremaine-Weinberg}} Method},
  author = {Debattista, Victor P.},
  year = 2003,
  month = jul,
  journal = {MNRAS},
  volume = {342},
  pages = {1194--1204},
  publisher = {OUP},
  issn = {0035-8711},
  doi = {10.1046/j.1365-8711.2003.06620.x}
}

@article{Dehnen_etal2022,
  title = {Measuring Bar Pattern Speeds from Single Simulation Snapshots},
  author = {Dehnen, Walter and Semczuk, Marcin and Schönrich, Ralph},
  year = 2022,
  month = nov,
  journal = {MNRAS},
  volume = {518},
  number = {2},
  pages = {2712--2718},
  issn = {0035-8711, 1365-2966},
  doi = {10.1093/mnras/stac3184},
  copyright = {https://academic.oup.com/journals/pages/open\_access/funder\_policies/chorus/standard\_publication\_model},
  langid = {english}
}

@article{Dehnen2002,
  title = {A {{Hierarchical O}}({{N}}) {{Force Calculation Algorithm}}},
  author = {Dehnen, Walter},
  year = 2002,
  month = jun,
  journal = {J. Comput. Phys.},
  volume = {179},
  pages = {27--42},
  issn = {0021-9991},
  doi = {10.1006/jcph.2002.7026}
}

@article{denBrok_etal2021,
  title = {Dynamical Modelling of the Twisted Galaxy {{PGC}} 046832},
  author = {{den Brok}, Mark and Krajnović, Davor and Emsellem, Eric and Brinchmann, Jarle and Maseda, Michael},
  year = 2021,
  month = dec,
  journal = {MNRAS},
  volume = {508},
  pages = {4786--4805},
  publisher = {OUP},
  issn = {0035-8711},
  doi = {10.1093/mnras/stab2852}
}

@article{deNicola_etal2022,
  title = {Accuracy and Precision of Triaxial Orbit Models – {{II}}. {{Viewing}} Angles, Shape, and Orbital Structure},
  author = {{de~Nicola}, Stefano and Neureiter, Bianca and Thomas, Jens and Saglia, Roberto P and Bender, Ralf},
  year = 2022,
  month = oct,
  journal = {MNRAS},
  volume = {517},
  number = {3},
  pages = {3445--3458},
  issn = {0035-8711, 1365-2966},
  doi = {10.1093/mnras/stac2852},
  copyright = {https://academic.oup.com/journals/pages/open\_access/funder\_policies/chorus/standard\_publication\_model},
  langid = {english}
}

@article{Erwin2018,
  title = {The Dependence of Bar Frequency on Galaxy Mass, Colour, and Gas Content - and Angular Resolution - in the Local Universe},
  author = {Erwin, Peter},
  year = 2018,
  month = mar,
  journal = {MNRAS},
  volume = {474},
  pages = {5372--5392},
  publisher = {OUP},
  issn = {0035-8711},
  doi = {10.1093/mnras/stx3117}
}

@article{Falcon-Barroso_Martig2021,
  title = {{{BAYES-LOSVD}}: {{A Bayesian}} Framework for Non-Parametric Extraction of the Line-of-Sight Velocity Distribution of Galaxies},
  shorttitle = {{{BAYES-LOSVD}}},
  author = {{Falcón-Barroso}, J. and Martig, M.},
  year = 2021,
  month = feb,
  journal = {A\&A},
  volume = {646},
  pages = {A31},
  publisher = {EDP},
  issn = {0004-6361},
  doi = {10.1051/0004-6361/202039624}
}

@article{Fragkoudi_etal2021,
  title = {Revisiting the Tension between Fast Bars and the {{ΛCDM}} Paradigm},
  author = {Fragkoudi, F. and Grand, R. J. J. and Pakmor, R. and Springel, V. and White, S. D. M. and Marinacci, F. and Gomez, F. A. and Navarro, J. F.},
  year = 2021,
  month = jun,
  journal = {A\&A},
  volume = {650},
  pages = {L16},
  publisher = {EDP},
  issn = {0004-6361},
  doi = {10.1051/0004-6361/202140320}
}

@article{Frankel_etal2022,
  title = {Simulated {{Bars May Be Shorter}} but {{Are Not Slower Than Those Observed}}: {{TNG50}} versus {{MaNGA}}},
  shorttitle = {Simulated {{Bars May Be Shorter}} but {{Are Not Slower Than Those Observed}}},
  author = {Frankel, Neige and Pillepich, Annalisa and Rix, Hans-Walter and {Rodriguez-Gomez}, Vicente and Sanders, Jason and Bovy, Jo and Kollmeier, Juna and Murray, Norm and Mackereth, Ted},
  year = 2022,
  month = nov,
  journal = {ApJ},
  volume = {940},
  number = {1},
  pages = {61},
  issn = {0004-637X, 1538-4357},
  doi = {10.3847/1538-4357/ac9972},
  langid = {english}
}

@article{Gadotti_etal2019,
  title = {Time {{Inference}} with {{MUSE}} in {{Extragalactic Rings}} ({{TIMER}}): Properties of the Survey and High-Level Data Products},
  shorttitle = {Time {{Inference}} with {{MUSE}} in {{Extragalactic Rings}} ({{TIMER}})},
  author = {Gadotti, Dimitri A and {Sánchez-Blázquez}, Patricia and {Falcón-Barroso}, Jesús and Husemann, Bernd and Seidel, Marja K and Pérez, Isabel and {de~Lorenzo-Cáceres}, Adriana and {Martinez-Valpuesta}, Inma and Fragkoudi, Francesca and Leung, Gigi and {van~de~Ven}, Glenn and Leaman, Ryan and Coelho, Paula and Martig, Marie and Kim, Taehyun and Neumann, Justus and Querejeta, Miguel},
  year = 2019,
  month = jan,
  journal = {MNRAS},
  volume = {482},
  number = {1},
  pages = {506--529},
  issn = {0035-8711},
  doi = {10.1093/mnras/sty2666}
}

@article{Garma-Oehmichen_etal2020,
  title = {{{SDSS-IV MaNGA}}: Bar Pattern Speed Estimates with the {{Tremaine-Weinberg}} Method and Their Error Sources},
  shorttitle = {{{SDSS-IV MaNGA}}},
  author = {{Garma-Oehmichen}, L. and {Cano-Díaz}, M. and {Hernández-Toledo}, H. and {Aquino-Ortíz}, E. and Valenzuela, O. and Aguerri, J. A. L. and Sánchez, S. F. and Merrifield, M.},
  year = 2020,
  month = jan,
  journal = {MNRAS},
  volume = {491},
  pages = {3655--3671},
  publisher = {OUP},
  issn = {0035-8711},
  doi = {10.1093/mnras/stz3101}
}

@article{Garma-Oehmichen_etal2022,
  title = {{{SDSS IV MaNGA}}: Bar Pattern Speed in {{Milky Way}} Analogue Galaxies},
  shorttitle = {{{SDSS IV MaNGA}}},
  author = {{Garma-Oehmichen}, L. and {Hernández-Toledo}, H. and {Aquino-Ortíz}, E. and {Martinez-Medina}, L. and Puerari, I. and {Cano-Díaz}, M. and Valenzuela, O. and {Vázquez-Mata}, J. A. and Géron, T. and {Martínez-Vázquez}, L. A. and Lane, R.},
  year = 2022,
  month = dec,
  journal = {MNRAS},
  volume = {517},
  pages = {5660--5677},
  publisher = {OUP},
  issn = {0035-8711},
  doi = {10.1093/mnras/stac3069}
}

@article{Gebhardt_etal2000a,
  title = {Axisymmetric, {{Three-Integral Models}} of {{Galaxies}}: {{A Massive Black Hole}} in {{NGC}} 3379},
  shorttitle = {Axisymmetric, {{Three-Integral Models}} of {{Galaxies}}},
  author = {Gebhardt, Karl and Richstone, Douglas and Kormendy, John and Lauer, Tod R. and Ajhar, Edward A. and Bender, Ralf and Dressler, Alan and Faber, S. M. and Grillmair, Carl and Magorrian, John and Tremaine, Scott},
  year = 2000,
  month = mar,
  journal = {ApJ},
  volume = {119},
  pages = {1157--1171},
  publisher = {IOP},
  issn = {0004-6256},
  doi = {10.1086/301240}
}

@article{Gebhardt_etal2003,
  title = {Axisymmetric {{Dynamical Models}} of the {{Central Regions}} of {{Galaxies}}},
  author = {Gebhardt, Karl and Richstone, Douglas and Tremaine, Scott and Lauer, Tod R. and Bender, Ralf and Bower, Gary and Dressler, Alan and Faber, S. M. and Filippenko, Alexei V. and Green, Richard and Grillmair, Carl and Ho, Luis C. and Kormendy, John and Magorrian, John and Pinkney, Jason},
  year = 2003,
  month = jan,
  journal = {ApJ},
  volume = {583},
  number = {1},
  pages = {92},
  publisher = {IOP Publishing},
  issn = {0004-637X},
  doi = {10.1086/345081},
  langid = {english}
}

@article{Gebhardt_etal2007,
  title = {The {{Black Hole Mass}} and {{Extreme Orbital Structure}} in {{NGC}} 1399},
  author = {Gebhardt, Karl and Lauer, Tod R. and Pinkney, Jason and Bender, Ralf and Richstone, Douglas and Aller, Monique and Bower, Gary and Dressler, Alan and Faber, S. M. and Filippenko, Alexei V. and Green, Richard and Ho, Luis C. and Kormendy, John and Siopis, Christos and Tremaine, Scott},
  year = 2007,
  month = dec,
  journal = {ApJ},
  volume = {671},
  number = {2},
  pages = {1321},
  publisher = {IOP Publishing},
  issn = {0004-637X},
  doi = {10.1086/522938},
  langid = {english}
}

@article{Geron_etal2023,
  title = {Galaxy {{Zoo}}: Kinematics of Strongly and Weakly Barred Galaxies},
  shorttitle = {Galaxy {{Zoo}}},
  author = {Géron, Tobias and Smethurst, Rebecca J. and Lintott, Chris and Kruk, Sandor and Masters, Karen L. and Simmons, Brooke and Mantha, Kameswara Bharadwaj and Walmsley, Mike and {Garma-Oehmichen}, L. and Drory, Niv and Lane, Richard R.},
  year = 2023,
  month = may,
  journal = {MNRAS},
  volume = {521},
  pages = {1775--1793},
  publisher = {OUP},
  issn = {0035-8711},
  doi = {10.1093/mnras/stad501}
}

@article{Grand_etal2017,
  title = {The {{Auriga Project}}: The Properties and Formation Mechanisms of Disc Galaxies across Cosmic Time},
  shorttitle = {The {{Auriga Project}}},
  author = {Grand, Robert J. J. and Gómez, Facundo A. and Marinacci, Federico and Pakmor, Rüdiger and Springel, Volker and Campbell, David J. R. and Frenk, Carlos S. and Jenkins, Adrian and White, Simon D. M.},
  year = 2017,
  month = may,
  journal = {MNRAS},
  volume = {467},
  pages = {179--207},
  publisher = {OUP},
  issn = {0035-8711},
  doi = {10.1093/mnras/stx071}
}

@article{Gultekin_etal2009a,
  title = {A {{Quintet}} of {{Black Hole Mass Determinations}}},
  author = {Gültekin, Kayhan and Richstone, Douglas O. and Gebhardt, Karl and Lauer, Tod R. and Pinkney, Jason and Aller, M. C. and Bender, Ralf and Dressler, Alan and Faber, S. M. and Filippenko, Alexei V. and Green, Richard and Ho, Luis C. and Kormendy, John and Siopis, Christos},
  year = 2009,
  month = apr,
  journal = {ApJ},
  volume = {695},
  pages = {1577--1590},
  publisher = {IOP},
  issn = {0004-637X},
  doi = {10.1088/0004-637X/695/2/1577}
}

@article{Guo_etal2019,
  title = {{{SDSS-IV MaNGA}}: Pattern Speeds of Barred Galaxies},
  shorttitle = {{{SDSS-IV MaNGA}}},
  author = {Guo, Rui and Mao, Shude and Athanassoula, E. and Li, Hongyu and Ge, Junqiang and Long, R. J. and Merrifield, Michael and Masters, Karen},
  year = 2019,
  month = jan,
  journal = {MNRAS},
  volume = {482},
  pages = {1733--1756},
  publisher = {OUP},
  issn = {0035-8711},
  doi = {10.1093/mnras/sty2715}
}

@article{Hafner_etal2000,
  title = {A Dynamical Model of the Inner {{Galaxy}}},
  author = {Häfner, Ralf and Evans, N. Wyn and Dehnen, Walter and Binney, James},
  year = 2000,
  month = may,
  journal = {MNRAS},
  volume = {314},
  pages = {433--452},
  publisher = {OUP},
  issn = {0035-8711},
  doi = {10.1046/j.1365-8711.2000.03242.x}
}

@article{Hamilton_etal2023,
  title = {Galactic {{Bar Resonances}} with {{Diffusion}}: {{An Analytic Model}} with {{Implications}} for {{Bar}}–{{Dark Matter Halo Dynamical Friction}}},
  shorttitle = {Galactic {{Bar Resonances}} with {{Diffusion}}},
  author = {Hamilton, Chris and Tolman, Elizabeth A. and Arzamasskiy, Lev and Duarte, Vinícius N.},
  year = 2023,
  month = aug,
  journal = {ApJ},
  volume = {954},
  number = {1},
  pages = {12},
  publisher = {The American Astronomical Society},
  issn = {0004-637X},
  doi = {10.3847/1538-4357/acd69b},
  langid = {english}
}

@article{Hernquist1990,
  title = {An {{Analytical Model}} for {{Spherical Galaxies}} and {{Bulges}}},
  author = {Hernquist, Lars},
  year = 1990,
  month = jun,
  journal = {ApJ},
  volume = {356},
  pages = {359},
  issn = {0004-637X},
  doi = {10.1086/168845}
}

@article{Iannuzzi_Athanassoula2015,
  title = {{{2D}} Kinematic Signatures of Boxy/Peanut Bulges},
  author = {Iannuzzi, Francesca and Athanassoula, E.},
  year = 2015,
  month = jul,
  journal = {MNRAS},
  volume = {450},
  pages = {2514--2538},
  publisher = {OUP},
  issn = {0035-8711},
  doi = {10.1093/mnras/stv764}
}

@article{Jin_etal2025a,
  title = {Recovering the Pattern Speeds of Edge-on Barred Galaxies via an Orbit-Superposition Method},
  author = {Jin, Yunpeng and Zhu, Ling and Tahmasebzadeh, Behzad and Mao, Shude and {van de Ven}, Glenn and Guo, Rui and Cai, Runsheng},
  year = 2025,
  month = aug,
  journal = {A\&A},
  volume = {700},
  pages = {A249},
  publisher = {EDP},
  issn = {0004-6361},
  doi = {10.1051/0004-6361/202555378}
}

@article{Khoperskov_etal2025,
  title = {Rediscovering the {{Milky Way}} with an Orbit Superposition Approach and {{APOGEE}} Data: {{I}}. {{Method}} Validation},
  shorttitle = {Rediscovering the {{Milky Way}} with an Orbit Superposition Approach and {{APOGEE}} Data},
  author = {Khoperskov, Sergey and {van de Ven}, Glenn and Steinmetz, Matthias and Ratcliffe, Bridget and Minchev, Ivan and Krajnović, Davor and Haywood, Misha and Di Matteo, Paola and Kacharov, Nikolay and Marques, Léa and Valentini, Marica and {de Jong}, Roelof S.},
  year = 2025,
  month = mar,
  journal = {A\&A},
  volume = {695},
  pages = {A220},
  publisher = {EDP},
  issn = {0004-6361},
  doi = {10.1051/0004-6361/202453304}
}

@article{Krajnovic_etal2009,
  title = {Determination of Masses of the Central Black Holes in {{NGC}} 524 and 2549 Using Laser Guide Star Adaptive Optics},
  author = {Krajnović, Davor and McDermid, Richard M. and Cappellari, Michele and Davies, Roger L.},
  year = 2009,
  month = nov,
  journal = {MNRAS},
  volume = {399},
  number = {4},
  pages = {1839--1857},
  issn = {0035-8711},
  doi = {10.1111/j.1365-2966.2009.15415.x}
}

@article{Li_etal2018,
  title = {Shape of {{LOSVDs}} in {{Barred Disks}}: {{Implications}} for {{Future IFU Surveys}}},
  shorttitle = {Shape of {{LOSVDs}} in {{Barred Disks}}},
  author = {Li, Zhao-Yu and Shen, Juntai and Bureau, Martin and Zhou, Yingying and Du, Min and Debattista, Victor P.},
  year = 2018,
  month = feb,
  journal = {ApJ},
  volume = {854},
  pages = {65},
  issn = {0004-637X},
  doi = {10.3847/1538-4357/aaa771}
}

@article{Lipka_Thomas2021,
  title = {A Novel Approach to Optimize the Regularization and Evaluation of Dynamical Models Using a Model Selection Framework},
  author = {Lipka, Mathias and Thomas, Jens},
  year = 2021,
  month = jul,
  journal = {MNRAS},
  volume = {504},
  pages = {4599--4625},
  issn = {0035-8711},
  doi = {10.1093/mnras/stab1092}
}

@article{Ma_etal2014,
  title = {{{THE MASSIVE SURVEY}}. {{I}}. {{A VOLUME-LIMITED INTEGRAL-FIELD SPECTROSCOPIC STUDY OF THE MOST MASSIVE EARLY-TYPE GALAXIES WITHIN}} 108 {{Mpc}}},
  author = {Ma, Chung-Pei and Greene, Jenny E. and McConnell, Nicholas and Janish, Ryan and Blakeslee, John P. and Thomas, Jens and Murphy, Jeremy D.},
  year = 2014,
  month = oct,
  journal = {ApJ},
  volume = {795},
  number = {2},
  pages = {158},
  publisher = {The American Astronomical Society},
  issn = {0004-637X},
  doi = {10.1088/0004-637X/795/2/158},
  langid = {english}
}

@article{Marel_etal1998,
  title = {Improved {{Evidence}} for a {{Black Hole}} in {{M32}} from {{HST}}/{{FOS Spectra}}. {{II}}. {{Axisymmetric Dynamical Models}}*},
  author = {van der Marel, Roeland P. and Cretton, N. and de Zeeuw, P. Tim and Rix, Hans-Walter},
  year = 1998,
  month = feb,
  journal = {ApJ},
  volume = {493},
  number = {2},
  pages = {613},
  publisher = {IOP Publishing},
  issn = {0004-637X},
  doi = {10.1086/305147},
  langid = {english}
}

@article{Marinacci_etal2018,
  title = {First Results from the {{IllustrisTNG}} Simulations: Radio Haloes and Magnetic Fields},
  shorttitle = {First Results from the {{IllustrisTNG}} Simulations},
  author = {Marinacci, Federico and Vogelsberger, Mark and Pakmor, Rüdiger and Torrey, Paul and Springel, Volker and Hernquist, Lars and Nelson, Dylan and Weinberger, Rainer and Pillepich, Annalisa and Naiman, Jill and Genel, Shy},
  year = 2018,
  month = nov,
  journal = {MNRAS},
  volume = {480},
  pages = {5113--5139},
  publisher = {OUP},
  issn = {0035-8711},
  doi = {10.1093/mnras/sty2206}
}

@article{McMillan_Dehnen2007,
  title = {Initial Conditions for Disc Galaxies},
  author = {McMillan, Paul J. and Dehnen, Walter},
  year = 2007,
  month = jun,
  journal = {MNRAS},
  volume = {378},
  pages = {541--550},
  issn = {0035-8711},
  doi = {10.1111/j.1365-2966.2007.11753.x}
}

@article{Mehrgan_etal2019,
  title = {A 40 {{Billion Solar-mass Black Hole}} in the {{Extreme Core}} of {{Holm 15A}}, the {{Central Galaxy}} of {{Abell}} 85},
  author = {Mehrgan, Kianusch and Thomas, Jens and Saglia, Roberto and Mazzalay, Ximena and Erwin, Peter and Bender, Ralf and Kluge, Matthias and Fabricius, Maximilian},
  year = 2019,
  month = dec,
  journal = {ApJ},
  volume = {887},
  pages = {195},
  publisher = {IOP},
  issn = {0004-637X},
  doi = {10.3847/1538-4357/ab5856}
}

@article{Meidt_etal2008,
  title = {Tests of the {{Radial Tremaine-Weinberg Method}}},
  author = {Meidt, Sharon E. and Rand, Richard J. and Merrifield, Michael R. and Debattista, Victor P. and Shen, Juntai},
  year = 2008,
  month = apr,
  journal = {ApJ},
  volume = {676},
  pages = {899--919},
  issn = {0004-637X},
  doi = {10.1086/527530}
}

@article{Merrifield_Kuijken1995,
  title = {The Pattern Speed of the Bar in {{NGC}} 936},
  author = {Merrifield, Michael R. and Kuijken, Konrad},
  year = 1995,
  month = jun,
  journal = {MNRAS},
  volume = {274},
  pages = {933--938},
  issn = {0035-8711},
  doi = {10.1093/mnras/274.3.933}
}

@article{Naiman_etal2018,
  title = {First Results from the {{IllustrisTNG}} Simulations: A Tale of Two Elements - Chemical Evolution of Magnesium and Europium},
  shorttitle = {First Results from the {{IllustrisTNG}} Simulations},
  author = {Naiman, Jill P. and Pillepich, Annalisa and Springel, Volker and {Ramirez-Ruiz}, Enrico and Torrey, Paul and Vogelsberger, Mark and Pakmor, Rüdiger and Nelson, Dylan and Marinacci, Federico and Hernquist, Lars and Weinberger, Rainer and Genel, Shy},
  year = 2018,
  month = jun,
  journal = {MNRAS},
  volume = {477},
  pages = {1206--1224},
  publisher = {OUP},
  issn = {0035-8711},
  doi = {10.1093/mnras/sty618}
}

@article{Nelson_etal2018,
  title = {First Results from the {{IllustrisTNG}} Simulations: The Galaxy Colour Bimodality},
  shorttitle = {First Results from the {{IllustrisTNG}} Simulations},
  author = {Nelson, Dylan and Pillepich, Annalisa and Springel, Volker and Weinberger, Rainer and Hernquist, Lars and Pakmor, Rüdiger and Genel, Shy and Torrey, Paul and Vogelsberger, Mark and Kauffmann, Guinevere and Marinacci, Federico and Naiman, Jill},
  year = 2018,
  month = mar,
  journal = {MNRAS},
  volume = {475},
  pages = {624--647},
  publisher = {OUP},
  issn = {0035-8711},
  doi = {10.1093/mnras/stx3040}
}

@article{Nelson_etal2019,
  title = {First Results from the {{TNG50}} Simulation: Galactic Outflows Driven by Supernovae and Black Hole Feedback},
  shorttitle = {First Results from the {{TNG50}} Simulation},
  author = {Nelson, Dylan and Pillepich, Annalisa and Springel, Volker and Pakmor, Rüdiger and Weinberger, Rainer and Genel, Shy and Torrey, Paul and Vogelsberger, Mark and Marinacci, Federico and Hernquist, Lars},
  year = 2019,
  month = dec,
  journal = {MNRAS},
  volume = {490},
  pages = {3234--3261},
  publisher = {OUP},
  issn = {0035-8711},
  doi = {10.1093/mnras/stz2306}
}

@article{Neureiter_etal2021,
  title = {{{SMART}}: A New Implementation of {{Schwarzschild}}'s {{Orbit Superposition}} Technique for Triaxial Galaxies and Its Application to an {{N-body}} Merger Simulation},
  shorttitle = {{{SMART}}},
  author = {Neureiter, B. and Thomas, J. and Saglia, R. and Bender, R. and Finozzi, F. and Krukau, A. and Naab, T. and Rantala, A. and Frigo, M.},
  year = 2021,
  month = jan,
  journal = {MNRAS},
  volume = {500},
  pages = {1437--1465},
  issn = {0035-8711},
  doi = {10.1093/mnras/staa3014}
}

@article{NFW1997,
  title = {A {{Universal Density Profile}} from {{Hierarchical Clustering}}},
  author = {Navarro, Julio F. and Frenk, Carlos S. and White, Simon D. M.},
  year = 1997,
  month = dec,
  journal = {ApJ},
  volume = {490},
  pages = {493--508},
  publisher = {IOP},
  issn = {0004-637X},
  doi = {10.1086/304888}
}

@article{Nowak_etal2007,
  title = {The Supermassive Black Hole in {{NGC4486a}} Detected with {{SINFONI}} at the {{Very Large Telescope}}},
  author = {Nowak, N. and Saglia, R. P. and Thomas, J. and Bender, R. and Pannella, M. and Gebhardt, K. and Davies, R. I.},
  year = 2007,
  month = aug,
  journal = {MNRAS},
  volume = {379},
  pages = {909--914},
  publisher = {OUP},
  issn = {0035-8711},
  doi = {10.1111/j.1365-2966.2007.11949.x}
}

@article{Nowak_etal2008a,
  title = {The Supermassive Black Hole of {{FornaxA}}},
  author = {Nowak, N. and Saglia, R. P. and Thomas, J. and Bender, R. and Davies, R. I. and Gebhardt, K.},
  year = 2008,
  month = dec,
  journal = {MNRAS},
  volume = {391},
  pages = {1629--1649},
  publisher = {OUP},
  issn = {0035-8711},
  doi = {10.1111/j.1365-2966.2008.13960.x}
}

@article{Pfenniger_etal2023,
  title = {Five Methods for Determining Pattern Speeds in Galaxies. {{I}}. {{Methods}}},
  author = {Pfenniger, Daniel and Saha, Kanak and Wu, Yu-Ting},
  year = 2023,
  month = may,
  journal = {A\&A},
  volume = {673},
  pages = {A36},
  issn = {0004-6361},
  doi = {10.1051/0004-6361/202245463}
}

@article{Pfenniger1984a,
  title = {The Velocity Fields of Barred Galaxies.},
  author = {Pfenniger, D.},
  year = 1984,
  month = dec,
  journal = {A\&A},
  volume = {141},
  pages = {171--188},
  publisher = {EDP},
  issn = {0004-6361},
  url = {https://ui.adsabs.harvard.edu/abs/1984A&A...141..171P}
}

@article{Pilawa_etal2024,
  title = {{{TriOS Schwarzschild Orbit Modeling}}: {{Robustness}} of {{Parameter Inference}} for {{Masses}} and {{Shapes}} of {{Triaxial Galaxies}} with {{Supermassive Black Holes}}},
  shorttitle = {{{TriOS Schwarzschild Orbit Modeling}}},
  author = {Pilawa, Jacob and Liepold, Emily R. and Ma, Chung-Pei},
  year = 2024,
  month = may,
  journal = {ApJ},
  volume = {966},
  pages = {205},
  publisher = {IOP},
  issn = {0004-637X},
  doi = {10.3847/1538-4357/ad3935}
}

@article{Pillepich_etal2018,
  title = {First Results from the {{IllustrisTNG}} Simulations: The Stellar Mass Content of Groups and Clusters of Galaxies},
  shorttitle = {First Results from the {{IllustrisTNG}} Simulations},
  author = {Pillepich, Annalisa and Nelson, Dylan and Hernquist, Lars and Springel, Volker and Pakmor, Rüdiger and Torrey, Paul and Weinberger, Rainer and Genel, Shy and Naiman, Jill P. and Marinacci, Federico and Vogelsberger, Mark},
  year = 2018,
  month = mar,
  journal = {MNRAS},
  volume = {475},
  pages = {648--675},
  publisher = {OUP},
  issn = {0035-8711},
  doi = {10.1093/mnras/stx3112}
}

@article{Pillepich_etal2019,
  title = {First Results from the {{TNG50}} Simulation: The Evolution of Stellar and Gaseous Discs across Cosmic Time},
  shorttitle = {First Results from the {{TNG50}} Simulation},
  author = {Pillepich, Annalisa and Nelson, Dylan and Springel, Volker and Pakmor, Rüdiger and Torrey, Paul and Weinberger, Rainer and Vogelsberger, Mark and Marinacci, Federico and Genel, Shy and {van der Wel}, Arjen and Hernquist, Lars},
  year = 2019,
  month = dec,
  journal = {MNRAS},
  volume = {490},
  pages = {3196--3233},
  publisher = {OUP},
  issn = {0035-8711},
  doi = {10.1093/mnras/stz2338}
}

@article{Richstone_Tremaine1985,
  title = {Dynamical Models of {{M}} 87 without a Central Black Hole.},
  author = {Richstone, D. O. and Tremaine, S.},
  year = 1985,
  month = sep,
  journal = {ApJ},
  volume = {296},
  pages = {370--378},
  publisher = {IOP},
  issn = {0004-637X},
  doi = {10.1086/163455}
}

@article{Richstone_Tremaine1988,
  title = {Maximum-{{Entropy Models}} of {{Galaxies}}},
  author = {Richstone, Douglas O. and Tremaine, Scott},
  year = 1988,
  month = apr,
  journal = {ApJ},
  volume = {327},
  pages = {82},
  publisher = {IOP},
  issn = {0004-637X},
  doi = {10.1086/166171}
}

@article{Rix_etal1997,
  title = {Dynamical {{Modeling}} of {{Velocity Profiles}}: {{The Dark Halo}} around the {{Elliptical Galaxy NGC}} 2434},
  shorttitle = {Dynamical {{Modeling}} of {{Velocity Profiles}}},
  author = {Rix, Hans-Walter and {de Zeeuw}, P. Tim and Cretton, Nicolas and {van der Marel}, Roeland P. and Carollo, C. Marcella},
  year = 1997,
  month = oct,
  journal = {ApJ},
  volume = {488},
  pages = {702--719},
  publisher = {IOP},
  issn = {0004-637X},
  doi = {10.1086/304733}
}

@article{Roshan_etal2021,
  title = {Fast Galaxy Bars Continue to Challenge Standard Cosmology},
  author = {Roshan, Mahmood and Ghafourian, Neda and Kashfi, Tahere and Banik, Indranil and Haslbauer, Moritz and Cuomo, Virginia and Famaey, Benoit and Kroupa, Pavel},
  year = 2021,
  month = nov,
  journal = {MNRAS},
  volume = {508},
  number = {1},
  pages = {926--939},
  issn = {0035-8711},
  doi = {10.1093/mnras/stab2553}
}

@article{Roshan_etal2025,
  title = {The {{Tremaine-Weinberg}} Method at High Redshifts},
  author = {Roshan, Mahmood and Habibi, Asiyeh and Aguerri, J. Alfonso L. and Cuomo, Virginia and Bottrell, Connor and Costantin, Luca and Corsini, Enrico Maria and Kim, Taehyun and Lee, Yun Hee and {Mendez-Abreu}, Jairo and Frosst, Matthew and {de Lorenzo-Cáceres}, Adriana and Morelli, Lorenzo and Pizzella, Alessandro},
  year = 2025,
  month = sep,
  journal = {A\&A},
  volume = {701},
  pages = {A160},
  publisher = {EDP},
  issn = {0004-6361},
  doi = {10.1051/0004-6361/202555547}
}

@article{Rusli_etal2013,
  title = {Depleted {{Galaxy Cores}} and {{Dynamical Black Hole Masses}}},
  author = {Rusli, S. P. and Erwin, P. and Saglia, R. P. and Thomas, J. and Fabricius, M. and Bender, R. and Nowak, N.},
  year = 2013,
  month = dec,
  journal = {ApJ},
  volume = {146},
  pages = {160},
  publisher = {IOP},
  issn = {0004-6256},
  doi = {10.1088/0004-6256/146/6/160}
}

@article{Saha_etal2018,
  title = {Building the {{Peanut}}: {{Simulations}} and {{Observations}} of {{Peanut-shaped Structures}} and {{Ansae}} in {{Face-on Disk Galaxies}}},
  shorttitle = {Building the {{Peanut}}},
  author = {Saha, Kanak and Graham, Alister W. and {Rodríguez-Herranz}, Isabel},
  year = 2018,
  month = jan,
  journal = {ApJ},
  volume = {852},
  pages = {133},
  publisher = {IOP},
  issn = {0004-637X},
  doi = {10.3847/1538-4357/aa9ed8}
}

@article{Sanchez_etal2012,
  title = {{{CALIFA}}, the {{Calar Alto Legacy Integral Field Area}} Survey. {{I}}. {{Survey}} Presentation},
  author = {Sánchez, S. F. and Kennicutt, R. C. and {Gil de Paz}, A. and {van de Ven}, G. and Vílchez, J. M. and Wisotzki, L. and Walcher, C. J. and Mast, D. and Aguerri, J. A. L. and {Albiol-Pérez}, S. and {Alonso-Herrero}, A. and Alves, J. and Bakos, J. and Bartáková, T. and {Bland-Hawthorn}, J. and Boselli, A. and Bomans, D. J. and {Castillo-Morales}, A. and {Cortijo-Ferrero}, C. and {de Lorenzo-Cáceres}, A. and Del Olmo, A. and Dettmar, R. -J. and Díaz, A. and Ellis, S. and {Falcón-Barroso}, J. and Flores, H. and Gallazzi, A. and {García-Lorenzo}, B. and González Delgado, R. and Gruel, N. and Haines, T. and Hao, C. and Husemann, B. and {Iglésias-Páramo}, J. and Jahnke, K. and Johnson, B. and Jungwiert, B. and Kalinova, V. and Kehrig, C. and Kupko, D. and {López-Sánchez}, Á. R. and Lyubenova, M. and Marino, R. A. and {Mármol-Queraltó}, E. and Márquez, I. and Masegosa, J. and Meidt, S. and {Mendez-Abreu}, J. and {Monreal-Ibero}, A. and Montijo, C. and Mourão, A. M. and {Palacios-Navarro}, G. and Papaderos, P. and Pasquali, A. and Peletier, R. and Pérez, E. and Pérez, I. and Quirrenbach, A. and Relaño, M. and {Rosales-Ortega}, F. F. and Roth, M. M. and {Ruiz-Lara}, T. and {Sánchez-Blázquez}, P. and Sengupta, C. and Singh, R. and Stanishev, V. and Trager, S. C. and Vazdekis, A. and Viironen, K. and Wild, V. and Zibetti, S. and Ziegler, B.},
  year = 2012,
  month = feb,
  journal = {A\&A},
  volume = {538},
  pages = {A8},
  issn = {0004-6361},
  doi = {10.1051/0004-6361/201117353}
}

@article{Schaye_etal2015,
  title = {The {{EAGLE}} Project: Simulating the Evolution and Assembly of Galaxies and Their Environments},
  shorttitle = {The {{EAGLE}} Project},
  author = {Schaye, Joop and Crain, Robert A. and Bower, Richard G. and Furlong, Michelle and Schaller, Matthieu and Theuns, Tom and Dalla Vecchia, Claudio and Frenk, Carlos S. and McCarthy, I. G. and Helly, John C. and Jenkins, Adrian and {Rosas-Guevara}, Y. M. and White, Simon D. M. and Baes, Maarten and Booth, C. M. and Camps, Peter and Navarro, Julio F. and Qu, Yan and Rahmati, Alireza and Sawala, Till and Thomas, Peter A. and Trayford, James},
  year = 2015,
  month = jan,
  journal = {MNRAS},
  volume = {446},
  pages = {521--554},
  publisher = {OUP},
  issn = {0035-8711},
  doi = {10.1093/mnras/stu2058}
}

@article{Schwarzschild1979,
  title = {A Numerical Model for a Triaxial Stellar System in Dynamical Equilibrium.},
  author = {Schwarzschild, M.},
  year = 1979,
  month = aug,
  journal = {ApJ},
  volume = {232},
  pages = {236--247},
  publisher = {IOP},
  issn = {0004-637X},
  doi = {10.1086/157282}
}

@article{Sellwood2014,
  title = {Secular Evolution in Disk Galaxies},
  author = {Sellwood, J. A.},
  year = 2014,
  month = jan,
  journal = {Reviews of Modern Physics},
  volume = {86},
  pages = {1--46},
  issn = {0034-6861},
  doi = {10.1103/RevModPhys.86.1}
}

@article{Semczuk_etal2024,
  title = {Pattern Speed Evolution of Barred Galaxies in {{TNG50}}},
  author = {Semczuk, Marcin and Dehnen, Walter and Schönrich, Ralph and Athanassoula, E.},
  year = 2024,
  month = dec,
  journal = {A\&A},
  volume = {692},
  pages = {A159},
  publisher = {EDP},
  issn = {0004-6361},
  doi = {10.1051/0004-6361/202451521}
}

@article{Sheth_etal2008,
  title = {Evolution of the {{Bar Fraction}} in {{COSMOS}}: {{Quantifying}} the {{Assembly}} of the {{Hubble Sequence}}},
  shorttitle = {Evolution of the {{Bar Fraction}} in {{COSMOS}}},
  author = {Sheth, Kartik and Elmegreen, Debra Meloy and Elmegreen, Bruce G. and Capak, Peter and Abraham, Roberto G. and Athanassoula, E. and Ellis, Richard S. and Mobasher, Bahram and Salvato, Mara and Schinnerer, Eva and Scoville, Nicholas Z. and Spalsbury, Lori and Strubbe, Linda and Carollo, Marcella and Rich, Michael and West, Andrew A.},
  year = 2008,
  month = mar,
  journal = {ApJ},
  volume = {675},
  number = {2},
  pages = {1141--1155},
  issn = {0004-637X},
  doi = {10.1086/524980},
  langid = {english}
}

@article{Smirnov_etal2021,
  title = {Face-on Structure of Barlenses and Boxy Bars: An Insight from Spectral Dynamics},
  shorttitle = {Face-on Structure of Barlenses and Boxy Bars},
  author = {Smirnov, Anton A. and Tikhonenko, Iliya S. and Sotnikova, Natalia Ya},
  year = 2021,
  month = apr,
  journal = {MNRAS},
  volume = {502},
  pages = {4689--4707},
  issn = {0035-8711},
  doi = {10.1093/mnras/stab327}
}

@article{Smirnov_Sotnikova2018,
  title = {What Determines the Flatness of {{X-shaped}} Structures in Edge-on Galaxies?},
  author = {Smirnov, Anton A. and Sotnikova, Natalia Ya},
  year = 2018,
  month = dec,
  journal = {MNRAS},
  volume = {481},
  pages = {4058--4076},
  issn = {0035-8711},
  doi = {10.1093/mnras/sty2423}
}

@article{Springel_etal2018,
  title = {First Results from the {{IllustrisTNG}} Simulations: Matter and Galaxy Clustering},
  shorttitle = {First Results from the {{IllustrisTNG}} Simulations},
  author = {Springel, Volker and Pakmor, Rüdiger and Pillepich, Annalisa and Weinberger, Rainer and Nelson, Dylan and Hernquist, Lars and Vogelsberger, Mark and Genel, Shy and Torrey, Paul and Marinacci, Federico and Naiman, Jill},
  year = 2018,
  month = mar,
  journal = {MNRAS},
  volume = {475},
  number = {1},
  eprint = {1707.03397},
  pages = {676--698},
  issn = {0035-8711, 1365-2966},
  doi = {10.1093/mnras/stx3304},
  archiveprefix = {arXiv}
}

@article{Tahmasebzadeh_etal2022,
  title = {Orbit-Superposition {{Dynamical Modeling}} of {{Barred Galaxies}}},
  author = {Tahmasebzadeh, Behzad and Zhu, Ling and Shen, Juntai and Gerhard, Ortwin and {van de Ven}, Glenn},
  year = 2022,
  month = dec,
  journal = {ApJ},
  volume = {941},
  number = {2},
  pages = {109},
  issn = {0004-637X, 1538-4357},
  doi = {10.3847/1538-4357/ac9df6},
  langid = {english}
}

@article{Tahmasebzadeh_etal2024,
  title = {Schwarzschild Modelling of Barred S0 Galaxy {{NGC}} 4371},
  author = {Tahmasebzadeh, Behzad and Zhu, Ling and Shen, Juntai and Gadotti, Dimitri A. and Valluri, Monica and Thater, Sabine and {van de Ven}, Glenn and Jin, Yunpeng and Gerhard, Ortwin and Erwin, Peter and Jethwa, Prashin and Zocchi, Alice and Lilley, Edward J. and Fragkoudi, Francesca and {de Lorenzo-Cáceres}, Adriana and {Méndez-Abreu}, Jairo and Neumann, Justus and Guo, Rui},
  year = 2024,
  month = oct,
  journal = {MNRAS},
  volume = {534},
  pages = {861--882},
  publisher = {OUP},
  issn = {0035-8711},
  doi = {10.1093/mnras/stae2109}
}

@inproceedings{Teuben1995,
  title = {The Stellar Dynamics Toolbox {{NEMO}}},
  shorttitle = {Astronomical Data Analysis Software and Systems {{IV}}},
  booktitle = {Astron. {{Data Anal}}. {{Softw}}. {{Syst}}. {{IV}}},
  author = {Teuben, P.},
  editor = {Shaw, R. A. and Payne, H. E. and Hayes, J. J. E.},
  year = 1995,
  month = jan,
  series = {Astronomical Society of the Pacific Conference Series},
  volume = {77},
  pages = {398}
}

@article{Thater_etal2023,
  title = {Effect of the Initial Mass Function on the Dynamical {{SMBH}} Mass Estimate in the Nucleated Early-Type Galaxy {{FCC}} 47},
  author = {Thater, Sabine and Lyubenova, Mariya and Fahrion, Katja and {Martín-Navarro}, Ignacio and Jethwa, Prashin and Nguyen, Dieu D. and {van de Ven}, Glenn},
  year = 2023,
  month = jul,
  journal = {A\&A},
  volume = {675},
  pages = {A18},
  publisher = {EDP},
  issn = {0004-6361},
  doi = {10.1051/0004-6361/202245362}
}

@misc{Thater_etal2026,
  title = {Supermassive Black Holes in Six Triaxial Galaxies: {{Insights}} from {{SINFONI}} and {{MUSE}} Observations},
  shorttitle = {Supermassive Black Holes in Six Triaxial Galaxies},
  author = {Thater, Sabine and Chaturvedi, Avinash and Krajnovic, Davor and Cappellari, Michele and Khochfar, Sadegh and Naab, Thorsten and Sarzi, Marc and {van de Ven}, Glenn},
  year = 2026,
  month = may,
  publisher = {arXiv},
  doi = {10.48550/arXiv.2605.28959}
}

@article{Thomas_etal2004,
  title = {Mapping Stationary Axisymmetric Phase-Space Distribution Functions by Orbit Libraries},
  author = {Thomas, J. and Saglia, R. P. and Bender, R. and Thomas, D. and Gebhardt, K. and Magorrian, J. and Richstone, D.},
  year = 2004,
  month = sep,
  journal = {MNRAS},
  volume = {353},
  pages = {391--404},
  issn = {0035-8711},
  doi = {10.1111/j.1365-2966.2004.08072.x}
}

@article{Thomas_etal2005,
  title = {Regularized Orbit Models Unveiling the Stellar Structure and Dark Matter Halo of the {{Coma}} Elliptical {{NGC}} 4807},
  author = {Thomas, J. and Saglia, R. P. and Bender, R. and Thomas, D. and Gebhardt, K. and Magorrian, J. and Corsini, E. M. and Wegner, G.},
  year = 2005,
  month = jul,
  journal = {MNRAS},
  volume = {360},
  pages = {1355--1372},
  publisher = {OUP},
  issn = {0035-8711},
  doi = {10.1111/j.1365-2966.2005.09139.x}
}

@article{Thomas_etal2007,
  title = {Dynamical Modelling of Luminous and Dark Matter in 17 {{Coma}} Early-Type Galaxies},
  author = {Thomas, Jens and Saglia, R. and Bender, R. and Thomas, D. and Gebhardt, K. and Magorrian, J. and Corsini, E. M. and Wegner, G.},
  year = 2007,
  month = dec,
  journal = {MNRAS},
  volume = {382},
  number = {2},
  pages = {657--684},
  issn = {0035-8711},
  doi = {10.1111/j.1365-2966.2007.12434.x}
}

@article{Thomas_Lipka2022,
  title = {A Simple Data-Driven Method to Optimize the Penalty Strengths of Penalized Models and Its Application to Non-Parametric Smoothing},
  author = {Thomas, Jens and Lipka, Mathias},
  year = 2022,
  month = aug,
  journal = {MNRAS},
  volume = {514},
  pages = {6203--6214},
  issn = {0035-8711},
  doi = {10.1093/mnras/stac1581}
}

@article{Tremaine_Weinberg1984,
  title = {A Kinematic Method for Measuring the Pattern Speed of Barred Galaxies.},
  author = {Tremaine, S. and Weinberg, M. D.},
  year = 1984,
  month = jul,
  journal = {ApJ},
  volume = {282},
  pages = {L5-L7},
  publisher = {IOP},
  issn = {0004-637X},
  doi = {10.1086/184292}
}

@article{Valluri_etal2004,
  title = {Difficulties with {{Recovering}} the {{Masses}} of {{Supermassive Black Holes}} from {{Stellar Kinematical Data}}},
  author = {Valluri, Monica and Merritt, David and Emsellem, Eric},
  year = 2004,
  month = feb,
  journal = {ApJ},
  volume = {602},
  number = {1},
  pages = {66},
  publisher = {IOP Publishing},
  issn = {0004-637X},
  doi = {10.1086/380896},
  langid = {english}
}

@article{Valluri_etal2005,
  title = {The {{Low End}} of the {{Supermassive Black Hole Mass Function}}: {{Constraining}} the {{Mass}} of a {{Nuclear Black Hole}} in {{NGC}} 205 via {{Stellar Kinematics}}},
  shorttitle = {The {{Low End}} of the {{Supermassive Black Hole Mass Function}}},
  author = {Valluri, Monica and Ferrarese, Laura and Merritt, David and Joseph, Charles L.},
  year = 2005,
  month = jul,
  journal = {ApJ},
  volume = {628},
  pages = {137--152},
  publisher = {IOP},
  issn = {0004-637X},
  doi = {10.1086/430752}
}

@article{vandenBosch_etal2008,
  title = {Triaxial Orbit Based Galaxy Models with an Application to the (Apparent) Decoupled Core Galaxy {{NGC}} 4365},
  author = {{van den Bosch}, Remco C. E. and {van de Ven}, Glenn and Verolme, E. K. and Cappellari, M. and {de Zeeuw}, P. T.},
  year = 2008,
  month = feb,
  journal = {MNRAS},
  volume = {385},
  number = {2},
  pages = {647--666},
  issn = {0035-8711},
  doi = {10.1111/j.1365-2966.2008.12874.x}
}

@article{vandenBosch_vandeVen2009,
  title = {Recovering the Intrinsic Shape of Early-Type Galaxies},
  author = {{van den Bosch}, Remco C. E. and {van de Ven}, Glenn},
  year = 2009,
  month = sep,
  journal = {MNRAS},
  volume = {398},
  number = {3},
  pages = {1117--1128},
  issn = {0035-8711},
  doi = {10.1111/j.1365-2966.2009.15177.x}
}

@misc{VandenHeuvel2024,
  title = {Naturalneighbours.Jl},
  author = {VandenHeuvel, Daniel John},
  year = 2024,
  month = jul,
  doi = {10.5281/zenodo.12683284},
  howpublished = {Zenodo}
}

@article{Vasiliev_Valluri2020,
  title = {A {{New Implementation}} of the {{Schwarzchild Method}} for {{Constructing Observationally Driven Dynamical Models}} of {{Galaxies}} of {{All Morphological Types}}},
  author = {Vasiliev, Eugene and Valluri, Monica},
  year = 2020,
  month = jan,
  journal = {ApJ},
  volume = {889},
  pages = {39},
  issn = {0004-637X},
  doi = {10.3847/1538-4357/ab5fe0}
}

@article{Wang_etal2012,
  title = {A New Model for the {{Milky Way}} Bar},
  author = {Wang, Yougang and Zhao, Hongsheng and Mao, Shude and Rich, R. M.},
  year = 2012,
  month = dec,
  journal = {MNRAS},
  volume = {427},
  pages = {1429--1440},
  publisher = {OUP},
  issn = {0035-8711},
  doi = {10.1111/j.1365-2966.2012.22063.x}
}

@article{Wang_etal2013,
  title = {A {{Schwarzschild}} Model of the {{Galactic}} Bar with Initial Density from {{N-body}} Simulations},
  author = {Wang, Yougang and Mao, Shude and Long, Richard J. and Shen, Juntai},
  year = 2013,
  month = nov,
  journal = {MNRAS},
  volume = {435},
  pages = {3437--3443},
  publisher = {OUP},
  issn = {0035-8711},
  doi = {10.1093/mnras/stt1537}
}

@article{Weinberg_Katz2007a,
  title = {The Bar-Halo Interaction - {{I}}. {{From}} Fundamental Dynamics to Revised {{N-body}} Requirements},
  author = {Weinberg, Martin D. and Katz, Neal},
  year = 2007,
  month = feb,
  journal = {MNRAS},
  volume = {375},
  pages = {425--459},
  publisher = {OUP},
  issn = {0035-8711},
  doi = {10.1111/j.1365-2966.2006.11306.x}
}

@article{Weinberg_Katz2007b,
  title = {The Bar-Halo Interaction - {{II}}. {{Secular}} Evolution and the Religion of {{N-body}} Simulations},
  author = {Weinberg, Martin D. and Katz, Neal},
  year = 2007,
  month = feb,
  journal = {MNRAS},
  volume = {375},
  pages = {460--476},
  publisher = {OUP},
  issn = {0035-8711},
  doi = {10.1111/j.1365-2966.2006.11307.x}
}

@article{Williams_etal2021,
  title = {Applying the {{Tremaine-Weinberg Method}} to {{Nearby Galaxies}}: {{Stellar-mass-based Pattern Speeds}} and {{Comparisons}} with {{ISM Kinematics}}},
  shorttitle = {Applying the {{Tremaine-Weinberg Method}} to {{Nearby Galaxies}}},
  author = {Williams, Thomas G. and Schinnerer, Eva and Emsellem, Eric and Meidt, Sharon and Querejeta, Miguel and Belfiore, Francesco and Bešlić, Ivana and Bigiel, Frank and Chevance, Mélanie and Dale, Daniel A. and Glover, Simon C. O. and Grasha, Kathryn and Klessen, Ralf S. and Kruijssen, J. M. Diederik and Leroy, Adam K. and Pan, Hsi-An and Pety, Jérôme and Pessa, Ismael and Rosolowsky, Erik and Saito, Toshiki and Santoro, Francesco and Schruba, Andreas and Sormani, Mattia C. and Sun, Jiayi and Watkins, Elizabeth J.},
  year = 2021,
  month = apr,
  journal = {AJ},
  volume = {161},
  pages = {185},
  publisher = {IOP},
  issn = {0004-6256},
  doi = {10.3847/1538-3881/abe243}
}

@article{Zakharova_etal2023,
  title = {B/{{PS}} Bulges and Barlenses from a Kinematic Viewpoint - {{I}}},
  author = {Zakharova, Daria and Tikhonenko, Iliya S. and Sotnikova, Natalia Ya and Smirnov, Anton A.},
  year = 2023,
  month = nov,
  journal = {MNRAS},
  volume = {525},
  pages = {6112--6129},
  issn = {0035-8711},
  doi = {10.1093/mnras/stad2662}
}

@article{Zakharova_etal2024,
  title = {B/{{PS}} Bulges and Barlenses from a Kinematic Viewpoint - {{II}}},
  author = {Zakharova, Daria and Tikhonenko, Iliya S. and Sotnikova, Natalia Ya and Smirnov, Anton A.},
  year = 2024,
  month = jan,
  journal = {MNRAS},
  volume = {527},
  pages = {3038--3053},
  issn = {0035-8711},
  doi = {10.1093/mnras/stad3468}
}

@article{Zhao_etal2020,
  title = {Barred {{Galaxies}} in the {{IllustrisTNG Simulation}}},
  author = {Zhao, Dongyao and Du, Min and Ho, Luis C. and Debattista, Victor P. and Shi, Jingjing},
  year = 2020,
  month = dec,
  journal = {ApJ},
  volume = {904},
  number = {2},
  pages = {170},
  issn = {0004-637X, 1538-4357},
  doi = {10.3847/1538-4357/abbe1b}
}

@article{Zhao1996b,
  title = {A Steady-State Dynamical Model for the {{COBE-detected Galactic}} Bar.},
  author = {Zhao, Hongsheng},
  year = 1996,
  month = nov,
  journal = {MNRAS},
  volume = {283},
  pages = {149--166},
  publisher = {OUP},
  issn = {0035-8711},
  doi = {10.1093/mnras/283.1.149}
}

@article{Zhu_etal2018,
  title = {Orbital Decomposition of {{CALIFA}} Spiral Galaxies},
  author = {Zhu, Ling and {van den Bosch}, Remco and {van de Ven}, Glenn and Lyubenova, Mariya and {Falcón-Barroso}, Jesús and Meidt, Sharon E. and Martig, Marie and Shen, Juntai and Li, Zhao-Yu and Yildirim, Akin and Walcher, C. Jakob and Sanchez, Sebastian F.},
  year = 2018,
  month = jan,
  journal = {MNRAS},
  volume = {473},
  number = {3},
  pages = {3000--3018},
  issn = {0035-8711},
  doi = {10.1093/mnras/stx2409}
}

@article{Zou_etal2019,
  title = {Testing the {{Tremaine-Weinberg Method Applied}} to {{Integral-field Spectroscopic Data Using}} a {{Simulated Barred Galaxy}}},
  author = {Zou, Yanfei and Shen, Juntai and Bureau, Martin and Li, Zhao-Yu},
  year = 2019,
  month = oct,
  journal = {ApJ},
  volume = {884},
  pages = {23},
  publisher = {IOP},
  issn = {0004-637X},
  doi = {10.3847/1538-4357/ab3f34}
}



\appendix

\section{Qualitative comparison of the dynamical pattern speed recovery with the Tremaine-Weinberg method}
\begin{figure*}
	\centering
	\includegraphics[width=0.75\linewidth]{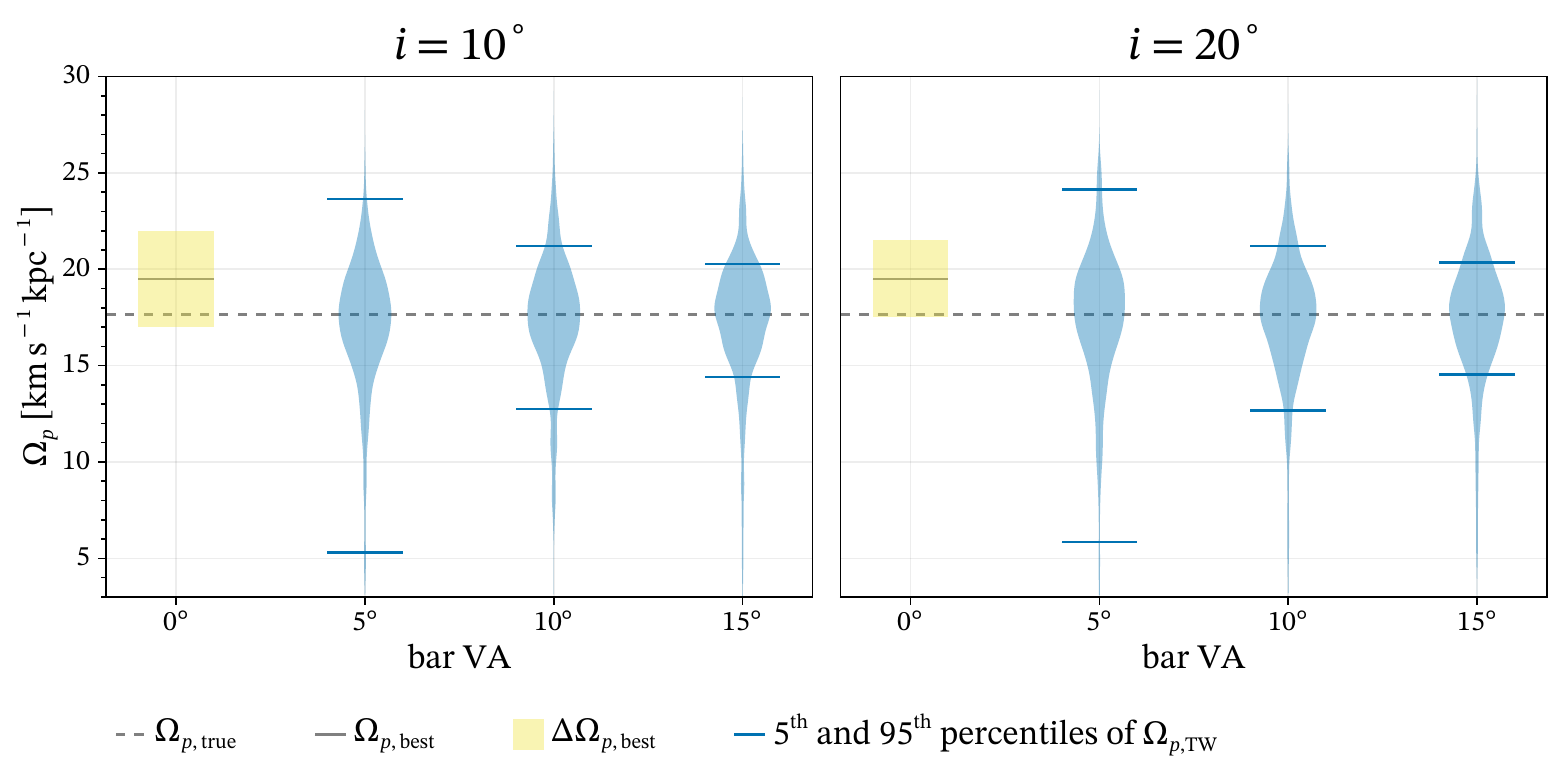}
	\caption{The comparison of the dynamical modeling pattern speed estimates (gray lines with yellow bands showing the uncertainties) for \texttt{I10} and \texttt{I20} mock data sets with Tremaine-Weinberg pattern speed estimates for the same inclination but different viewing angles of the bar (blue violins). We show 5th and 95th percentiles of the TW $\Omega_p$ distributions with solid blue lines. The reference N-body simulation value $17.7\ \text{km}\,\text{s}^{-1}\,\text{kpc}^{-1}$ is indicated with a dashed gray horizontal line.}
	\label{fig:TW-dyn-comp}
\end{figure*}

{It is an interesting question how well our results hold up with respect to the more conventional and well-studied Tremaine-Weinberg (TW) method.
In all three mock tests, we assume that the viewing angle of the bar is $0^{\circ}$, which means that strictly speaking it would be impossible to compare the results with TW, because the integrals in the expression for the pattern speed would vanish.
However, an exact side-on orientation is rather unlikely, meaning that we can just run the analysis for some small value of the bar viewing angle and compare its performance with our dynamical estimates. 
The exact face-on case would also be beyond the scope of TW due to $\sin i = 0$, therefore we consider only \texttt{I10} and \texttt{I20} cases in this Appendix.}

{We project the N-body particle data at $i=10^{\circ},20^{\circ}$ and $\text{bar VA} = 5^\circ, 10^\circ, 15^\circ$ and then construct mock mean velocity maps with MUSE-like pixel resolution (see \cref{sssec:mock-kinematics}). 
The $x$-axis of the map matches the major axis of the disc while the $z$-axis coincides with the LOS. We therefore position the pseudo-slits at different values of $y$ with a width of $\Delta y=0.2\ \text{kpc}$
and compute the surface-brightness-weighted $\langle v_\text{LOS}\rangle$ and $\langle x \rangle$. Following \citet{Merrifield_Kuijken1995}, we obtain $\Omega \sin i$ as the slope of a linear fit of $\langle v_\text{LOS}\rangle$ against $\langle x \rangle$. }

{The largest source of uncertainty in the final pattern speed value comes from the misalignment between the slits and the major axis of the disc \citep{Debattista2003,Zou_etal2019,Roshan_etal2025}. We model these errors by running 1000 Monte-Carlo (MC) simulations with \Nbody particles rotated in the sky plane by a normally distributed $\delta \text{PA}$ with a $\sigma_{\delta\text{PA}} = 1^\circ$ (see e.g. tab.~1 in \citealp{Geron_etal2023} for typical PA errors) before the mock maps are constructed. We also add a Gaussian noise
  to the mean velocities in each MC iteration simular to the noise that we see in GH fits of our kinematic data (\cref{fig:kinematic-profiles-errors-bestfit-i20}).
The obtained distributions of $\Omega_p$ compared to the dynamical modelling estimates are shown in \cref{fig:TW-dyn-comp}.}

{Even for these quite idealized measurements, where we assume that the inclination is known exactly and the errors of other parameters are as good as they can possibly be, the scatter of TW estimates rapidly increases when the bar viewing angle is approaching $0^\circ$.
  For example, for $\text{bar VA} = 5^{\circ}$ and $i=10^{\circ}$ the $[5\%;95\%]$ interval for the measured $\Omega_p$ would reach almost $[5;24]\ \text{km}\,\text{s}^{-1}\,\text{kpc}^{-1}$. This result is expected and well described in TW studies \citep[see e.g. fig.~16 of][]{Zou_etal2019}.
The dynamical modelling estimate, by comparison, is much more robust with respect to the orientation (see also the discussion in \citealp{Vasiliev_Valluri2020} or \citealp{Tahmasebzadeh_etal2024}), which implies that the dynamical modelling can be used to complement the existing approaches where their results can not be considered reliable.}

\section{Marginal distributions for \texttt{I0} and \texttt{I10} cases}
\label{sec:i0i10-par-recovery}

In this section, we provide marginal parameter distributions similar to \cref{fig:marginaldists-i20} presented in \cref{ssec:marginal-dists} for the rest of our inclinations cases for completeness.
In the exact face-on case, the 1-dimensional marginal distribution of dark halo scaling factor is essentially flat, with a tentative minimum at $s_{\text{DM}} = 0.75$, while at $i=10^\circ$ the range of the $\chi^2$ is larger. As outlined in \cref{ssec:marginal-dists}, the recovery of halo scaling is largely affected by the noise in the wings of the \acronym{LOSVD}s and becomes increasingly more challenging when the inclination approaches $0$, because the \acronym{LOSVD}s become more narrow. This also results in the 2D $(\Upsilon, s_{\text{DM}})$ distribution becoming increasingly degenerate with decreasing inclination, which contributes to the larger uncertainty values for $s_{\text{DM}}$ in \cref{fig:3dsearch-relerr-i0} compared to the $i=20^{\circ}$ case.

\begin{figure*}
	\centering
	\includegraphics[width=\linewidth]{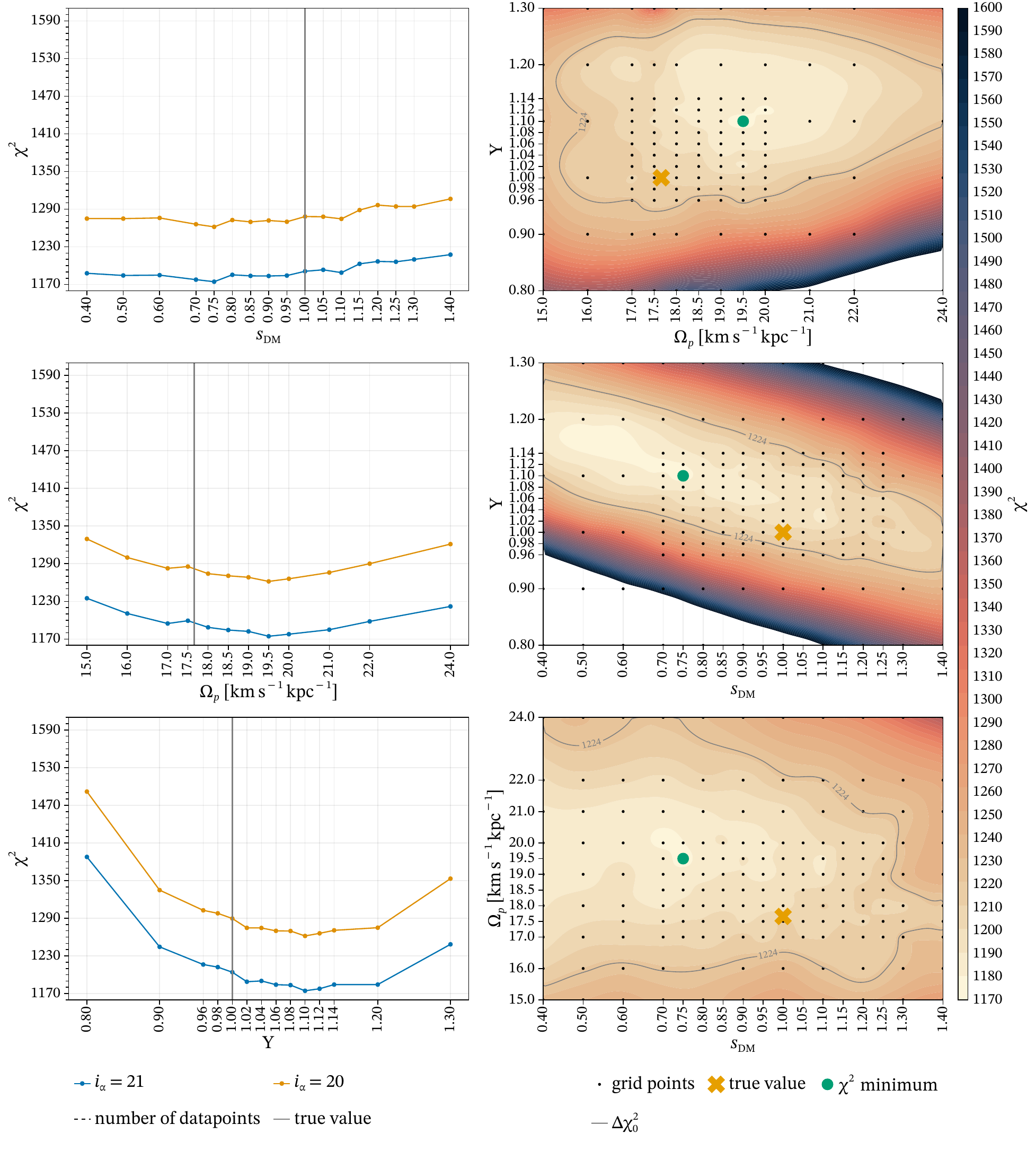}
	\caption{Same as \cref{fig:marginaldists-i20} but for $i=0^\circ$.}
	\label{fig:marginaldists-i10}
\end{figure*}

\begin{figure*}
	\centering
	\includegraphics[width=\linewidth]{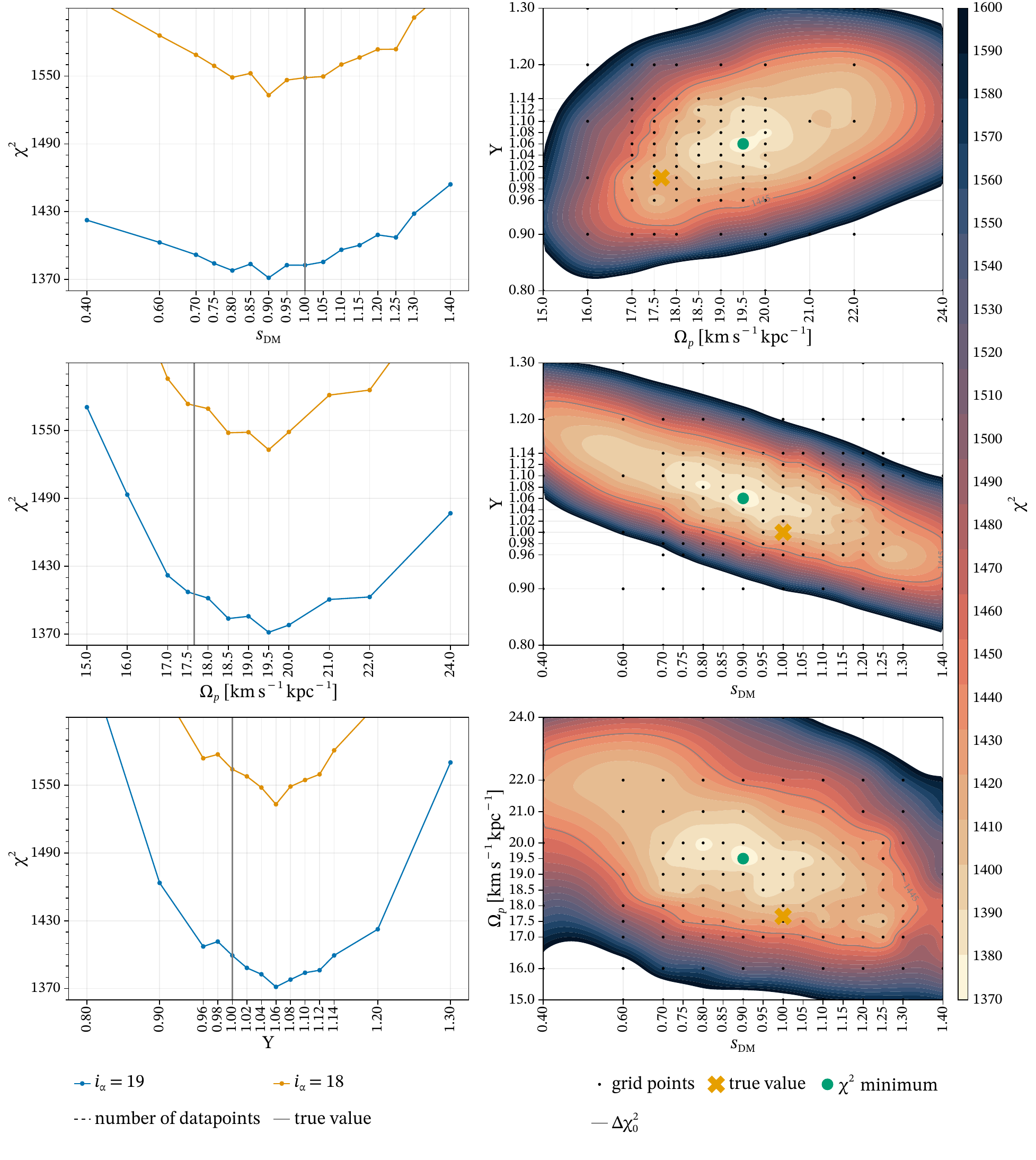}
	\caption{Same as \cref{fig:marginaldists-i20} but for $i=10^\circ$.}
	\label{fig:marginaldists-i0}
\end{figure*}

\section{GH parameter maps for \texttt{I0} and \texttt{I10} cases}
\label{sec:gh-i0-i10}

{For completeness, we also include the maps of Gauss-Hermite moments as well as their residual profiles (similar to \cref{fig:kinematic-maps-bestfit-i20} and \cref{fig:kinematic-profiles-errors-bestfit-i20} respectively) for the rest of the modelled inclinations (\texttt{I10} and \texttt{I0}). The LOS velocities and dispersions match the simulation data very accurately despite the noise in the mock data. The $h_4$ however is slightly overestimated for the outer Voronoi bins in the exact face-on case due to the width of the LOSVD peak being comparable to the resolution of the velocity grid. This resolution effect can also be seen in the uncertainty profiles of GH parameters in \cref{fig:kinematic-profiles-errors-bestfit-i0}, however this effect is not pronounced in the bar region, therefore it does not affect our discussion in \cref{ssec:face-on-patter-speed-recovery}.}

\begin{figure*}
  \centering
  \includegraphics[width=\linewidth]{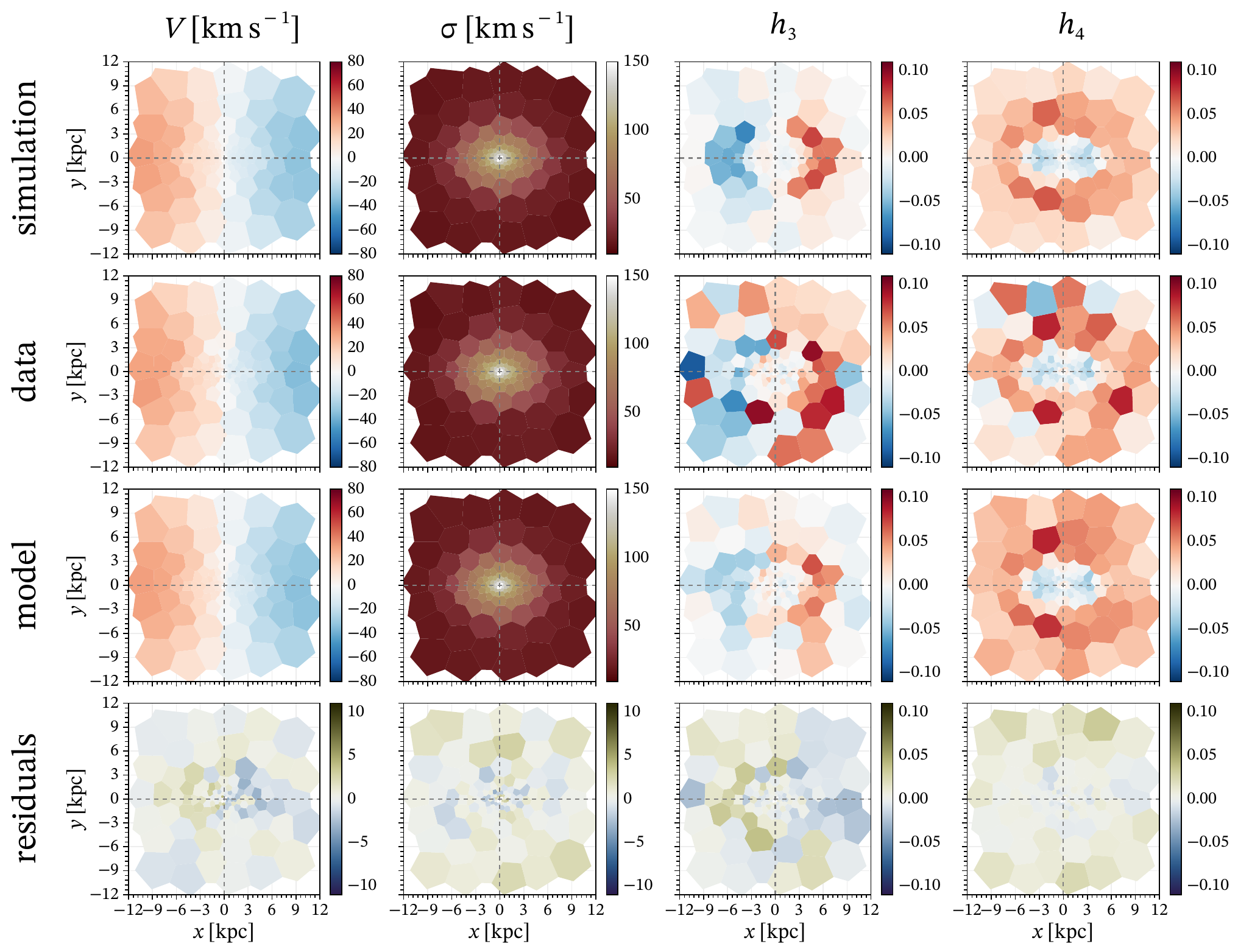}
  \caption{Maps of the Gauss-Hermite parameters ($V$, $\sigma$, $h_3$, $h_4$) and their residuals for inclination $10^{\circ}$ (similar to \cref{fig:kinematic-maps-bestfit-i20}).}
  \label{fig:kinematic-maps-bestfit-i10}
\end{figure*}

\begin{figure*}
  \centering
  \includegraphics[width=\linewidth]{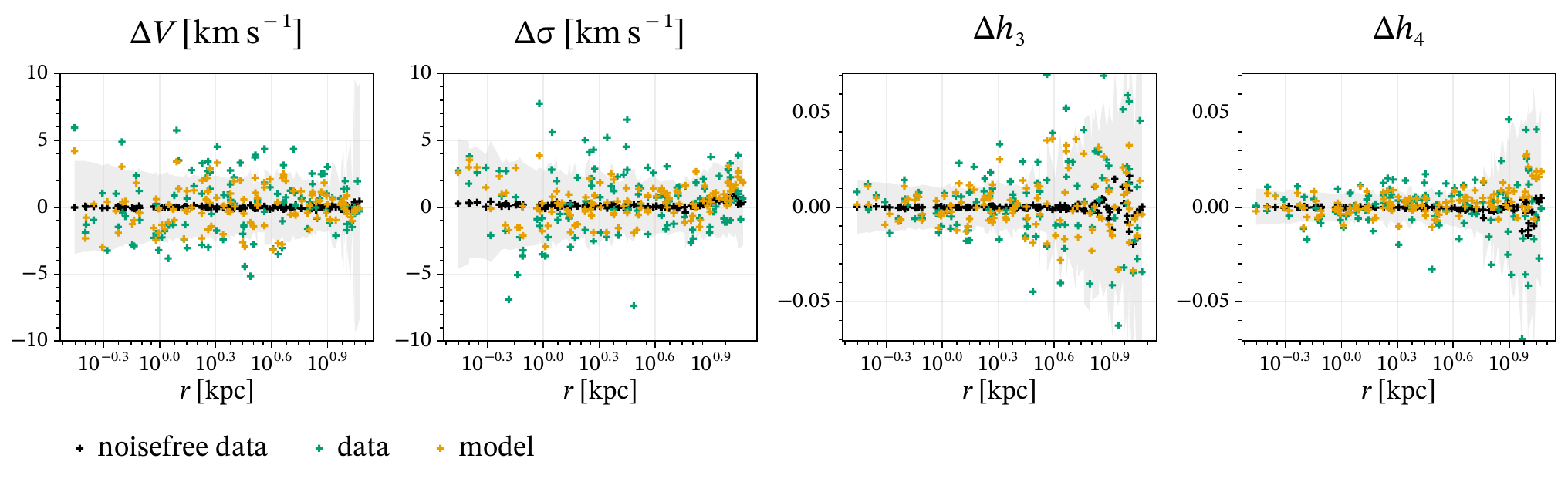}
	\caption{Profiles of the Gauss-Hermite residuals ($\Delta V$, $\Delta \sigma$, $\Delta h_3$, $\Delta h_4$) with respect to the true N-body values for inclination $10^{\circ}$ (similar to \cref{fig:kinematic-profiles-errors-bestfit-i20}). }
  \label{fig:kinematic-profiles-errors-bestfit-i10}
\end{figure*}

\begin{figure*}
  \centering
  \includegraphics[width=\linewidth]{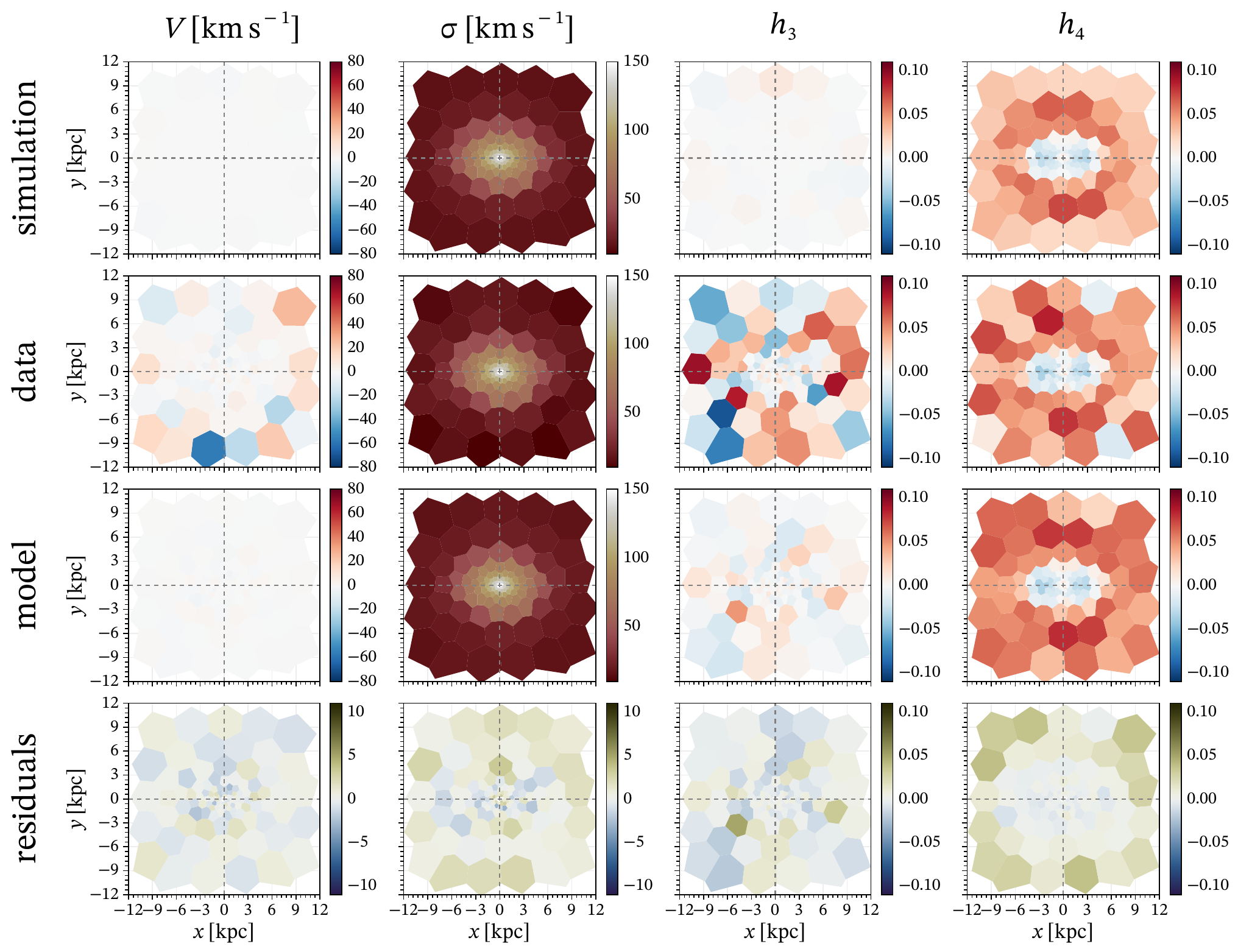}
  \caption{Maps of the Gauss-Hermite parameters ($V$, $\sigma$, $h_3$, $h_4$) and their residuals for inclination $0^{\circ}$ (similar to \cref{fig:kinematic-maps-bestfit-i20}).}
  \label{fig:kinematic-maps-bestfit-i0}
\end{figure*}

\begin{figure*}
  \centering
  \includegraphics[width=\linewidth]{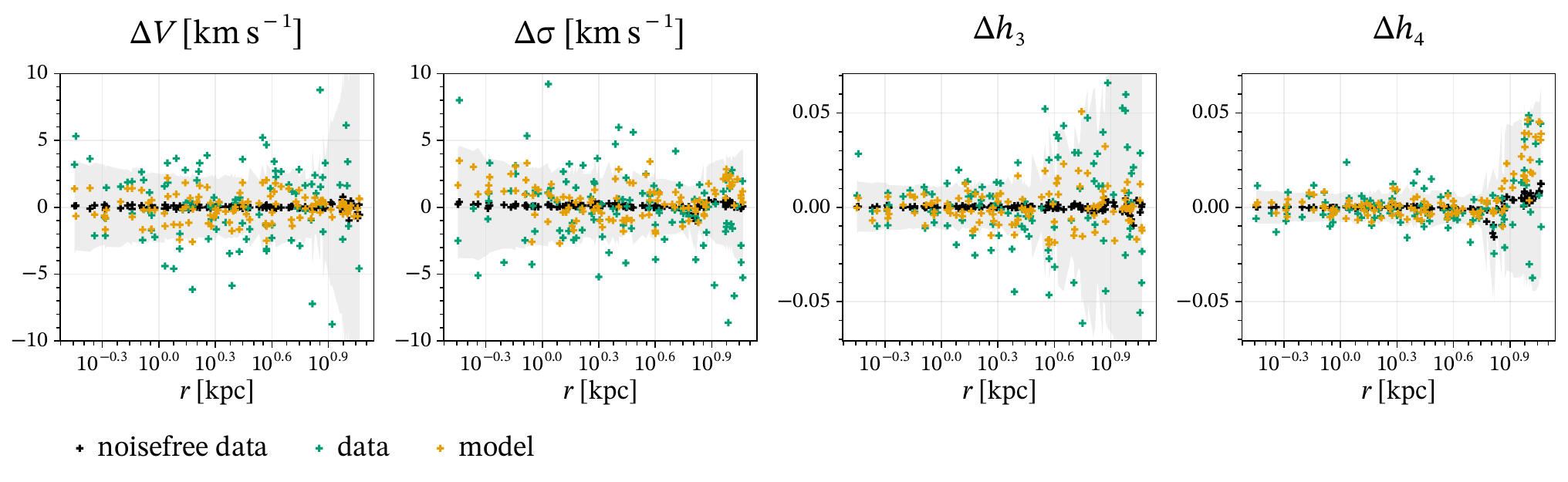}
	\caption{Profiles of the Gauss-Hermite residuals ($\Delta V$, $\Delta \sigma$, $\Delta h_3$, $\Delta h_4$) with respect to the true N-body values for inclination $0^{\circ}$ (similar to \cref{fig:kinematic-profiles-errors-bestfit-i20}). }
  \label{fig:kinematic-profiles-errors-bestfit-i0}
\end{figure*}


\bsp	
\label{lastpage}
\end{document}